\documentclass[fleqn,usenatbib]{mnras}

\usepackage{newtxtext,newtxmath}

\usepackage[T1]{fontenc}

\DeclareRobustCommand{\VAN}[3]{#2}
\let\VANthebibliography\thebibliography
\def\thebibliography{\DeclareRobustCommand{\VAN}[3]{##3}\VANthebibliography}

\usepackage{graphicx}	
\usepackage{amsmath}	
\usepackage[utf8]{inputenc}
\usepackage[T1]{fontenc}
\usepackage[justification=centering]{caption}
\usepackage{xcolor}
\newcommand{\preetish}[1]{#1}

\title[Galaxy size-halo radius relation]{Stellar mass dependence of galaxy size-dark matter halo radius relation probed by Subaru-HSC survey weak lensing measurements}

\author[P. K. Mishra et al.]{
Preetish K. Mishra,$^{1}$\thanks{E-mail: preetish@iucaa.in (PKM)}
Divya Rana,$^{1}$\thanks{E-mail: divyar@iucaa.in (DR)}
Surhud More$^{1,2}$\thanks{E-mail: surhud@iucaa.in (SM)}
\\
$^{1}$Inter-University Centre for Astronomy and Astrophysics, Ganeshkhind, Pune 411007, India\\
$^{2}$Kavli Institute for the Physics and Mathematics of the Universe (WPI), University of Tokyo, 5-1-5, Kashiwanoha, 2778583, Japan
}

\date{Accepted XXX. Received YYY; in original form ZZZ}

\pubyear{2022}

\begin{document}
\label{firstpage}
\pagerange{\pageref{firstpage}--\pageref{lastpage}}
\maketitle

\begin{abstract}
We investigate the stellar mass-dependence of the galaxy size-dark matter halo radius relation for low redshift galaxies using weak gravitational lensing measurements. Our sample consists of $\sim$38,000 galaxies more massive than $10^{8}{\rm M}_{\odot}h^{-2}$ and within $z<0.3$ drawn from the overlap of GAMA survey DR4 and HSC-SSP PDR2. We divide our sample into a number of stellar mass bins and measure stacked weak lensing signals. We model the signals using a conditional stellar mass function to infer the stellar mass-halo mass relation. We fit a single S\'ersic model to HSC $i$-band images of our galaxies and obtain their three-dimensional half-light radii. We use these measurements to construct a median galaxy size-mass relation. We then combine these relations to infer the galaxy size-halo radius relation. \preetish{We confirm that this relation appears linear given the statistical errors, i.e. the ratio of galaxy size to halo radius remains constant over two orders of magnitudes in stellar mass above $\sim 10^{9} {\rm M}_{\odot}h^{-2}$. Extrapolating the stellar mass-halo mass relation below this limit, we see an indication of a decreasing galaxy size-halo radius ratio with the decline in stellar mass. At stellar mass $\sim 10^{8} {\rm M}_{\odot}h^{-2}$ the ratio becomes 30\% smaller than its value in linear regime. The possible existence of a such trend in dwarf galaxy sectors calls for either modification in models employing a constant fraction of halo angular momentum transferred to explain sizes of dwarfs
or else points towards our lack of knowledge about dark matter haloes of low-mass galaxies.}

\end{abstract}




\section{Introduction}
According to our current understanding of galaxy formation, galaxies form via the cosmic infall of gas on to dark matter haloes \citep{White1978}. This gas settles at the center of the dark matter halo potential and cools down to form the stars in a galaxy. Therefore, we expect that some of the basic properties of galaxies will be set by the properties of dark matter haloes themselves. Over cosmic time the properties of galaxies evolve continuously and result in a diverse population of galaxies that we observe today. Galaxy formation models aim to accurately describe the properties of galaxies and their evolution as a function of cosmic time via a number of analytic/semi-analytic \citep[see e.g.,][]{White1991, Kauffmann1993, Cole1994, Somerville1999, Guo2013, Hernquues2015}, empirical \citep[see e.g.,][]{Peacock2000, Bullock2002, Berlind2003, Yang2003, Conroy2009, Behroozi2013, Moster2013, Puebla2017, Behroozi2019}, and, numerical approaches \citep[see e.g.,][]{Genel2014, Vogelsberger2014, Schaye2015, Nelson2019}. In each of these models, dark matter haloes act as the prime formation sites of galaxies. It follows that these galaxy formation models predict connections between the properties of galaxies and dark matter haloes. Therefore, by studying these galaxy-halo connections we can test and revise existing models of galaxy formation.

One such prediction is regarding the sizes of galaxies and their host dark matter haloes. Classic galaxy formation models by \cite{Fall1980} and \cite{Mo1998} use the angular momentum conservation principle to assign sizes to disc galaxies. In the simplest of cases, these models predict the size of galaxies to be directly proportional to the spin and the size of their host dark matter haloes i.e., $r_{\rm gal}\sim\lambda R_{\rm halo}$. \cite{Kravtsov2013} used the abundance matching ansatz \citep[see for eg.,][]{Kravtsov2004, Vale2004, Conroy2006} to link nearby galaxies (within $z \sim 0.1$) observed in SDSS as well as in HST and VLT large programs to dark matter haloes whose abundance can be measured in cosmological simulations. \cite{Kravtsov2013} finds a linear relation between galaxy half-mass radius ($r_{\rm 1/2}$) and halo radius ($R_{\rm 200c}$) in the form $r_{1/2}$ = 0.015 $R_{\rm 200c}$ for the galaxies ranging all the way from low-mass dwarfs to high-mass early type galaxies.  This linear relationship was also found to be valid in a wide redshift range (0$\leq z \leq$3) by using the abundance matching technique in \cite{Huang2017}.  This study also reports that early and late-type galaxies follow parallel linear galaxy size-halo radius relations. The observed linearity of the relationship between galaxy size and halo radius was further strengthened by results from \cite{Somerville2018} who used the abundance matching ansatz to link galaxies from the Galaxy And Mass Assembly \citep[GAMA, ][]{Driver2011} and the Cosmic Assembly Near Infrared Deep Extragalactic Legacy
(of 0.01 as compared to 0.014) than that of central galaxies.Survey \citep[CANDELS, ][]{Grogin2011} surveys to dark matter haloes. They find that the ratio of galaxy 3D half-mass radius to halo virial radius is roughly independent of stellar mass and stays constant around $\sim$0.018 at $z\sim$0.1. For the low-mass galaxies, this size ratio was also found to mildly decrease over cosmic time (from $z$=3 to $z$=0.1). No evolution was found for the size ratio of early-type galaxies and their host haloes. \cite{Rodriguez2021} used group finder algorithms to link dark matter halo mass and their size to the sizes of central and satellite galaxies separately. Their result for galaxy size-halo radius relation for central galaxies agrees with \cite{Kravtsov2013} results. The satellites were also found to exhibit a linear galaxy size-halo radius relation albeit with a 30 percent smaller normalization (of 0.01 as compared to 0.014) than that of central galaxies. \preetish{Similar linear galaxy size-halo radius relation was also reported by \cite{Rohr2022} for low-mass ($\sim 10^{7 - 9}$ M$_{\odot}$) central galaxies using the FIRE suite of simulations.}

The existence of a universal linear scaling relation between galaxy size and halo radius has been used in previous studies to support a galaxy formation scenario where the sizes of galaxies are set by the spin and the extent of their host dark matter haloes. Many works, however, disagree with halo spin being an important ingredient in predicting galaxy sizes. Using the hydro-dynamical simulation EAGLE \citep{Schaye2015}, \cite{Desmond2017} find only a weak correlation between galaxy size and halo spin while \cite{Jiang2019} did not find linear scaling between 3d galaxy size and halo radius in NIHAO and VELA simulations. According to \cite{Jiang2019}, the halo concentration instead of galaxy spin plays an important role in setting galaxy sizes.  They find a relation between galaxy effective radius ($R_{\rm e}$), halo virial radius ($R_{\rm vir}$), and concentration ($c$) in the following form $R_{\rm e} = 0.02(c/10)^{-0.7}R_{\rm vir}$. \cite{Zanisi2020} modeled the observed scatter in the sizes of SDSS galaxies by assuming the validity of the galaxy size - halo radius connection coming from the works of \cite{Mo1998}, \cite{Kravtsov2013}, and \cite{Jiang2019}. They concluded that galaxy formation models where sizes are set by halo spin, fail to reproduce the observed scatter in galaxy size. The authors proposed a model where the angular momentum of galaxies rather than the halo spin is instrumental in steering galaxy size. Recently, \cite{Zhang2022} explored the galaxy-halo radius relation by linking SDSS-DR7 galaxies to dark matter haloes from the constrained simulation ELUCID using abundance matching. They find a power law relation between galaxy half mass radius ($r_{\rm 1/2}$) and halo virial radius in the form $r_{\rm 1/2} \propto R_{\rm vir}^{0.55}$. This relation is found to depend on the halo mass in a way that the ratio $r_{\rm 1/2}/R_{vir}$ decreases with an increase in halo mass. They also find no significant dependence of galaxy size on halo spin or concentration. 

The above results show that the galaxy-halo connection can be used to test and refine our understanding of galaxy formation. However, as is evident from the discussion in the previous paragraphs, there is no consensus on the nature of the relationship between galaxy size and halo radius. It must be noted that the majority of results on this topic have been obtained either using the abundance matching ansatz or are from hydrodynamical simulations. The abundance matching ansatz is a useful technique to link galaxies and (sub)haloes but involves a number of modeling assumptions. The prediction from (sub)halo abundance matching depends on which (sub)halo and galaxy properties are used to form a link between them. It has been shown that sub-halo abundance matching (SHAM) models using different primary properties such as peak halo mass, peak circular velocity etc., predict different galaxy clustering of different amplitudes (see \cite{Wechlser2018} and references therein). Similarly, when galaxies are matched to haloes using galaxy luminosity as the primary galaxy property, one obtains the same luminosity-halo mass relation for red and blue galaxies. However, when abundance matching is done using stellar mass as primary galaxy property one gets different stellar mass-halo mass relation for red and blue galaxies \citep{Yang2007}. On the other hand, even the hydrodynamical simulations have not been able to successfully reproduce the complex distribution of galaxy structure that we observe today \citep{Gomez2019, Zanisi2021}. 

In light of these issues, there is a  clear need to probe the relationship between the sizes of galaxies and their host haloes in a more direct and observational manner.  

In this work, we link the sizes of galaxies drawn from  Galaxy And Mass
Assembly \citep[GAMA, ][]{Driver2011} and Subaru HSC-SSP \citep{Aihara2018} surveys to the dark matter halo radius using weak gravitational (galaxy-galaxy) lensing. The presence of matter associated with the foreground galaxy causes distortion in the shapes of background galaxies due to gravitational lensing. The coherent shape distortion of background source galaxies depends on the combined (visible and dark) matter distribution of foreground lens galaxies. These distortions in the background galaxy are usually tiny for a single galaxy and buried in the noise arising due to the intrinsic shapes of galaxies. However, these can be statistically measured by stacking the lensing signal around a statistical sample of galaxies. By modeling these stacked weak lensing signals, one can assign average halo properties to a typical/average galaxy. Weak lensing is one of the most direct methods of probing the invisible matter in the universe and has been shown to be successful in forming galaxy-halo connections by a number of studies \citep{Mandelbaum2006, Leauthaud2012, Valender2014, Hudson2015, Uitert2016, Sonnenfeld2019, Dvornik2020, Wang2021}. 

The organisation of this paper is as follows. We start by describing the data and the selection of galaxy sample that we use in section \ref{sec2}. We then provide an outline of the methods used for the measurements of galaxy sizes and dark matter halo properties in section \ref{sec3}. We present the results of our study in section \ref{sec4} and discuss its limitations and possible implication for galaxy formation picture in section \ref{sec5}. We present a final summary and conclusion for this work in section \ref{sec5}. Throughout this work, we have assumed a flat $\Lambda$CDM cosmology with parameters ($\Omega_{\rm m}$, $\Omega_{\Lambda}$, $\sigma_8$, $h$, $n_{\rm s}$) = (0.26, 0.74, 0.77, 0.72, 0.95) which is consistent with WMAP7 \citep{Komatsu2011}, similar to that assumed by \citet{Zu2015}. Unless stated all measurements related to galaxy photometry are done in the $i$ band. The units of mass and size are  $h^{-1}{\rm M}_{\odot}$ and kpc respectively.

\section{Data and sample selection} \label{sec2}

 \begin{figure*}
     \centering
     \includegraphics[width = 0.475\textwidth]{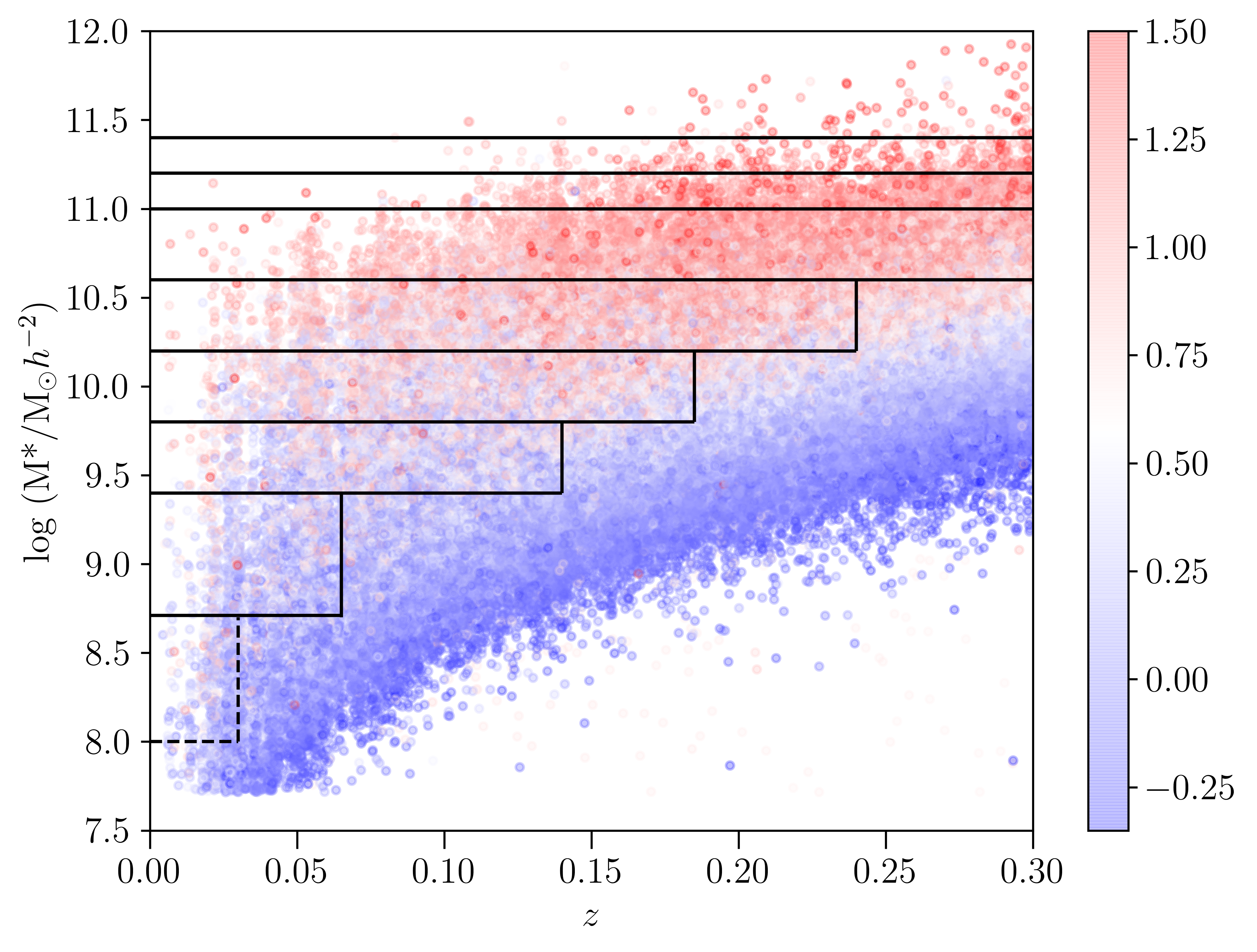}
     \includegraphics[width = 0.49\textwidth]{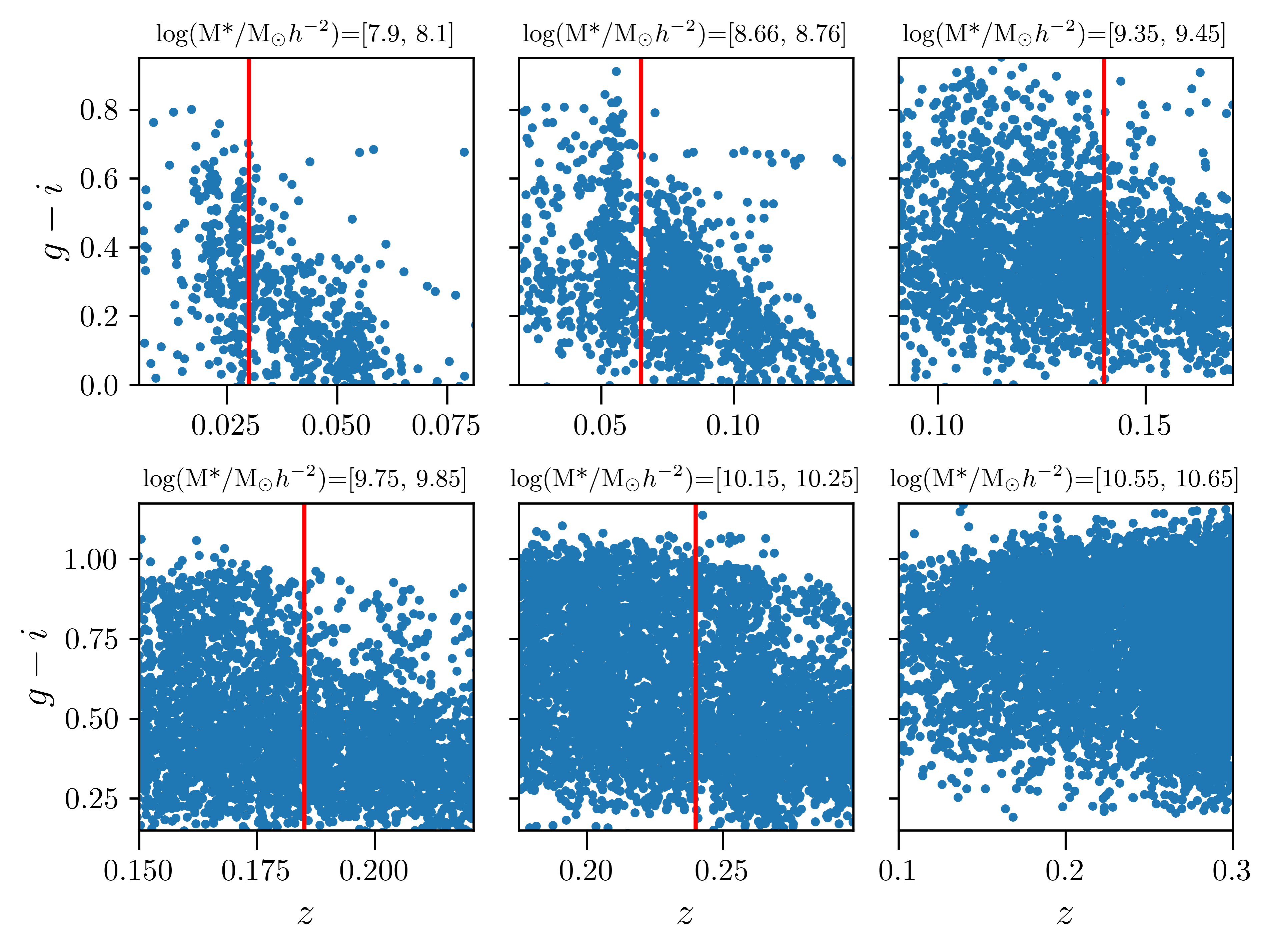}
     \caption{\preetish{Left:} Stellar mass vs redshift scatter plot of our parent sample galaxies. Each point represents a galaxy that is colour coded by its restframe $g-i$ colour coming from stellar emission. Galaxies in bins of stellar mass-redshift space marked by black lines are kept for further analysis. We have not performed weak lensing analysis on galaxies belonging to the lowest stellar mass bin marked with dashed lines. \preetish{Right:} Scatter plot of rest frame $g-i$ colour vs. redshift for galaxies in narrow bins of $\sim$0.1-0.2 dex centered at stellar mass bin edges corresponding to lower six stellar mass bin edges in the plot to the left. The vertical red line is the limiting redshift beyond which we start missing out on redder galaxies.} 
     \label{fig1}
 \end{figure*}

We have used data from the GAMA survey DR4\footnote{http://www.gama-survey.org/dr4/} \citep{Driver2022} and HSC-SSP survey PDR2\footnote{https://hsc-release.mtk.nao.ac.jp/doc/index.php/sample-page/pdr2/} \citep{Aihara2019} for this work. GAMA is a flux-limited (to $r$ band mag 19.8) highly complete spectroscopic survey of $\sim$ 286 square degrees of sky carried using the AAOmega spectrograph at the AAT. The fourth data release of the GAMA survey contains about 300,000 galaxies and information on their redshifts, stellar mass, morphology, environment, and many other quantities of interest. The HSC-SSP is a wide-field imaging survey which has observed $\sim1100$ sq. degrees of sky in multiple optical filters using the Hyper-Suprime Cam camera mounted at the prime focus of the 8.2m Subaru telescope \citep{Aihara2018}. The second data release of HSC-SSP \citep{Aihara2019} covers $\sim300$ square degrees of the sky in $grizy$ broadband filters. The wide layer of this survey reaches a limiting depth of $\sim$ 26 magnitude in the $i$-band along with a median seeing of 0.6 arcsec, making it one of the deepest and sharpest wide-field sky surveys. The imaging data products as well as catalogues on the photometric redshifts and galaxy shapes have been made available in this data release. We outline our sample selection methodology in the paragraphs below. 

In order to select the sample for our study, we started with galaxies present in GAMA Galaxy Group Catalogue \citep[G3C,][]{Robotham2011}. The G3C catalogue is a Friends-of-Friends (FoF) group catalogue run on galaxies in the GAMA survey equatorial and G02 fields. Within these survey regions, all galaxies having high-quality spectra and lying within the redshift range $0.003 < z < 0.6$ are included in the G3C catalogue. As our first selection cut, we choose galaxies having redshifts $z \leq 0.3$ and residing in sky region Dec$>-6$ degrees. The second criterion stems from the fact that the GAMA survey is highly complete only to the north of $-6$ degrees. The selected galaxies were then cross-matched to the GAMA survey catalogues \textsc{StellarMassesv19} and \textsc{StellarMassesG02SDSSv24} containing stellar masses and rest frame $g-i$ colours of galaxies derived from multiband SED fitting \citep{Taylor2011}. 

In the left panel of Figure \ref{fig1}, we plot our galaxies in the stellar mass vs redshift plane. The galaxies are colour coded according to their rest frame $g-i$ colours coming from stellar emission. One can see from this plot that the higher redshift end of our sample is dominated by blue galaxies. This happens because the mass-to-light ratio of blue/star-forming galaxies is less than that of red/passive galaxies. Hence among equally massive galaxies, blue galaxies tend to be more luminous and hence can be seen at farther distances by a flux-limited survey. At the same time, the fraction of equally massive yet less luminous red galaxies drops with increasing redshift. This results in flux-limited surveys being under-represented by red galaxies, especially at the flux limit \citep[see e.g.,][]{More2011, Zu2015, Wright2017, Vazquez-Mata2020}. 

We attempt to avoid such a bias by dividing our sample into a number of bins created in stellar mass-redshift space. We follow \cite{Zu2015} and subdivide our sample into eight stellar mass bins with edges at $\log (M_*/h^{-2} {\rm M}_{\odot}$) = [8, 8.71, 9.4, 9.8, 10.2, 10.6, 11, 11.2, 11.4].  The choice of these bins ensures enough galaxies in each of these mass bins to get a decent signal-to-noise ratio for the stacked weak lensing signal. Next, we define a narrow stellar mass bin of $\sim$0.1-0.2 dex in width centered around each of the aforementioned stellar mass bin edges, and then plot the $g-i$ color versus redshift for these narrow bins. We show such scatter plots corresponding to six bin edges $\log (M_*/h^{-2} {\rm M}_{\odot}$) = [8, 8.71, 9.4, 9.8, 10.2, 10.6] in the right panel of Figure \ref{fig1}. We visually identify redshifts beyond which we start missing redder galaxies, i.e. those with higher values of $g-i$ colour. The corresponding redshift cuts of z = [0.03, 0.065, 0.14, 0.18, 0.24] guarantee that the given stellar mass bins will contain an unbiased proportion of blue and red galaxies. We show these redshift limits as red vertical lines in the right panel of Figure \ref{fig1}. For stellar masses greater than $\log (M_*/h^{-2} {\rm M}_{\odot}$)=10.6, we do not see such a drop in red galaxies primarily because massive galaxies are usually red. This means that for the massive end of our galaxy sample we can go up to the sample redshift limit of $z = 0.3$. \preetish{Our estimates of redshift limits are roughly consistent with the results of \citet{Wright2017} and \citet{Vazquez-Mata2020} who employed a similar strategy to estimate these limiting redshifts.} We use these redshift threshold values to construct eight bins in stellar-mass redshift space shown as black horizontal and vertical lines in the left panel of Figure~\ref{fig1}. We chose to retain only those galaxies in our parent sample which reside within these bins defined in stellar mass - redshift space. Imposing this selection cut results in a sample of $\sim$ 73,000 galaxies which we refer to as our reduced sample. We do not directly estimate the dark matter halo properties of galaxies below the stellar mass limit of $10^{8.71}$ h$^{-2} {\rm M}_{\odot}$ due to noisy weak lensing signals, although we measure their sizes in our work. To distinguish such galaxies from the rest of the sample, we have marked the lowest stellar mass bin with dashed lines in the left hand panel of Figure \ref{fig1}.

The final step in sample selection for our work comes from the area overlap between the GAMA survey and the first-year shape catalogue of the HSC-SSP survey \citep{Mandelbaum2018}. The catalogue provides information on galaxy shapes measured using the re-Gaussianization \citep{Hirata2003} PSF correction technique. The shapes are measured on the coadded $i$-band images of galaxies from an internal data release S16A of HSC-SSP survey. This measurement technique has been well studied in the past using the data from the SDSS survey \citep{2005_Mandelbaum,2012_Reyes,2013_Mandelbaum}. The S16A data release consists of shape measurements for galaxies in six different fields - HECTOMAP, GAMA09H, WIDE12H, GAMA15H, XMM, and VVDS covering an area over $136.9~{\rm deg}^2$ with an effective galaxy number density of $21.5\, {\rm arcmin}^{-2}$ and a median redshift of $0.8$. We use the S16A data from the HSC fields - GAMA09H, WIDE12H, GAMA15H, XMM, and VVDS which overlap with GAMA. The galaxy shapes are measured as ellipticities $(e_{\rm 1},e_{\rm 2}) = (e\cos2\phi, e\sin2\phi)$ where $e = (a^2-b^2)/(a^2 + b^2)$ where $a$ and $b$ denote the semi-major and semi-minor axes, respectively, while $\phi$ denotes the position angle of the major axis in the equatorial coordinate system \citep{Bernstein2002}. The shape catalogue also provides additive bias corrections $(c_1, c_2)$, multiplicative bias corrections $(m)$, rms $e_{\rm rms}$ values of ellipticities for the intrinsic shapes and shape measurement error $\sigma_e$ required for unbiased shear estimation. These corrections are calibrated using image simulations ran using the open-source package - GALSIM \citep{2015_rowe} which mimic the observing conditions of the HSC survey \citep{2018_Mandlebaum}. The $e_{\rm rms}$ and $\sigma_e$ are further used to assign the weights $w_{\rm s} = (e^2_{\rm rms} + \sigma_e^2)^{-1}$ to the individual galaxies to enable a minimum variance estimator for the weak lensing signal \citep[see for more details][]{2018_Mandlebaum}. We also apply various quality cuts on the shape catalog data as described in \citet{Mandelbaum2018} for conducting weak lensing studies. Further, we use the full redshift distribution $P(z)$ provided by the HSC survey for each galaxy in our shape catalogue data \citep[][]{2018_Tanaka}. In our analysis, we are using the redshifts estimated using the classical template fitting code \textsc{Mizuki} \citep{2015_Tanaka} along with the quality cuts described in Sec. \ref{sec_esd} to obtain a secure sample of the background galaxies. We have checked that our measurements are consistent even if we use a different photometric redshift estimation code available from HSC.

Our aim is to establish a link between our sample of galaxies and their dark matter content measured using weak gravitational lensing. Therefore, we retain only those galaxies from the reduced sample which fall within the sky coverage of first-year shape catalogue of HSC-SSP survey. This task was accomplished by creating the required healpix sky coverage map of the shape catalogue using the publicly available python based software package \textsc{HEALPY}\citep{Healpy}. This gave us our final sample of $\sim$38,000 galaxies at a median redshift of $z$ = 0.16.

\section{Methods} \label{sec3}

\subsection{Galaxy Size measurements}
We measure the sizes of galaxies by fitting a single S\'ersic light profile. The S\'ersic model has been used extensively by past studies to model galaxy light profiles. This profile has the following mathematical form:
\begin{eqnarray}
I(r) &=& I_{e} \exp \left(-b_n\left[\left( \frac{r}{ R_{e}} \right)^{{1/n}} - 1 \right]\right)\,, 
\end{eqnarray} and is described by parameters corresponding to the S\'ersic index ($n$), half-light radius ($R_{\rm e}$) and intensity at half-light radius $I_{\rm e}$. The parameter $b_{\rm n}$ is related to the S\'ersic index ($n$) such that $b_{\rm n} = 1.9992n - 0.3271$, following approximation from \cite{Capaccioli1989} which is valid for $0.5 < n < 10$. We fit the S\'ersic model to $i$-band images of our sample galaxies using the software \textsc{GALFIT} \citep{Peng2010}, which is capable of fitting 2D analytical functions of light profiles directly to galaxy images. As input \textsc{GALFIT} requires users to provide a cut-out image of the target galaxy, specify the point spread function, mask out all sources in the cutout image except the target, and a rough initial guess for the model parameters related to the structure and brightness of the target galaxy. This involves significant pre-processing work from the user’s side before the actual fitting is performed using \textsc{GALFIT}. We briefly describe these pre-processing steps and our setup for running \textsc{GALFIT} in the following paragraphs.  

 \begin{figure*}
     \centering
     \includegraphics[width = 0.48\textwidth]{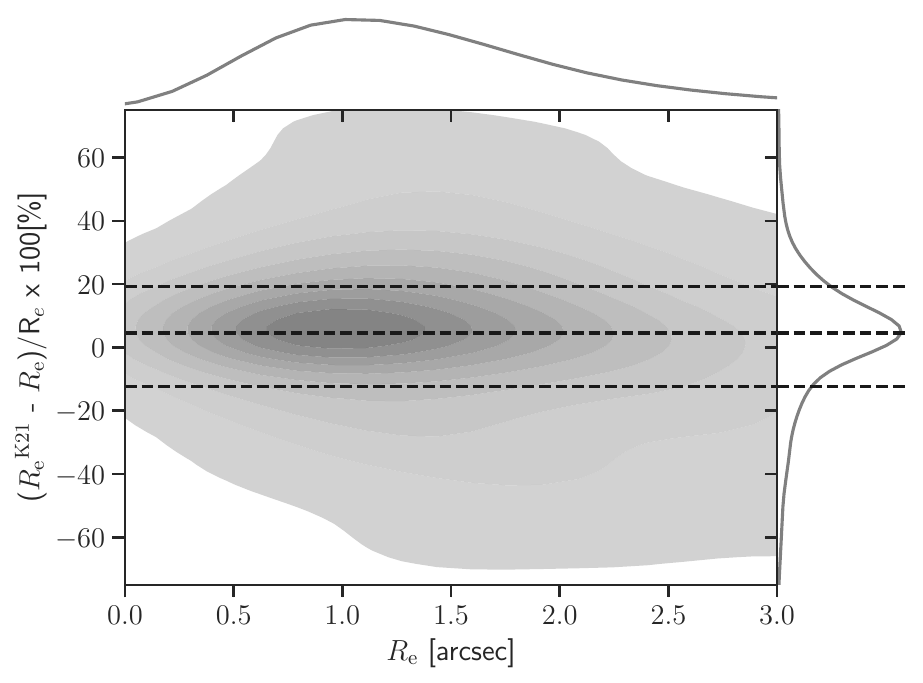}
     \includegraphics[width = 0.48\textwidth]{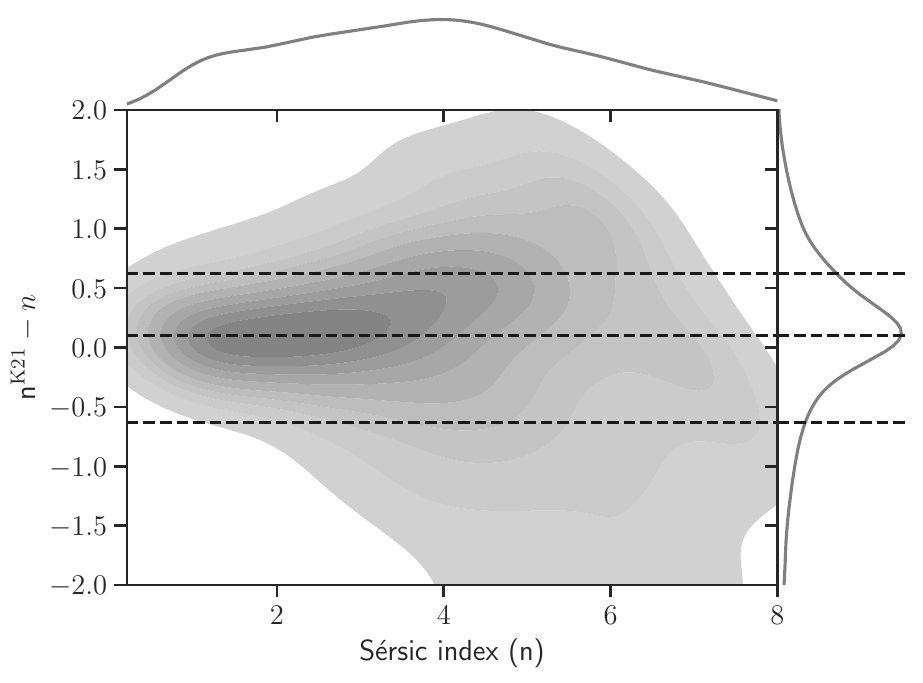}
     \caption{{\bf Left:} The percentage deviation of our measurements of half-light radius ($R_{\rm e}$) from those coming from \protect\cite{Kawinwanichakij2021} (K21) as a function of $R_{\rm e}$ shown in a joint plot of two dimensional and one-dimensional KDEs. The three dashed black lines from bottom to up mark the 16-50-84 percentile of percentage deviation which are at -12\%, 4.5\%, and 19\% respectively. {\bf Right:} Difference in S{\'e}rsic index measured by us ($n$) from the same measured by \protect\cite{Kawinwanichakij2021} (K21) as a function of $n$ shown in a joint plot of two dimensional and one-dimensional KDEs. The three dashed black lines from bottom to up mark the 16-50-84 percentile of percentage deviation which are at -0.63, 0.1, and 0.62 respectively.} 
   \label{fig2}
 \end{figure*}

Our starting point is the image cutout service provided by the HSC-SSP database which allows users to download cutouts by uploading a list of targets. The users can specify the data release version, filters, and cutout sizes to suit their requirements. For modeling galaxy light, \textsc{GALFIT} usually requires an image cutout large enough that it contains sufficient pixels free from any source for a good estimation of the sky background. To determine the cutout size, we first performed a cross-match in the sky positions (within 1 arcsec) of our sample galaxies to the galaxies listed in the \textsc{pdr2\_wide\_forced} catalog in HSC-SSP PDR2 database having stellar masses greater than $10^8 {\rm M}_{\odot}$ and within (photometric) redshift of 0.3 as determined by classical template fitting code \textsc{Mizuki} \citep{2015_Tanaka}. The \textsc{pdr2\_wide\_forced} is a catalogue of forced photometry measurements performed on galaxies and it lists their various photometric measurements related to flux and size in all broadband filters of Subaru-HSC. \preetish{This cross-match resulted in a sample of 34,035 galaxies out of our total sample of 37,815 galaxies. These galaxies have available $i$-band Kron radius measurements from the \textsc{pdr2\_wide\_forced} catalog and were subjected to size measurements. The Kron radius is defined as the first moment of the surface brightness light profile. It is argued that an aperture twice the size of the Kron radius captures more than 90\% of galaxy light thus making it a reasonably good, although crude estimate for the size of galaxies. We had missed 3780 galaxies in our cross-match because of the photometric redshift and stellar mass selection cuts imposed on the \textsc{pdr2\_wide\_forced} catalog. This may happen because often the photometric redshift estimates can significantly differ from the true redshifts. As a result, stellar masses derived using photometric redshifts can also differ from their true value. For these missing galaxies, we collected information on the Kron and Petrosian radius measurements from input catalogs \textsc{InputCatAv07}, \textsc{G02SDSSInputCat} and \textsc{G02CFHTLSInputCat} from the GAMA survey database. We used the crude size estimates to create a square coadd image cutouts in $i$-band centered at our target galaxies and with a side length of 40 times their Kron or Petrosian radii. The large image size ensures that we are capturing the diffuse outer wings of early-type galaxies in our sample. The image cutout comes along with the variance image containing pixel-wise variance of measurements of the coadded image. We also downloaded the coadded $i$-band PSF using the PSF picker utility of the HSC-SSP PDR2 database at the location of the galaxy.}

Next, we identify the sources we want to model using \texttt{GALFIT} and mask out all other sources. We use the python based open source software \textsc{astropy} \citep{astropy} and \textsc{photutils} \citep{Bradley2020} for this purpose. \textsc{Photutils} detects sources from an image above some user-specified threshold. We have chosen this threshold by first identifying pixels free from any sources using sigma-clipping and then obtaining the median and standard deviation counts of these background pixels. Any region having counts at least 1.5$\sigma$ above the background median and having at least 5 connected pixels was treated as detection in our scheme. \textsc{Photutils} also creates a segmentation map where pixels of each detected source are labeled with a single unique positive number. All the pixels below the detection threshold are assigned a value of zero in the segmentation map. By design, the cutouts are centered on our sample galaxies and thus correspond to the central segment in the segmentation map. In the segmentation map, we assign a value of zero to the central segment and the rest of the source segments are used as a mask image required by \textsc{GALFIT}.  

We use \textsc{photutils} to obtain basic measurements on fluxes, position angle, and the ellipticity of detected sources. It also gives a basic measurement of detected sources treating their light as a 2D Gaussian function and returns values of the 1-sigma standard deviation along the major and minor axes. As an initial starting point for \textsc{GALFIT}, we use 5 times the square root of the product of 1-sigma major and minor axis as a guess value for the size of our target galaxies.

In our sample, there are also cases where one or more bright galaxies are sitting very near to, or are overlapping with our target galaxies, as identified in the segmentation maps. In this situation masking these close bright neighbours is not a good option, instead, we opted to fit these neighbours simultaneously along with our targets. If the fiducial size sum of the target and a neighbour is greater than the distance between them, then such a neighbour is labeled as an overlapping galaxy. Any overlapping galaxy which is no fainter than the target by 2 mags is simultaneously modeled along with the target. Galaxies fainter than this, even if they are overlapping have been masked in our scheme. 
 
The whole pre-processing work outlined above was automated using scripts written in Python. The image cutout, image mask, error image (square root of variance image), PSF, and the parameter files containing guess values of free parameters are provided as input to GALFIT in order to fit a single S{\'e}rsic model to each galaxy in our sample. In addition, we also use prior constraints on the parameter space that GALFIT is allowed to search. This prevents the software from searching for solutions in abnormal/unphysical ranges of parameter space. We constrain the S{\'e}rsic index, $n$, to lie within the range $[0.2, 8]$ and the half-light radius, \preetish{$R_{\rm e}$ to lie within $[1.5, 750]$ pixels.} The stated S{\'e}rsic index upper and lower limits have been used in a number of past works on the automated fitting of galaxies \cite{Simard2011, Wel2012, Meert2015, Kawinwanichakij2021}. It has been argued that galaxies rarely have S{\'e}rsic index values outside these ranges. The lower limit on $R_{\rm e}$ is nearly equal to 80\% of half width at half maxima (HWHM) of $i$-band median seeing (0.6 arcsec = 3.6 pixels). \cite{Gadotti2009} has shown that reliable fitting of S{\'e}rsic light profile can be done on structures having true $R_{\rm e}$ greater than 80\% of HWHM. \preetish{The upper limit on $R_{\rm e}$ of 750 pixels translates to 125.25 arcsec for Subaru-HSC having a pixel scale of 0.167 arcsec/pixel. This limit is around 15 times greater than the Kron radius of the largest galaxy in our sample.} 

We ran the GALFIT following the above setup on our sample galaxies and collected the output in tabular form. The bad fits usually piled up at the limits of parameter constraints which are then discarded i.e., \preetish{we removed galaxies outside the range of 0.2$<n<$8 and $R_{\rm e}$ = (1.5, 750) pixels.} Additionally, we removed galaxies having an axis ratio too low (b/a$\leq$0.1) or if the model centroid moved more than 3 pixels from the original guess value as it might be a result of improper masking of a nearby compact source. \preetish{We also removed fits with a value of reduced chi-square greater than 5, ensuring that we are not including terrible models in our size measurements. After the removal of the failures and poor fits, we end up with a sample of 32055 galaxies with reliable structural measurements which are about $\sim$ 85\% of our 37,815 target galaxies.} A similar success rate is also obtained by works employing automated fitting on a large sample of galaxies \citep[see e.g.,][]{Meert2015, Lange2015}.

\preetish{We compared our measurements of galaxy half-light radius ($R_{\rm e}$) and S{\'e}rsic index ($n$) with the measurements of the same from \cite{Kawinwanichakij2021} who have studied the size-mass relation of HSC-SSP PDR2 galaxies in the redshift range $0.2<z<0.8$. A 1 arcsec cross-match on the sky positions of our sample with the sample of  \cite{Kawinwanichakij2021} (obtained via private communication) yielded around 2,000 common galaxies. We refer to these 2,000 galaxies as a comparison sample. We show the percentage deviation of our measurements of half-light radius from those coming from \cite{Kawinwanichakij2021} as a function of our measured $R_{\rm e}$ in a joint plot of two-dimensional and one-dimensional KDEs in the left panel of Figure \ref{fig2}. The median deviation of our size measurements with respect to the same in \cite{Kawinwanichakij2021} is only about 4.5\% with the majority (16-84 percentile) of comparison samples displaying deviation at a level of 12-19\%. In a similar manner we display the difference in our estimates of S{\'e}rsic index with those from \cite{Kawinwanichakij2021} in the right panel of Figure \ref{fig2}. The median difference in S{\'e}rsic index measurements is about 0.1 with the majority (16-84 percentile) of comparison samples displaying deviation around 0.6.} Based on these figures and numbers, we can say that there is a fair agreement between our measurements of galaxy structure with \cite{Kawinwanichakij2021}. Small differences in galaxy size and S{\'e}rsic index measurements can arise due to differences in the methodology of light profile fitting. Previous works have shown that using different software to model galaxy images of the same quality can lead to 5-10\% changes \citep[see e.g.,][]{Meert2015}. Moreover, \cite{Kawinwanichakij2021} have studied the systematic uncertainty of their size measurement using simulated images and corrected their size measurements of observed galaxies accordingly. For bright objects (with $18<i<24$), they find the median deviation of measured sizes from the true sizes of simulated galaxies to be about 5\% or better. We have not applied any such correction which also might be a reason for slight differences in measured structural parameters. However, a difference in size measurements at a 5\% level is unlikely to significantly change our results.

\subsection{Dark matter halo properties}
\label{sec_esd}

\subsubsection{Measurements}
\label{sec_esd_meas}

The effect of weak gravitational lensing due to the dark matter distribution around the galaxies imparts a shear on the images of the background (source) galaxies \citep[for recent reviews see][]{2015_Kilbinger, 2018ARA&A_Mandelbaum}. The stacked weak gravitational lensing signal measured using these distortions is given by
\begin{equation}
\Delta\Sigma(R)  \equiv  \bar{\Sigma}(<R) - \langle \Sigma(R)\rangle = \Sigma_{\rm crit} \gamma_t \,,
\end{equation}
where $R$ denotes the projected comoving galaxy-centric distance $R$, $\Delta\Sigma(R)$ the excess surface density (ESD), $\bar{\Sigma}(<R)$  is the average projected surface matter density within a projected distance $R$, $\langle \Sigma(R)\rangle$ is the azimuthally averaged surface matter density at the same distance, $\gamma_{\rm t}$ denotes the average tangential shear. The critical surface density $\Sigma_{\rm crit}$ is a geometrical lensing efficiency factor for a lens-source pair, such that  
\begin{equation}
    \Sigma_{\rm crit} = \frac{c^2}{4\pi G} \frac{D_{\rm a}(z_{
    \rm s})}{(1+z_{\rm l})^2 D_{\rm a}(z_{\rm l}) D_{\rm a}(z_{\rm l},z_{\rm s})}\,.
\end{equation}
with  $D_{\rm a}(z_{\rm s}), D_{\rm a}(z_{\rm l})$ and $D_{\rm a}(z_{\rm l}, z_{\rm s})$ as the angular diameter distance for the source, lens and between the lens-source pair. The factor $(1+z_{\rm l})^2$ in the denominator in the result of our use of comoving coordinates \citep{2006_Mandelbaum}. 

We follow the methodology described in \citet{Mandelbaum2018} to compute the ESD profile $\Delta \Sigma (R_i)$ for our lensing sample. We use ten logarithmically spaced projected comoving galaxy centric radial bins $R_i$ ranging from 0.03 to 1.4 ${\rm h^{-1} Mpc}$ and the corresponding $\Delta \Sigma (R_i)$ is 
\begin{multline}
    \Delta \Sigma(R_i) = \frac{1}{1 + m} \left[ \frac{\sum_{{\rm ls}\in R_i} w_{\rm ls} e_{t,{\rm ls}} \langle \Sigma_{\rm crit}^{-1} \rangle^{-1}}{2\mathcal{R} \sum_{{\rm ls}\in R_i} w_{\rm ls}} \right. \\ - \left. \frac{\sum_{{\rm ls}\in R_i} w_{\rm ls} c_{t,{\rm ls}} \langle \Sigma_{\rm crit}^{-1} \rangle^{-1}}{\sum_{{\rm ls}\in R_i} w_{\rm ls}} \right]\,.\label{eq:dsig}
\end{multline}
Here $e_{t,{\rm ls}}$ is the tangential component of the ellipticites  and $c_{t,{\rm ls}}$ is the tangential component of additive bias along with a weight $w_{\rm ls} = w_{\rm s}\langle \Sigma^{-1}_{\rm crit}  \rangle^2$ for each lens-source pair. The summation is over all lens-source pairs at the projected radial separation of $R_i$ and given the probability redshift distribution $P(z_{\rm s})$ for each source galaxy, the average inverse critical surface density $\langle \Sigma^{-1}_{\rm crit}  \rangle$ can be calculated as
\begin{equation}
    \langle \Sigma_{\rm crit}^{-1} \rangle = \frac{4\pi G (1+z_{\rm l})^2}{c^2} \int_{z_{\rm l}}^\infty \frac{D_{\rm a} (z_{\rm l}) D_{\rm a} (z_{\rm l},z_{\rm s})}{ D_{\rm a} (z_{\rm s})} P(z_{\rm s}) \, d z_{\rm s}\,.
\end{equation}
The factor $(1+m)$ in the denominator of Eq.~\ref{eq:dsig} corrects for the multiplicative bias and can be estimated using $m = \sum_{{\rm ls} \in R_i} w_{\rm ls} m_{\rm s} / \sum_{{\rm ls} \in R_i} w_{\rm ls}$. The shear responsivity factor $\mathcal{R}$ is used to correct the response of ellipticities to the applied shear and expressed in terms of $e_{\rm rms}$ \citep{Bernstein2002} as
\begin{equation}
        \mathcal{R} = \frac{\sum_{{\rm ls} \in R_i} \left(1 -  w_{\rm ls} e^2_{{\rm rms},{\rm ls}}\right)}{\sum_{{\rm ls} \in R_i} w_{\rm ls}}\,.
\end{equation}

We select a secure sample of background sources using galaxies having $\int_{z_{\rm max} + z_{\rm diff}}^\infty P(z_{\rm s})\, dz_{\rm s} > 0.99$ where we use $z_{\rm max}=0.3$ as the maximum redshift of our lensing sample and apply an additional offset of $z_{\rm diff}=0.2$ for a cleaner background selection. We further apply a quality cut of $\texttt{photo\_z\_risk\_best\_value} < 0.5$. We correct for the multiplicative bias $m_{\rm sel}$ related to selection bias arising from a sharp cut on the resolution of galaxies, $R_2 \geq 0.3$, applied during the construction of the shape catalogue. This bias is estimated as $m_{\rm sel} = A\,P(R_2 =0.3)$, where $P(R_2 =0.3)$ is the lens-source weighted probability in each radial bin $R_i$ at the resolution threshold and $A=0.00865$ is a constant multiplicative factor \citep[for more details see.][]{2018_Mandlebaum}.

Our quality cuts should result in a secure sample of source galaxies lying behind the lens galaxies. In order to test for any residual galaxies which are correlated with the lens galaxies and lie in the same redshift range, we compute the boost parameter $C(R_i)$ \citep[for example][]{2004_Hirata, 2005_Mandelbaum, 2015_Miyatake} as the ratio between weighted lens-source pairs and normalized randoms-source pairs in the radial bin $R_{\rm i}$, 
\begin{equation}
    C(R_i) = \frac{N_{\rm r} \sum_{{\rm ls}\in R_i} w_{\rm ls}}{N_{\rm l} \sum_{{\rm rs}\in R_i} w_{\rm rs}}\,.
    \label{boost_eqn.}
\end{equation}

Here $w_{\rm rs}$ are weights for the randoms-source pair with $N_{\rm l}$ number of lenses and $N_{\rm r}$ number of randoms. Any contamination of our source sample arising due to systematics in the $P(z_{\rm s})$ estimates will dilute the $\Delta \Sigma(R_{\rm i})$ measurements. We calculate the random contribution using 100 different random realizations, each having the same numbers as our lenses. These randoms are uniformly distributed in the GAMA survey geometry following the same star mask as our lenses and matching the redshift distribution as our lensing sample by construction. The boost factors we obtain for our measurements are all within 2 percent, which shows the effectiveness of our source selection strategy. We have verified that boost factor changes of order 5 percent do not change any of the conclusions presented in this paper.

\begin{table*}
\caption{Sample selection for weak gravitaitonal lensing analysis in seven stellar mass bins. The table provides bin widths, average stellar mass $\log(\bar{M}^*/{\rm M}_{\odot}h^{-2})$, number of galaxies and median redshift in each stellar mass bin. These selection bins are shown in the scatter plot between stellar mass and redshift in Figure \ref{fig1}.}
\label{weak_bins}
\resizebox{\textwidth}{!}{

\begin{tabular}{llllllll}
\hline
 $\log(M^*/{\rm M}_{\odot}h^{-2}$) & (8.71, 9.40) & (9.40, 9.80) & (9.80, 10.20) & (10.20, 10.60) & (10.60, 11.00) & (11.00, 11.20) & (11.20, 11.39)\\ \hline
 $\log(\bar{M}^*/{\rm M}_{\odot}h^{-2}$) & 9.06 & 9.62 & 10.01 & 10.39 & 10.76 & 11.08 & 11.28\\
 Number of galaxies & 1269 & 4904 & 9572 & 12274 & 8675 & 834 & 215\\
 Median redshift & 0.05  & 0.11  & 0.14  & 0.18  & 0.25  & 0.26  & 0.25  \\
 \hline
\end{tabular}
}
\end{table*}

Apart from the boost parameter, we also studied the effect on $\Delta \Sigma$ from the systematics in $\Sigma_{\rm crit}$ computated using the photometric redshifts of the source galaxies. We use eqn.~5 in \citet{2008_Mandelbaum} to calculate the photo-z bias $b(z_{\rm l})$ for lens at redshift $z_{\rm l}$ using
\begin{equation}
    \frac{\Delta \Sigma}{\Delta \Sigma^\mathcal{T}} = 1 + b(z_{\rm l})  = \frac{\sum_{\rm s} w_{\rm ls} \langle \Sigma^{-1}_{{\rm crit},{\rm ls}}\rangle^{-1} \left(\Sigma^{\mathcal{T}}_{{\rm crit}, {\rm ls}}\right)^{-1}}{\sum_{\rm s} w_{\rm ls}}\,,
\end{equation}
where the computation of quantities with superscript $\mathcal{T}$ uses the true redshift of the source galaxy and the summation is over all source galaxies. The computation of these quantities requires source galaxies with known spectroscopic redshifts. For this purpose we use the robust photometric redshift estimates available for sources in the COSMOS region based on 30-band photometry \citep[][]{2009_Ilbert} and we use these galaxies for estimating this bias. Following the literature \citep[for example][]{2012_Nakajima, 2019_Miyatake, 2019_Murata}, we include a weight $w_{\rm som}$ provided by the HSC team in $w_{\rm ls}$ to match the colour and magnitude distributions of COSMOS galaxies with our source galaxy sample. We then calculate the average bias for our lens selection using the lens weights given by eqn. 23 in \citet{2012_Nakajima}. We found an average photo-z bias which is of the order of 1 percent for our lens selection bins. This is negligible compared to the statistical uncertainties in our signals.   

We also remove any scale dependent systematic in our signal by subtracting the signal around random points \citep[for e.g.][]{2004_Sheldon, 2005_Mandelbaum, 2017_Singh}. We use different random realizations for computing an average value for the ESD signal systematics. We also checked the systematics using the cross component of the ESD measurements and signal around random points as described in Appendix \ref{apx_null_wl}.

One can expect covariance between the signal measurements at different radial bins, as the same source galaxies can be at a different projected distance away from different lens galaxies. In this work, we use the shape noise covariance computed using 500 random rotations of the shapes of the source galaxies for each of our stellar mass bins. The random rotations can be used to compute the covariance contribution from the intrinsic shapes of the sources. We refer to Appendix \ref{apx_wl_cov} for details about how the shape noise component dominates the error budget for the weak lensing signal.

The various panels of Figure \ref{fig:esd_sig} show our measured weak lensing signal in the seven different stellar mass bins as blue solid points with errors. These stellar mass bins are given in Table \ref{weak_bins} and shown in Figure \ref{fig1}. The weak lensing signals are well measured in each of these bins with varying signal-to-noise ratio as mentioned in the bottom left of each bin . As expected and seen previously in the literature \citep[see e.g.,][]{Zu2015}, we see that the weak lensing signal increases in amplitude as we go to higher stellar masses reflective of the changing relation between the halo masses of galaxies in different mass bins. Although the signal approximately looks like a power law with slope $-1$, we can also see evidence for deviations from it given the high signal to noise ratio. Such deviations are expected from the different radial behaviour of the 1-halo and 2-halo contributions around central and satellite galaxies. In what follows, we describe how we model these measurements.

\subsubsection{Halo occupation distribution model}
We use the conditional stellar function formalism \citep[see for e.g.,][]{2008_Yang,2009_Yang,Zu2015,Uitert2016} within the framework of the halo model \citep[e.g.,][]{2000_Seljak, 2002_Cooray, 2013_bosch} for inferring the halo occupation distribution of our sample of galaxies. The conditional stellar mass function (CSMF), $\phi(M^*|M) dM^*$, is the average number of galaxies with stellar masses in the range $M^*\pm dM^*/2$ residing in haloes of mass $M$. We split the CSMF into two terms, one corresponding to central galaxies that reside at the centers of the dark matter haloes, $\phi_c(M^*|M)$, and the second corresponding to satellite galaxies, $\phi_s(M^*|M)$, such that,
\begin{equation}
    \phi(M^*|M) = \phi_{\rm c}(M^*|M) + \phi_{\rm s}(M^*|M) \,. 
\end{equation}
Following \citet[][]{2009_Yang}, we express $\phi_{\rm c}(M^*|M)$ as a log-normal distribution with a mean $\log M^*_c$, which depends upon halo mass. We parameterize $M^*_c(M)$ as a double power law as a function of the halo mass $M$.  This relation thus corresponds to $\langle \log M^* \rangle(M)$, commonly called the stellar mass to halo mass (SMHM) relation for the central galaxies. We use a modified Schechter function to describe the satellites CSMF $\phi_{\rm s}(M^*|M)$. These two contributions can be parameterized as
\begin{align}
    \phi_{\rm c}(M^*|M)\,dM^* &= \frac{\log e}{\sqrt{2\pi}\sigma_0}\exp\left( -\frac{\left(\log M^* - \log M^*_c \right)^2}{2\sigma_0^2} \right) \frac{dM^*}{M^*}\, ,\label{eq:csmf_cen}\\
    M^*_{\rm c}(M) &= M_0 \frac{(M/M_1)^{\gamma_1}}{[1 + (M/M_1)]^{\gamma_1 - \gamma_2}}\, ,\\
    \phi_{\rm s}(M^*|M)\,dM^* &= \phi^*_s \left( \frac{M^*}{M^*_s}\right)^{(\alpha_s + 1)}\exp\left[ -\left( \frac{M^*}{M^*_s}\right)^2\right] \frac{dM^*}{M^*} \,.
\end{align}
Here $\sigma_0$ denotes the the scatter in our SMHM relation given by Eqn \ref{eq:csmf_cen} and we use $M^*_{\rm s}(M) = 0.562 M^*_{\rm c}(M)$ based on \citet[][]{2009_Yang}. In this work, we assume $\sigma_0$ and $\alpha_{\rm s}$ to be independent of halo mass as motivated from the past studies \citep[][]{2008_Yang,2009_Yang,Leauthaud2012, Uitert2016, More2009a, More2009, 2013_More,2013_Cacciato} and $\phi^*_{\rm s} (M)$ is expressed in terms of halo mass $M_{\rm 12} = M/(10^{12} h^{-1}{\rm M}_{\odot})$ as
\begin{equation}
\log \phi^*_{\rm s}(M) = b_0 + b_1 \log M_{\rm 12} + b_2(\log M_{\rm 12})^2\,.
\end{equation}
The CSMF model we use consists of a total of nine model parameters - $\log M_0$, $\log M_1$, $\gamma_1$, $\gamma_2$, $\sigma_0$, $b_0$, $b_1$, $b_2$ and $\alpha_{\rm s}$. 

The halo occupation distribution $\langle N|M\rangle$ specifies the average number of galaxies in a stellar mass bin of $M^*_1\leq M^* \leq M^*_2$ residing in haloes of fixed mass M, and it can be calculated using $\phi(M^*|M)$ as
\begin{equation}
    \langle N|M\rangle = \int_{M^*_1}^{M^*_2} \phi(M^*|M) \, dM^*\,.
\end{equation}

The weak gravitational lensing signal $\Delta \Sigma(R,z)$ can be written in terms of the projected matter density $\Sigma(R,z)$ which is related to the galaxy-matter correlation function $\xi_{\rm gm}(r,z)$ for a given redshift $z$ by
\begin{equation}
    \Sigma(R,z) = 2 \, \bar{\rho}_{\rm m}\int_0^\infty \left(1 + \xi_{\rm gm}\left( r, z\right)\right)\,d\pi \,.
\end{equation}
Here the distance $r=\sqrt{R^2 + \pi^2}$ and the integration is carried out along the line of sight direction $\pi$. The galaxy matter correlation function $\xi_{\rm gm}(r,z)$ is the Fourier transform of galaxy-matter cross spectra $P_{\rm gm}(k,z)$,
\begin{equation}
    \xi_{\rm gm}(r,z) = \frac{1}{2\pi^2}\int_0^\infty P_{\rm gm} (k,z) \frac{\sin kr}{kr} k^2 dk\,,
\end{equation}
and we can split the galaxy matter cross power spectrum $P_{\rm gm}(k,z)$ as a contribution from one-halo (1h) and two-halo (2h) terms for both central and satellite galaxies as
\begin{equation}
    P_{\rm gm}(k,z) = P^{\rm 1h}_{\rm cm}(k,z) + P^{\rm 1h}_{\rm sm}(k,z) + P^{\rm 2h}_{\rm cm}(k,z) + P^{\rm 2h}_{\rm sm}(k,z)\,.
\end{equation}
Each of these terms can be expressed in a compact form given by
\begin{align}
    P^{\rm 1h}_{\rm xm} (k,z) &= \int dM \, \mathcal{H}_{\rm x} (k, M,z) \mathcal{H}_{\rm m} (k, M,z) n(M,z)\, ,\\
    P^{\rm 2h}_{\rm xm} (k,z) &= \int dM_{\rm 1} \, \mathcal{H}_{\rm x} (k, M_{\rm 1},z) n(M_{\rm 1},z)\\
    &\times \int dM_2\, \mathcal{H}_{\rm m} (k, M_{\rm 2},z) n(M_{\rm 2},z) Q(k|M_{\rm 1}, M_{\rm 2}, z)\, ,\\
\end{align}
where x can either be centrals (c) or satellites (s). The quantity $n(M,z)$ is the halo mass function at redshift $z$ and $Q(k|M_{\rm 1}, M_{\rm 2}, z)$ is the halo-halo cross power spectrum between haloes of mass $M_{\rm 1}$ and $M_{\rm 2}$ as described in details in \citet{2013_bosch}. We further define
 \begin{align}
      \mathcal{H}_{\rm c}(k,M,z) &= \frac{\langle N_{\rm c}|M \rangle}{\bar{n}_{\rm g}(z)} \, ,\\ 
     \mathcal{H}_{\rm m}(k,M,z) &= \frac{M}{\bar{\rho}_{\rm m}}\Tilde{u}_{\rm h}(k,M,z) 
 \end{align}
 and,
 \begin{equation}
     \mathcal{H}_{\rm s}(k,M,z) = \frac{\langle N_{\rm s}|M \rangle}{\bar{n}_{\rm g}(z)} \Tilde{u}_{\rm s}(k,M,z) \, .
 \end{equation}
The functions $\Tilde{u}_{\rm h}(k,M,z)$ and  $\Tilde{u}_{\rm s}(k,M,z)$ represents the Fourier transform of the normalized dark matter profile and the normalized satellite number density profile. We assume that dark matter in the halo is distributed in the form of the NFW profile \citep{1996_Navarro} and that the satellite galaxies also follow the same distribution \citep[for e.g.][]{2013_bosch, 2013_More, 2013_Cacciato}. By default we assume no miscentering between the central galaxy and the true halo center. However we have also explored the effects of allowing for a fraction $p_{\rm off}$ of offcentered centrals galaxies along with a Gaussian miscentering kernel of width $r_{\rm off}$ in units of $r_{\rm 200m}$ \citep[][]{2015_More} and found no significant changes to our conclusions.

We also include an additional parameter $c_{\rm fac}$ to include any deviation of the concentration of halos compared to the concentration-mass relation given by \citet{Maccio2007}. We use a Gaussian prior on $c_{\rm fac}$ with unit mean and 0.2 width as adopted in the literature \citep[for e.g.][]{2013_More, 2013_Cacciato}. We also model the baryonic contribution $\Delta\Sigma^{\rm b}(R,z)$ to the $\Delta\Sigma(R,z)$ profile as a points mass model with a parameter $a_{\rm p}$, such that 
\begin{equation}
    \Delta\Sigma^{\rm b}(R,z) = \frac{a_{\rm p} \bar{M}^*}{\pi R^2}\,.
\end{equation}
Here the $\bar{M}^*$ is the mean stellar mass (in $h^{-1}{\rm M}_\odot$) at the median redshift $z$ for each stellar mass bin. Therefore, our final modelling consists of eleven parameters $\Theta = (\log M_0, \log M_1, \gamma_1, \gamma_2, \sigma_0, b_0, b_1, b_2, \alpha_{\rm s}, c_{\rm fac}, a_{\rm p})$. We use the public python package - \texttt{AUM}\footnote{https://github.com/surhudm/aum} \citep[][]{2013_bosch,2013_More, 2013_Cacciato} and carry out the required modifications to include modelling based on the CSMF. We carry out a Bayesian analysis to infer the posterior probability $P(\Theta|\mathcal{D})$ of our model parameters given the data $\mathcal{D}$, with priors $P(\Theta)$, such that
\begin{equation}
    P(\Theta|\mathcal{D}) \propto P(\mathcal{D}|\Theta)P(\Theta)\,. 
\end{equation}
Here $P(\mathcal{D}|\Theta)$ is the Gaussian likelihood of the data $\mathcal{D}$ given the model parameter $\mathcal{M}(\Theta)$ and defined as
\begin{equation}
    P(\mathcal{D}|\Theta) \propto \exp\left[-\frac{\chi^2}{2} \right] P(\Theta)
\end{equation}
where,
\begin{equation}
    \chi^2 = [\mathcal{D} - \mathcal{M}(\Theta)]^{T} C^{-1} [\mathcal{D} - \mathcal{M}(\Theta)]\, .
\end{equation}

The noise in the estimate of the covariance matrix from a finite number of realizations can result in a biased estimate for the inverse covariance matrix $C^{-1}$. To correct for such a bias, we apply the Hartlap factor to $C^{-1}$ \citep[see eqn. 17 in ][]{2007_Hartlap}. We use the publicly available Python package - \texttt{emcee} \citep{2013_Mackey} which implements the affine invariant sampler of \citet[][]{2010_Goodman}, in order to sample the posterior of our model parameters given the data.  
\begin{table}
    \centering
    \begin{tabular}{lll}

    \hline
    \hline
    Parameters & Priors & Constraints \\
    \hline
     ${\rm \log M_{0}}$ & Flat[8,11.5] &$10.51^{+0.19}_{-0.21}$\vspace{0.1cm}\\
	 ${\rm \log M_1}$ & Flat[9.5,15] & $11.92^{+0.29}_{-0.22} \vspace{0.1cm} $\\
	 ${\rm \gamma_1}$ & Flat[1,4] &$2.87^{+0.83}_{-0.92} \vspace{0.1cm} $\\
	 ${\rm \gamma_2}$ & Flat[0.01,2] &$0.38^{+0.08}_{-0.10} \vspace{0.1cm} $\\
	 ${\rm \sigma_0}$ & Flat[0.05,0.5] &$0.11^{+0.04}_{-0.04} \vspace{0.1cm} $\\
	 ${\rm b_0}$ & Flat[-2,2] &$-0.35^{+0.42}_{-1.00} \vspace{0.1cm} $\\
	 ${\rm b_1}$ & Flat[-2,2] &$-0.52^{+1.18}_{-0.59} \vspace{0.1cm} $\\
	 ${\rm b_2}$ & Flat[-2,2] &$0.43^{+0.17}_{-0.29} \vspace{0.1cm} $\\
	 ${\rm \alpha_s}$ & Flat[-2,-1] &$-1.35^{+0.17}_{-0.21} \vspace{0.1cm} $\\
	 ${\rm c_{fac}}$ & Gauss(1.0,0.2) &$1.20^{+0.19}_{-0.17} \vspace{0.1cm} $\\
	 ${\rm a_p}$ & Flat[0.5,5] &$2.41^{+0.56}_{-0.58} \vspace{0.1cm} $\\
    \hline
    \end{tabular}
    \caption{{\it Parameter constraints:}  The table provides the priors along with the constriants on our weak lensing model parameters. The Flat denotes the uniform prior and Gauss($\mu,\sigma$) represents the Gaussian prior with mean $\mu$ and width $\sigma$. The parameter constraints column provides the median with $1\sigma$ around the median errors.}
    \label{tab:modparam}
\end{table}

\begin{figure*}
    \centering
    \includegraphics[width =\textwidth]{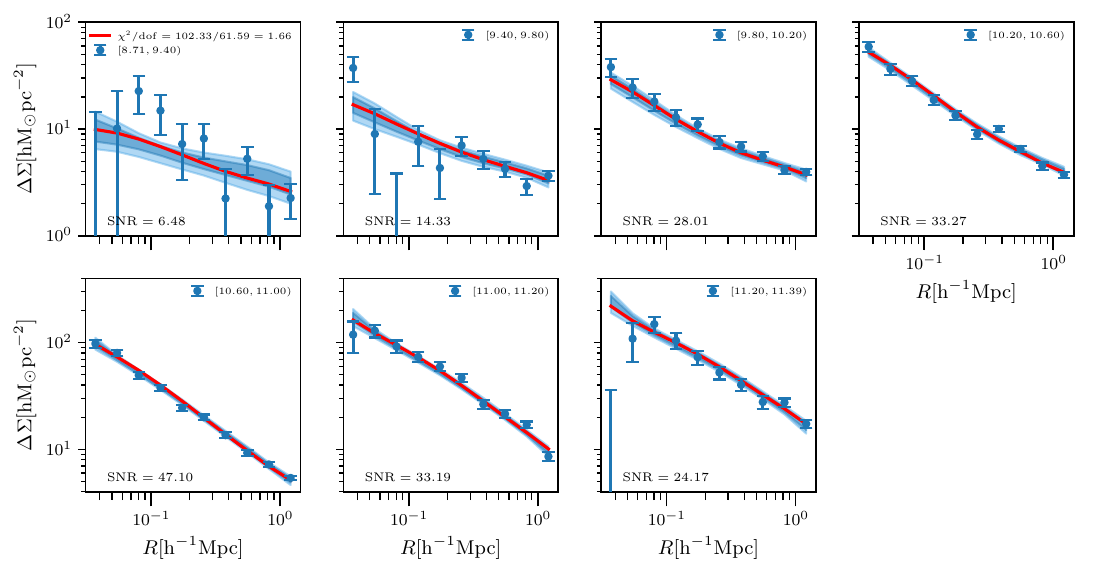}
    \caption{{\it Weak lensing signals:} The blue datapoint with errors in each panel represents the weak lensing signal measurements for the lenses in the stellar mass bin indicated at the bottom right corner. The red solid line represents the best fit model predictions for the joint fit to all seven stellar mass bins having $\chi^2$ value at the ${\rm dof}$ estimated using eqn 29 in \citet{2019_Raveri} as given in top of the first panel. The dark and light blue shaded regions corresponds to the $1\sigma$ and $2\sigma$ around the median model predicitions. The associated constraints on the model parameters are provided in the Table \ref{tab:modparam}.} 
    \label{fig:esd_sig}
\end{figure*}

We carry out a joint fit to the weak lensing measurements presented in each of the panels of Figure \ref{fig:esd_sig}. In Table \ref{tab:modparam}, we list the priors we assume on the different parameters along with the posterior distributions we obtain. The triangle plot corresponding to the various parameter degeneracies can be found in Appendix \ref{apx_wl_corner}. In Figure \ref{fig:esd_sig}, the solid red line corresponds to the best fit model prediction which has a $\chi^2 = 102.33$ for degrees of freedom of ${\rm dof} = 61.59$. The degrees of freedom were estimated by using eq 29 from \citet[][]{2019_Raveri}. The blue dark and light shaded regions denote the 1 and 2-$\sigma$ credible intervals around the median model predictions. In general, we find that our model provides a good statistical description of the data. In the last bin, we do see some evidence for the presence of mis-centering. Therefore, we have checked the effects of miscentering between galaxies and the host halo center. We used a Gaussian off-centering kernel for central galaxies in the halo model as shown in eqn 9 in \citet{2015_More} and found no change in our final results.

The constraints on the scatter in our SMHM relation $\sigma_0 = 0.11\pm 0.04$ is in good agreement with other results from abundance matching technique in galaxy groups \citep{2009_Yang}, using lensing and stellar mass function in galaxies \citep{Uitert2016} and joint modelling of clustering along with lensing \citep[][]{2018Dvornik}. We also find similar agreement in the inferred parameter $\alpha_{\rm s} = -1.35^{+0.17}_{-0.21}$, which characterizes the slope at the low stellar mass end of satellite CSMF, as well as the satellite fractions obtained in these studies. We compare our SMHMR  with the earlier results in Section \ref{sec:mstel_mhalo}.

\section{Results} \label{sec4}

We describe the main results of our work in this section. We start with a note on our adopted definitions of galaxy size, halo mass, and halo radius. Following \cite{Kravtsov2013}, we adopt the three dimensional (3D) half-light radius ($r_{1/2}$) as our galaxy size estimate and relate it to the measured projected two-dimensional (2D) half-light radius ($R_{\rm e}$) of galaxies. We assume that stars in late-type galaxies are in the disc, and hence the half-light radius $r_{1/2}$ is equal to $R_{\rm e}$. For early-type galaxies, the light distribution is more spheroidal and we use the relation $r_{1/2} = 1.34 R_{\rm e}$. \preetish{This relation is accurate for a wide variety of spheroidal light distribution described by S\'ersic profiles \citep{Neto1999} and has been verified in hydrodynamical simulation-based studies \citep{Graaff2022}}. We classify galaxies into early and late-type categories based on their S\'ersic index values. We labeled galaxies having S\'ersic index ($n$) values of $n\geq2.5$ as early-type while those with $n<2.5$ are termed as late-type. This S\'ersic index demarcation at $n=2.5$ to separate the galaxy population into early and late types has also been used in many previous studies of galaxy sizes \citep{Shen2003, Conselice2014, Lange2015}.   

For dark matter halo radius, we adopt the spherical overdensity radius $R_{\rm 200c}$ for a halo of mass $M_{\rm 200c}$. This radius encloses an average density of 200 times the critical density of the Universe ($\rho_{\rm c}$), such that $M_{\rm 200c}$ = (4$\pi$/3)$\rho_{\rm c} R_{\rm 200c}^{3}$. Even though our initial inference of halo properties from weak lensing follows a different halo mass definition ($M_{\rm 200m}$ (defined with respect to the mean matter density of the universe), we converted them to $M_{\rm 200c}$ in order to directly compare our results of the galaxy size-halo radius relation with previous works by \cite{Kravtsov2013} and \cite{Huang2017}. For this purpose, we use \textsc{COLOSSUS} \citep{Diemer2018} which is a python-based software capable of performing various calculations related to cosmology, the large-scale structure of the Universe, and the properties of dark matter haloes. 

In the following subsections, we present our result on the relationship between galaxy size - stellar mass, stellar mass-halo mass, and, the galaxy size-halo radius relations. 

\begin{figure*}
    \centering
    \includegraphics[width = 0.48\textwidth]{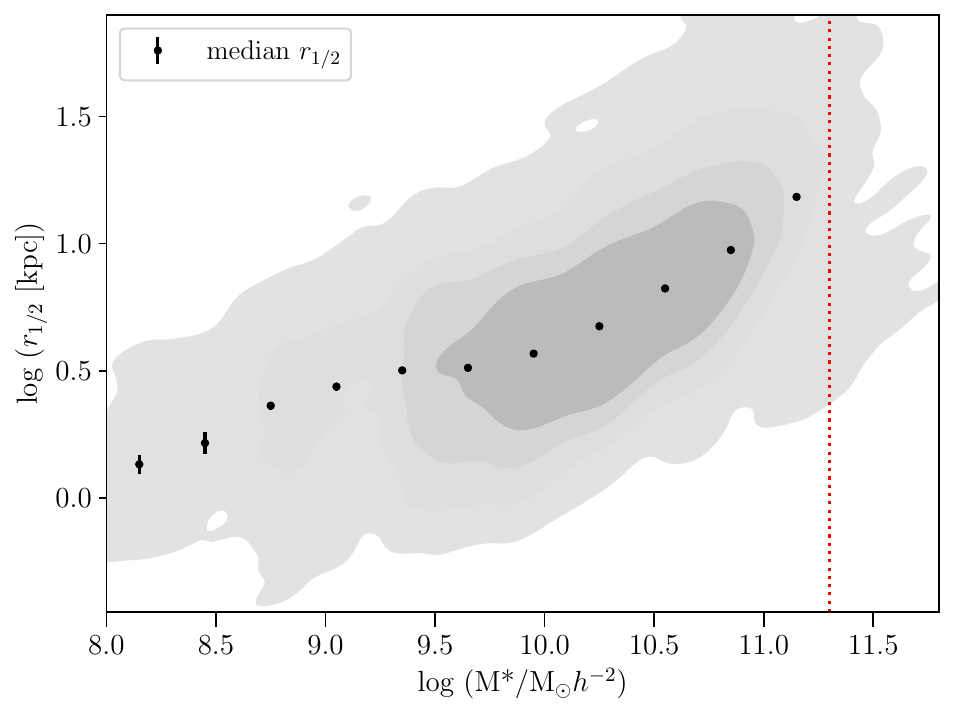}
    \includegraphics[width = 0.48\textwidth]{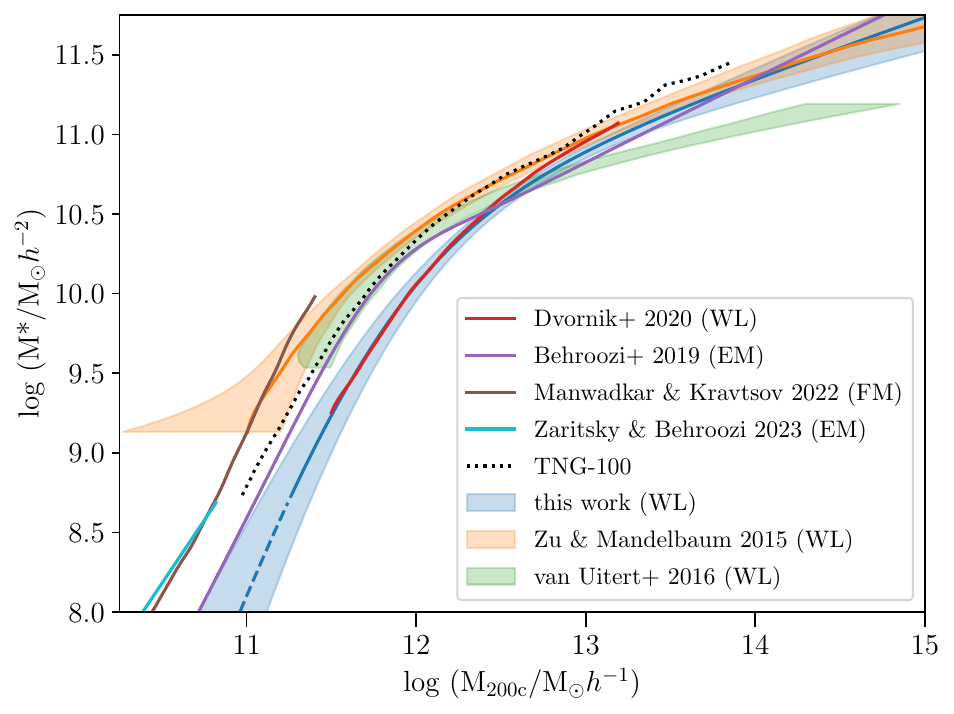}
    \caption{{\bf Left:} Distribution of sample galaxies in the size-stellar mass plane. The grey shaded contours enclose the 1$\sigma$-3$\sigma$ population in steps of 0.5$\sigma$. Black dots denote the median galaxy size ($r_{1/2}$) along with the Poisson error on the median displayed with the error bars. The red vertical line at $\log (M_*/h^{-2} {\rm M}_{\odot}$) = 11.3 mark the boundary of stellar mass beyond which our size-mass estimations are unreliable. {\bf Right:} Comparison of stellar mass-halo mass relation (SMHMR) obtained by us with the same from past studies. The median SMHMR  inferred by us and its extrapolation below the stellar mass of $10^{8.71} h^{-2} {\rm M}_{\odot}$ is shown by solid and dashed blue lines respectively. The blue-shaded region indicates 1$\sigma$ error on the median. A similar scheme is adopted for other SMHMRs shown in this figure where solid lines denote the best-fit/median relation and shaded regions are 1$\sigma$ errors.} 
    \label{fig4}
\end{figure*}

\begin{table*}
\centering

\caption{ Summary of median galaxy size ($r_{\rm 1/2}$) and galaxy counts in stellar mass bins used to construct galaxy size-stellar mass relation.}
\label{table 1}
\resizebox{\textwidth}{!}{

\begin{tabular}{llllllllllll}
\hline
 $\log(M^*/{\rm M}_{\odot}h^{-2}$)  & [8, 8.3) & [8.3, 8.6) & [8.6, 8.9) &[8.9, 9.2) & [9.2, 9.5) & [9.5, 9.8) & [9.8, 10.1) & [10.1, 10.4) & [10.4, 10.7) & [10.7, 11) & [11, 11.3)   \\ \hline
 Number of galaxies & 83 & 68 & 422 & 429 & 1339 & 3269 & 6412 & 8119  & 6949  & 4151  & 714  \\
 Median ($\log r_{1/2}$) [kpc]&0.131  &0.215  &0.362  &0.437  &0.501  &0.511  &0.567  &0.675  &0.823  &0.975  &1.184 \\
 \hline
\end{tabular}%
}

\end{table*}

\subsection{Galaxy size-stellar mass relation}
\label{sec:mstel_mhalo}
In order to construct the size-mass relation for galaxies, we first converted our two-dimensional galaxy size ($R_{\rm e}$) measurements into our adopted galaxy size definition of three-dimensional half-light radius ($r_{1/2}$) following the prescription described in paragraphs above. Next, we divided our sample into a number of stellar mass bins having a width of 0.3 dex and computed the median galaxy size associated with each bin. The plot on the right panel of Figure \ref{fig4} shows the distribution of our sample in galaxy size-stellar mass plane. The grey contours mark 1$\sigma$-3$\sigma$ of the population in steps of 0.5$\sigma$. The black dots with error bars denote the median value of the three-dimensional half-light radius ($r_{1/2}$) and its standard error: 1.253$\sigma/\sqrt(N)$ where $\sigma$ is the standard deviation and $N$ is the number of observations \citep{Williams2001}. The red vertical line marks the stellar mass of $10^{11.3} {\rm M}_{\odot}$. \preetish{We restrict our analysis of sizes to galaxies that have stellar masses below this limit. Above this mass limit, we have noticed inaccuracies in (SDSS photometry derived) stellar mass estimates for massive compact galaxies in the GAMA G02 field in \textsc{StellarMassesG02SDSS} catalog. The presence of these objects causes shredding of size-mass distribution beyond the stellar mass of $10^{11.4} {\rm M}_{\odot}$ as seen in the right panel of Figure \ref{fig4} that ultimately leads to median galaxy size-stellar mass relation to drop at the highest stellar mass end. This issue can be fixed by using stellar mass estimates from CFHT photometry derived \textsc{StellarMassesG02CFHTLS} catalog (available only for the G02 field) but that will make our sample inhomogeneous and can become a source of many systematic errors.}  

The number of galaxies and the median value of three-dimensional half-light radius ($r_{1/2}$) associated with the stellar mass bins used for further analysis are listed in Table \ref{table 1}. We find a monotonic increase in galaxy size with increasing stellar mass. However, the galaxy size-stellar mass relation shows a distinct break around stellar mass of $10^{9.35} h^{-2} {\rm M}_{\odot}$. Above this limit, the average slope $\frac{d \log r_{\rm 1/2}}{d \log M^*}$ of galaxy size-stellar mass relation steepens consistently with increasing stellar mass, as has also been noticed by previous studies \citep{Shen2003, Lange2015, Roy2018}. Below this stellar mass limit, however, the median galaxy size drops faster with decreasing stellar mass than the simple extrapolation of the relation observed beyond this limit. \preetish{Such a break or point of inflection has also been reported previously when the galaxy size-mass relation is studied using deep imaging data in nearby \citep{Trujillio2020}, intermediate \citep{Roy2018} or high-redshift \citep{Wel2023} universe. It is important to note that such a feature in galaxy size-mass relation is real and not an artifact of 2D to 3D size conversion as is shown in Appendix A of this paper.}

\subsection{Stellar mass-halo mass relation}
We obtain the stellar mass-halo mass relation of central galaxies from the CSMF modeling of the weak lensing signals which is described in detail in Section \ref{sec3}. This inferred stellar mass-halo mass relation (SMHMR) is shown in the right hand panel of Figure \ref{fig4} in blue. The blue line is the median SMHMR and the blue shaded region denotes the 1$\sigma$ error around the median. It must be noted that this relation was derived using only galaxies more massive than $10^{8.71} h^{-2}  {\rm M}_{\odot}$ but we extrapolate the resulting stellar mass-halo mass relation below this stellar mass limit to make an inference about dark matter haloes of lowest stellar mass galaxies in our sample. We have shown this extrapolation as a dashed right hand panel of Figure \ref{fig4}. \preetish{We also compare our stellar mass-halo mass relation with a number of existing relations obtained from weak lensing \citep{Zu2015, Uitert2016, Dvornik2020}, forward modeling \citep{Manwadkar2022}, empirical modeling \citep{Behroozi2019, Zaritsky2023}, and hydrodynamical simulation TNG \citep{Nelson2019}.} In all the cases the solid lines represent the best fit or median SMHMRs and shaded regions are error and scatter. Whenever we do not show the scatter for results from previous works it is because either this information is not available or in some cases, the (large) scatter reduces the clarity of this figure. We refer the reader to the original source of these relations to obtain extra information not conveyed in this plot. As was mentioned earlier at the start of this section, we have used the software \textsc{COLOSSUS} to convert the halo mass definition in these studies to be consistent with the definition we use, whenever necessary.   

We note that our stellar mass-halo mass relation is qualitatively consistent with the ones from the literature, especially at the high halo mass end. \preetish{At halo masses lower than $10^{12.5} {\rm M}_{\odot}$, we observe a difference of $\sim$ 0.5 dex in stellar mass at fixed halo mass between our SMHMR and the same derived using the weak lensing method measurements \cite{Zu2015} from SDSS, \cite{Uitert2016} from GAMA combined with KiDS, and by model-based estimates of \cite{Behroozi2019}, \cite{Manwadkar2022} and \cite{Zaritsky2023}.} We do not understand the exact origin of this difference; possibilities include differences in sample selection, redshift range probed, and the methodology. The use of different stellar population synthesis models and dust models may change stellar mass measurements by $\sim$0.1 dex \citep{Behroozi2019}. Moreover, the methodology of galaxy flux measurements can also significantly impact the estimation of stellar masses. \cite{Behroozi2019} have shown that the difference in stellar masses computed using SDSS cmodel mags vs. those estimated with galaxy flux extracted using a parametric model (S\'ersic, S\'ersic+exponential) fitting can be in the order of $\sim$ 0.1 dex at low ($<10^{11} {\rm M}_{\odot}$) and $\sim$ 0.45 dex at the high ($10^{12} {\rm M}_{\odot}$) stellar mass end. Such differences in stellar mass estimates may directly translate into differences in stellar mass-halo mass relations.  

However, we also observe the consistency of our stellar mass - halo mass relation with the same derived using 1-dimensional analysis of weak lensing signals of GAMA galaxies using KIDS survey data by \cite{Dvornik2020}. Additionally, there is an increasingly better agreement with our SMHMR with the same coming from the empirical modeling scheme of \cite{Behroozi2019} at lower stellar masses within the errors. This gives us confidence in the validity of our weak lensing derived stellar mass - halo mass relation and we proceed to use it to infer the connections between the sizes of galaxies and their dark matter haloes. Later in this paper, we also show the qualitative agreement between galaxy size-halo radius relation derived using our SMHMRs and the ones mentioned above. 

\subsection{Galaxy size-halo radius relation}

\begin{figure}
    \centering
    \includegraphics[width = 0.48\textwidth]{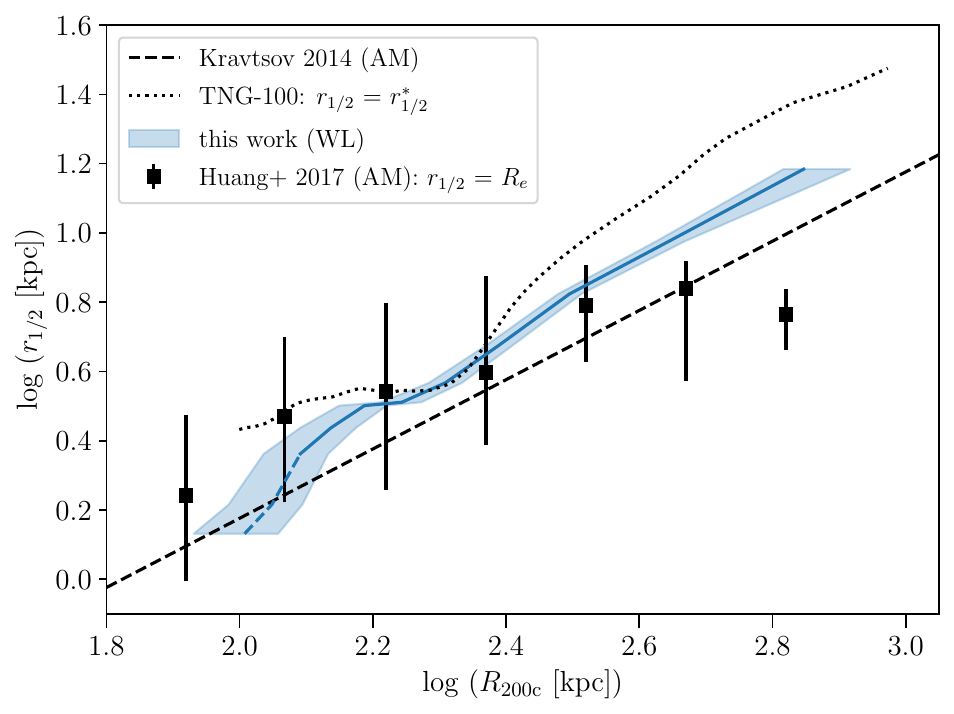}
    \caption{The median galaxy size-halo radius relation is shown in the blue line with shaded blue regions denoting error on the median. The solid and dashed blue lines mark results obtained using our inferred SMHMR and its extrapolation to stellar masses below $10^{8.71} h^{-2}  {\rm M}_{\odot}$ respectively. A similar scheme is adopted for all other plots in this manuscript.  We compare galaxy size-halo radius relation obtained by us with the results from abundance matching (AM) works of \citet[][black dashed line]{Kravtsov2013}, \citet[][black squares with scatter]{Huang2017}, and hydrodynamical simulation TNG-100 (black dotted line). Note that the size definition adopted in  result \protect\cite{Huang2017} and TNG-100 is two dimensional half-light radius ($R_{\rm e}$) and half-stellar mass radius ($r^*_{1/2}$). \preetish{Also, note that the displayed result from \protect\cite{Kravtsov2013} is an average characterization of the linear galaxy size-halo radius relation and not a statistical fit. Readers are referred to the main text of this paper and \protect\cite{Kravtsov2013} for further details. All other plots in this paper showing galaxy size-halo radius relation from \protect\cite{Kravtsov2013} should be read with this caveat in mind.}} 
    \label{fig5}
\end{figure}

\begin{figure*}
    \centering
    
    \includegraphics[width = 0.48\textwidth]{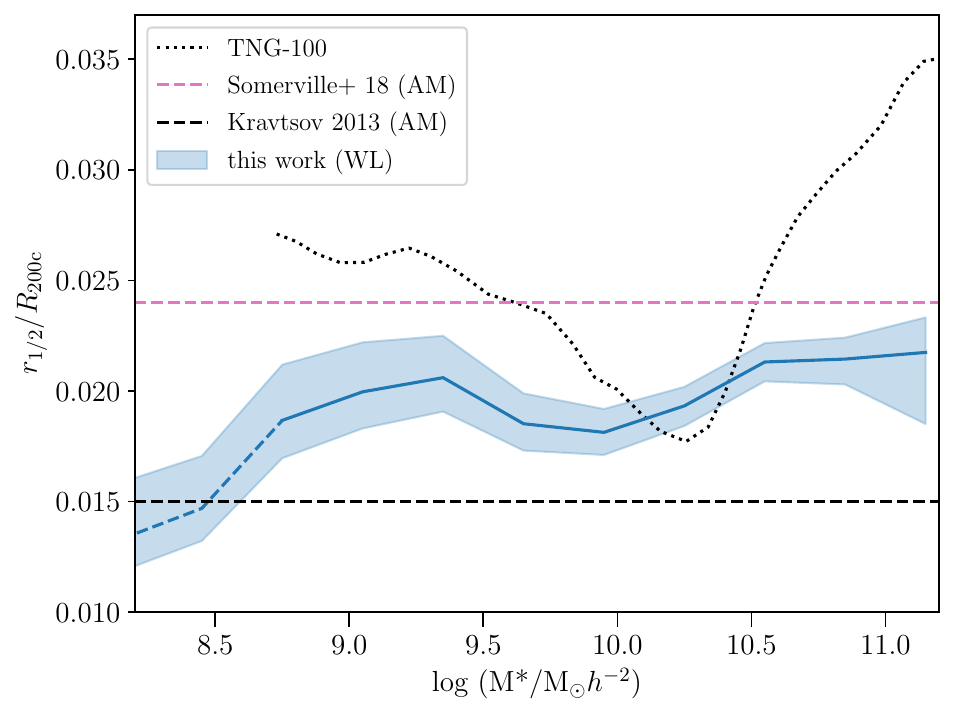}
    \includegraphics[width = 0.48\textwidth]{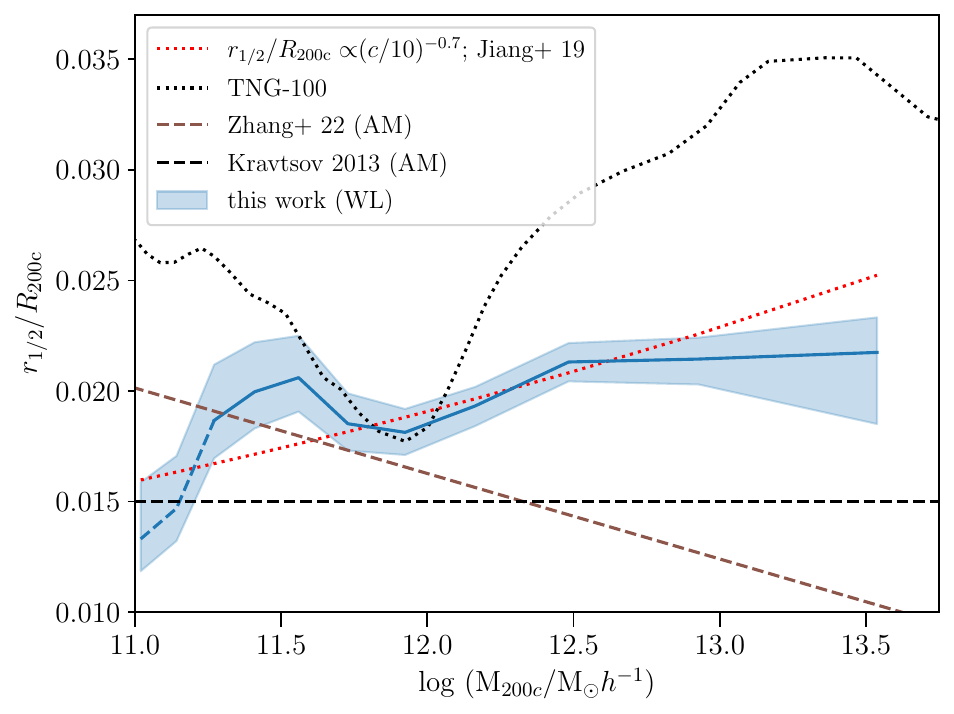}
    \caption{{\bf Left:} Ratio of galaxy size to halo radius as a function of stellar mass. The blue line and shaded regions indicate the median $r_{1/2}$/$R_{\rm 200c}$ and error on the median respectively. The solid and dashed blue lines mark results obtained using our inferred SMHMR and its extrapolation to lower stellar masses respectively. The average relation of galaxy-halo radius ratio from \protect\cite{Somerville2018} and the one obtained from TNG-100 is represented by a pink dashed line and a black dotted line respectively. For reference, we also over plot \protect\cite{Kravtsov2013} relation in the black dashed line in this figure. {\bf Right:} Similar to the figure on the left but now the stellar mass axis is replaced by dark matter halo mass. The markers and colour scheme remain the same as in the left figure except. We overplot the relationship between  $r_{1/2}$/$R_{\rm 200c}$ vs halo mass obtained by \protect\cite{Zhang2022} as brown dashed line. The red dashed line denotes the galaxy-halo radius ratio of the form $r_{1/2}$/$R_{200}$ = 0.0125 $(c/10)^{-0.7}$ similar to the result of \protect\cite{Jiang2019} plotted as a function of halo mass.} 
    \label{fig6}
\end{figure*}

We now combine the results of the galaxy size-stellar mass relation and the stellar-mass halo mass relation to link sizes of galaxies and their dark matter haloes. We have an estimate of the median galaxy size in a number of stellar mass bins of width 0.3 dex spanning the stellar mass range log($M^*$/${\rm M}_{\odot}h^{-2}$) = [8, 11.3] as shown in Table \ref{table 1}. We assign these median galaxy sizes to the central stellar mass values of the bins. Corresponding to these particular stellar mass values, we get halo mass estimates from the stellar mass-halo mass relation. This method allows us to probe the relationship between median galaxy size and dark matter halo radius with stellar mass acting as an intermediary. Note that our derived stellar mass halo mass relation is strictly applicable to central galaxies only. On the other hand, we have made no distinction between central and satellite galaxies while making the galaxy size-stellar mass relation. This is because past studies \citep{Huertas-Company2013, Wang2020, Rodriguez2021} have shown that galaxy size-stellar mass relations for central and satellite do not significantly differ from one another. Therefore, the full galaxy size-stellar mass relation can be taken as a representative of the same relation for central galaxies.

Using the method described above, we obtain the relationship between median galaxy size and halo radius. We plot this relation in Figure \ref{fig5} using a blue line. The shaded blue region denotes the 1$\sigma$ error on the median. We display the results on galaxy size-halo radius relation based on the abundance matching work of \cite{Kravtsov2013} and \cite{Huang2017} which are shown using a black dashed line and square boxes with error bars denoting 16th–84th percentile scatter on galaxy size, respectively. We also plot the relationship between galaxy size and halo radius at redshift $z\sim0.1$ from the hydrodynamical simulation TNG-100 \citep{Nelson2019} in a dotted black line. We obtain this relation from the median relationship between the stellar half-mass radius of central galaxies and their host dark matter halo masses in TNG simulation accessed through their web-based group catalog data interface\footnote{https://www.tng-project.org/data/groupcat/}. Note that the size definition in \cite{Huang2017} and TNG-100 are two-dimensional half-light radius and three-dimensional stellar half-mass radius, respectively. The conversion from two-dimensional size to three-dimensional size following our convention is likely to cause a maximum shift of results from \cite{Huang2017} only by 0.12 dex along the galaxy size axis. We mark the differences in the size definition in the caption and legend of this plot.

\preetish{Looking at Figure \ref{fig5}, we find that the median galaxy size-halo radius relation that we obtain displays linear behavior for halo radius above log ($R_{\rm 200c}$[kpc]) = 2.2, in qualitative consistency with the conclusion of \cite{Kravtsov2013}, albeit with a higher intercept on the galaxy size axis. However, below this limit, the galaxy size-halo radius relation shows an indication of deviation from linearity in the form of gradual down bending as one goes to lower halo radii. We find our results to be roughly consistent with \cite{Huang2017} in the intermediate halo radius range but deviate away from it in the (lower) higher halo radius ends where they (we) observe a downturn of this relation. \cite{Huang2017} speculate this turnover could be driven either by biases in the size measurements of high mass galaxies or a breakdown of their abundance matching scheme for group and cluster scale haloes. The relation between the sizes of galaxies and haloes from TNG-100 appears to be shifted upwards along the size axis as compared to our result. This could be due to the fact that the massive galaxies in TNG-100 tend to have more extended stellar mass distributions as compared to observed galaxies \citep{Ardila2021}. However, we caution that this could be also a result of differences in the stellar half-mass radius and the half-light radius measured from light.} The measurement of galaxy sizes using mock images from the TNG simulation in the $i$-band would have undoubtedly led to a better comparison, but performing such a comparison is beyond the scope of this work. The main takeaway from the result presented in Figure \ref{fig5} is the linearity of the median galaxy size-halo radius relation for galaxies over two orders of magnitude in stellar mass and an indication of a nonlinear relation for low stellar mass galaxies residing in smaller mass haloes.

\preetish{We explore the possible nonlinear behavior by probing the link between the ratio of median galaxy size ($r_{1/2}$) to halo radius ($R_{\rm 200c}$) and its dependence on the masses of galaxies and haloes in Figure \ref{fig6}.} In case of a linear galaxy size-halo radius relation of form, $r_{1/2}$ = constant x $R_{\rm 200c}$, the ratio $r_{1/2}$/$R_{\rm 200c}$ will simply be the slope of this relation. We plot the size of galaxies and haloes as a function of stellar mass in the left panel of Figure \ref{fig6}. The blue line denotes the median relation between the size ratio $r_{1/2}$/$R_{\rm 200c}$ and galaxy stellar mass while the blue shaded region represents its 1-sigma error. We compare this trend with the result from \cite{Somerville2018} who find the ratio of galaxy size to halo viral radius to be roughly equal to 0.018 on average at redshift $z\sim0.1$. We translate their results by converting the halo viral radius to our adopted definition of halo radius of $R_{\rm 200c}$ using COLOSSUS and then plot the same in the left hand panel of Figure \ref{fig6} in pink dashed line. \preetish{We also display the approximate determination of galaxy-halo radius ratio (equal to 0.15) from \cite{Kravtsov2013} as a black dashed line for reference. It must be noted that the galaxy sample used by \cite{Kravtsov2013} was a compilation of galaxies from different surveys along with their often differently measured sizes. Due to these factors bringing in heterogeneity in their sample, the ratio 0.15 of galaxy-to-halo size as quoted by \cite{Kravtsov2013} is an average value aimed at characterizing the linear galaxy size-halo radius relation and not a statistical fit. Similarly, note that we display the average galaxy size-halo radius relation from literature \citep[][in Figure~\ref{fig6} and later]{Somerville2018, Zhang2022, Jiang2019} as a line without errors to compare with our results. We do this because many times the statistical error on these results is not explicitly quoted. We refer interested readers to the original papers for further details regarding the errors on the galaxy size-halo radius relation found by previous works in the literature.  }

\preetish{The left hand panel of Figure \ref{fig6} examines the stellar mass dependence of galaxy size-halo radius relation. Looking at this plot, we confirm that galaxy size-halo radius relation is roughly linear over two orders of magnitudes in stellar mass above $\sim 10^{9} {\rm M}_{\odot}h^{-2}$. Below this limit, however, the ratio of galaxy size to halo radius seems to be gradually declining. There is about a 30\% decline in the ratio as one moves from the stellar mass of $\sim 10^{9} {\rm M}_{\odot}h^{-2}$ to $\sim 10^{8.15} {\rm M}_{\odot}h^{-2}$.} We also display the ratio of the half-mass radius of central galaxies and the radius of their host haloes in the TNG-100 simulation. This was obtained by combining the stellar mass-halo mass relation and the half-mass radius-stellar mass relation for central galaxies in TNG-100. We see that TNG-100 predicts the galaxy-halo radius ratio to first decrease, reach a minimum and then increase again as one moves from low to higher stellar masses. This is in conflict with our results as well as those from previous studies. However, as was pointed out earlier the apples-to-apples comparison of results will require us to perform mock observation on simulated galaxy images which is beyond the scope of this work.

The right hand panel of Figure \ref{fig6} shows the ratio $r_{1/2}$/$R_{\rm 200c}$ as a function of the dark matter halo mass. Colours and symbols carry the same meaning as in the plot on the left. Here we show the dependence of the galaxy-halo radius ratio on the halo mass and compare it with the predictions from \cite{Zhang2022} and \cite{Jiang2019}. Using abundance matching ansatz to link SDSS galaxies to dark matter haloes in ELUCID simulation, Zhang et al(2022) report that the ratio of the galaxy–halo radius ratio $r_{1/2}$/$R_{\rm 200c}$ decreases with increasing halo mass. They parameterize this via a best-fitting linear relation of the form $r_{1/2}$/$R_{\rm 200c}$ = 0.01 x (4.7 - 0.29 $\log M_{vir}$). We have converted their virial halo mass ($\log M_{vir}$) estimate into our adopted definition of halo mass ($M_{\rm 200c}$) and have re-plotted the aforementioned relation as a brown dashed line in this plot. Our results do not agree with those presented by \cite{Zhang2022}. \preetish{Instead of the predicted decrease of $r_{1/2}$/$R_{\rm 200c}$, we find that our computed value of galaxy-halo radius ratio shows an increase by $\sim$ 30\% as we move from our lowest mass haloes till a halo mass log($M_{\rm 200c}$/${\rm M}_{\odot}h^{-2}$) = 11.6. Above this halo mass limit, the curve flattens and can be considered constant for roughly two orders of magnitude in halo mass.} The mismatch between our results and those from \cite{Zhang2022} could be due to differences in sample selection for the construction of galaxy size-stellar mass relation. \cite{Zhang2022} use a flux-limited sample of SDSS galaxies within a redshift range of $0.1\leq z\leq0.2$ that is selected without consideration of galaxy colors. As discussed in Section 2, using a simple flux-limited sample can result in an overrepresentation of blue/star-forming galaxies at fixed stellar mass. The median size of blue/starforming galaxies is higher compared to red/quenched galaxies at fixed stellar mass. This can bring significant changes to the galaxy size-stellar mass relation and simultaneously affect the galaxy size-halo radius relation. Lastly, our work also differs from \cite{Zhang2022} with respect to the methodology of connecting galaxies to dark matter haloes.  

\cite{Jiang2019} have studied the relation of galaxy size and halo properties using NIHAO and VELA simulations. They find that galaxy half-light radius ($R_{e}$) to be related with halo virial radius ($R_{vir}$) in following form: $R_{e}$ = 0.02 $(c/10)^{-0.7} R_{vir}$ where $c$ is the concentration of dark matter halo. According to this prediction, the galaxy-halo radius ratio should scale proportional to $(c/10)^{-0.7}$. We use the concentration-halo mass relation by \cite{Maccio2007} to infer the concentration corresponding to our measured halo masses and plot a galaxy-halo radius relation of the form $r_{1/2}$/$R_{\rm 200c}$ = 0.0125 $(c/10)^{-0.7}$ as a red dashed line in the right hand panel of Figure \ref{fig6}. Note that we have changed the normalization of relation proposed by \cite{Jiang2019} to better compare their prediction with our result on the variation of galaxy-halo radius ratio with halo mass. We can see that assuming a concentration-dependent relation of the form mentioned above leads to a monotonic rise in $r_{1/2}$/$R_{\rm 200c}$ with increasing halo mass. Even though there seems to be a rough qualitative agreement between our results and the concentration-dependent galaxy-halo radius relation in intermediate halo masses, we see indications for a deviation from this prediction at the low and high halo mass ends.   

\preetish{The results presented in Figures \ref{fig5} and \ref{fig6} show evidence that the median galaxy size-halo radius relation is roughly linear for galaxies more massive than $\sim 10^{9} {\rm M}_{\odot}h^{-2}$. We also find an indication of a different non-linear stellar mass-dependent relationship between the sizes of galaxies less massive than $\sim 10^{9}{\rm M}_{\odot}h^{-2}$ and their host dark matter halo radii. Interestingly, this is also the stellar mass range where galaxies follow different size-mass relations as compared to their more massive counterparts. The observed indication of the stellar mass dependence of the galaxy size-halo radius relation has not been seen in previous works. We discuss various issues that could lead to an indication of such non-linearity and the implications of its existence in the next section.} 

\begin{figure*}
    \centering
    
    \includegraphics[width = 0.48\textwidth]{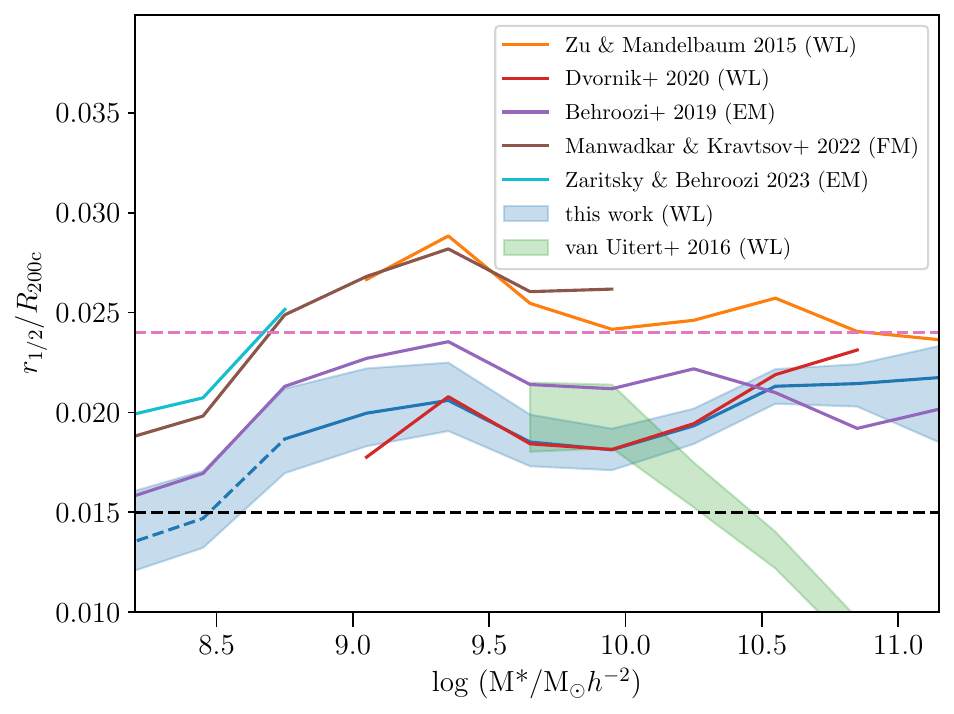}
    \includegraphics[width = 0.48\textwidth]{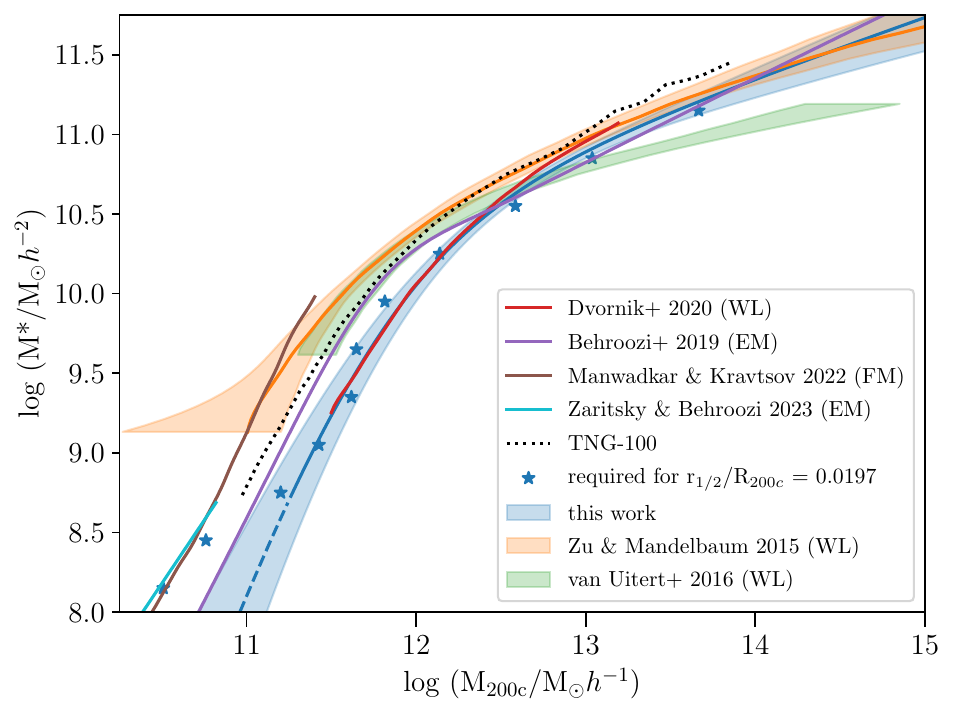}
    \caption{{\bf Left:} Affect of adopting different SMHMR on the trend of median $r_{1/2}$/$R_{\rm 200c}$ with stellar mass. The different colour represent the median galaxy-halo radius relation as a function of stellar mass remade with SMHMRs displayed in Figure \protect\ref{fig4} same colours. Over plotted is our result (in blue) along with results from \protect\cite{Somerville2018} and \protect\cite{Kravtsov2013}. {\bf Right:} Similar to the right panel plot of \protect\ref{fig4} but now overplotted the prediction on halo masses (blue stars) if there existed a universal linear galaxy size-halo radius relation of the form $r_{1/2}$ = 0.0197 $R_{\rm 200c}$.}
    \label{fig7}
\end{figure*}

\section{Discussion} \label{sec5}

The linearity of the galaxy size-halo radius relation that we infer for intermediate to high mass galaxies, its consistency with previous works and the indication we obtain for a departure from this linearity for low mass galaxies opens up interesting avenues to understand galaxy formation. However, prior to interpreting these results physically, we discuss a number of factors that can potentially affect these results.

We derived our galaxy size-halo radius relation by combining the galaxy size-stellar mass relations and the stellar mass halo mass relation (SMHMR). The SMHMR is thus an important ingredient and any error in its inference could possibly alter our results. Although we obtain a qualitative agreement between our obtained SMHMR and those existing in literature obtained from a variety of methods, there exist slight differences with respect to the slope, the position of the knee, and the halo masses of low mass galaxies which can be seen in Figure \ref{fig4}. To understand the impact of these differences on our conclusions, in Figure \ref{fig7}, we use a variety of different SMHMRs (shown in Figure \ref{fig4}) obtained using various methodologies on different data sets and present the inferred galaxy size and halo radius relations. \preetish{Two of these SMHMRs, those coming from works of \cite{Manwadkar2022} and \citep{Zaritsky2023} are specifically aimed at studying the dark matter content of dwarf galaxies.} Our results are shown as blue shaded regions in this figure for comparison. We notice that the use of different SMHMRs does not change the qualitative nature of our results. The ratio of the galaxy size to halo radius still shows an increase at the low mass end, and a subsequent flattening as one moves from a low to a higher stellar mass range, although the average value of this ratio can be shifted. There is also a hint of decreasing median value of galaxy size-halo radius ratio with stellar mass for galaxies more massive than $10^{10.6} {\rm M}_{\odot}h^{-2}$. But this seems to depend on the shape of SMHMRs at high halo mass end; shallower slopes typically translate to a prominent decrease of galaxy size-halo radius ratio with increasing stellar mass. This effect is most prominent for the galaxy size-halo radius determined with the SMHMR of \cite{Uitert2016}. The reader must note that due to the differences in the data sets, the entirely flux-limited sample selection, and the methodology adopted to obtain these SMHMRs, an exact comparison of the effect of using different SMHMRs on galaxy size-halo radius connection is not possible. Although given the rough agreement on galaxy size-halo radius relation by the majority of other SMHMRs, we do not think that differences in the inferred SMHMR can significantly alter the qualitative conclusions drawn from our results, especially the observed increase in galaxy-halo radius ratio as a function of stellar mass for dwarf/low-mass galaxies.

The other source of uncertainty in our results could be our lack of knowledge about the halo masses of low-mass dwarf galaxies. Our SMHMR is obtained by modeling weak lensing signals for sample galaxies more massive than $10^{8.7} {\rm M}_{\odot}h^{-2}$. We extrapolated our SMHR to assign halo masses to galaxies less massive than this limit even though we do not have weak lensing constraints on their halo masses. This limitation arises from the lack of enough low mass galaxies in large spectroscopic surveys such as GAMA to enable sufficient signal-to-noise weak lensing signals around them. This factor has forced many past works on galaxy-halo connections via weak lensing targeting only galaxies more massive than  $10^{8.7-9.5} {\rm M}_{\odot}h^{-2}$.

If our halo mass estimates for low-mass dwarf galaxies are off then it will significantly impact our results on galaxy size-halo radius relation at the low stellar mass end. Although there are methods available to infer the dark matter content of dwarf galaxies such as via modeling rotation curve data, systematic effects could potentially play an important role (see \cite{Leauthaud2020} and references therein). Given these issues, we predict the halo masses of these galaxies if the galaxy size-halo radius has a linear relation. \preetish{In order to do so, we notice that above the stellar mass of $10^{9} {\rm M}_{\odot}h^{-2}$ the size ratio $r_{1/2}$/$R_{\rm 200c}$ has a mean value of 0.0197. Assuming that there is a universal linear relation between the galaxy size ($r_{1/2}$) and halo radius ($R_{\rm 200c}$) of the form $r_{1/2}$ = 0.0197 $R_{\rm 200c}$, we calculate the halo masses of galaxies in stellar mass bins of the width of 0.3 dex spanning our sample stellar mass range.} We plot these re-estimated halo masses as blue stars in the stellar mass-halo mass plane in the right panel of Figure \ref{fig7}. For comparison, we also over-plot the SMHMRs obtained by us and those taken from literature in the same figure. One can notice from this figure that a universal and linear relationship between galaxy size-halo radius would require significantly lower values of halo masses for dwarfs than what we currently infer based on our model. This difference in halo masses can be as large as almost an order of magnitude for $\sim 10^{8} {\rm M}_{\odot}h^{-2}$ stellar mass galaxies. Hopefully, future weak lensing studies of dwarf galaxies using deeper surveys will be able to shed more light on this issue.  

\preetish{A final consideration comes from the surface brightness incompleteness from the GAMA survey which could manifest itself as missing out on low surface brightness (LSB) galaxies. The LSB galaxies usually have a larger effective radius than their equally massive high surface brightness counterparts. Therefore, if we are missing low-mass LSB galaxies in our sample then it could be a reason for declining $r_{1/2}$/$R_{\rm 200c}$ as one moves from higher to lower stellar masses. Indeed the presence of low surface brightness (LSB) galaxies can bias our size-mass relation. However, there are a number of observational works that argue that such galaxies contribute very little in terms of numbers, at least to a stellar-mass limit $10^{8} {\rm M}_{\odot}$ which is similar to the stellar mass limit of our sample.}

\preetish{\cite{Williams2016} have searched for LSB galaxies in GAMA fields using deep VISTA survey $z$ band images. Their aim was to see if the presence of LSB galaxies affected the stellar mass function measurement of \citep{Baldry2012} which was done using GAMA galaxies having stellar masses greater than  $10^{8} {\rm M}_{\odot}$. \cite{Williams2016} found 346 LSB galaxies in the GAMA survey fields, however, none were brighter than the GAMA survey flux limit of $r=19.8$ and hence are not contaminant to the GAMA survey galaxy sample. Recently, \citep{Greene2022} have inferred the stellar mass function of the LSB galaxies in relation to HSB ($\mu<24.3$ mag/arcsec$^2$; similar to GAMA survey surface brightness limit) galaxies. They find that the number density of LSB galaxies at the stellar mass of $10^{8} {\rm M}_{\odot}$ is approximately 1-2 orders of magnitude less than HSB galaxies. They conclude that the LSB galaxies make up a small fraction of the galaxy population at this mass. A similar result was also found by \cite{Jones2018} who compared the stellar mass function of HI-selected ultra-diffuse galaxies with HSB galaxies in GAMA survey. Based on these results, it seems unlikely that the trend of declining $r_{1/2}$/$R_{\rm 200c}$ with decreasing stellar mass originates from missing LSB galaxies in our sample.}

Having stated the above caveats, which can potentially impact our results, we proceed to discuss the possible implication of our results for galaxy formation models. 

The classic disc galaxy formation models of \cite{Fall1980} and \cite{Mo1998} use the angular momentum conservation principle to assign sizes to galaxies. In their model, the infalling baryons acquire a fraction of the angular momentum of the dark matter haloes and settle to form a disc with scale length ($R_d$) given by $ R_{d} = 1/\sqrt{2}. f_{\rm j} . \lambda . R_{\rm 200c} $, where $\lambda$ is the halo spin and $f_{\rm j}$ = $\frac{j_{gal}}{j_{DM}}$ is the ratio of specific angular momentum of disc galaxy to that of the dark matter haloes. In the simplest of cases, the baryons are thought to retain a fixed amount of angular momentum of their host haloes which translates to a constant value of $f_{\rm j}$. A number of observational works on galaxy kinematics offer support to this assumption \citep{Burkert2016}. It implies then that galaxy size is linearly related to halo radius on an average via $R_d \sim$ constant x $ R_{\rm 200c}$ because the average value of halo spin $\langle \lambda \rangle$ is also constant in the range 0.02-0.05 and shows no strong dependence on halo mass \citep{Burkert2016}. This linearity is also seen in the abundance matching works of \cite{Kravtsov2013}, \cite{Huang2017}, and \cite{Somerville2018}. \preetish{Our results do agree on the rough linearity of galaxy size-halo radius relation for intermediate to high-mass galaxies but also indicate a stellar mass-dependent relation in the dwarf galaxy sector.} This might imply that classic models assigning galaxy sizes based on halo spin may potentially require modification. 
 
\begin{figure}
    \centering
    \includegraphics[width = 0.48\textwidth]{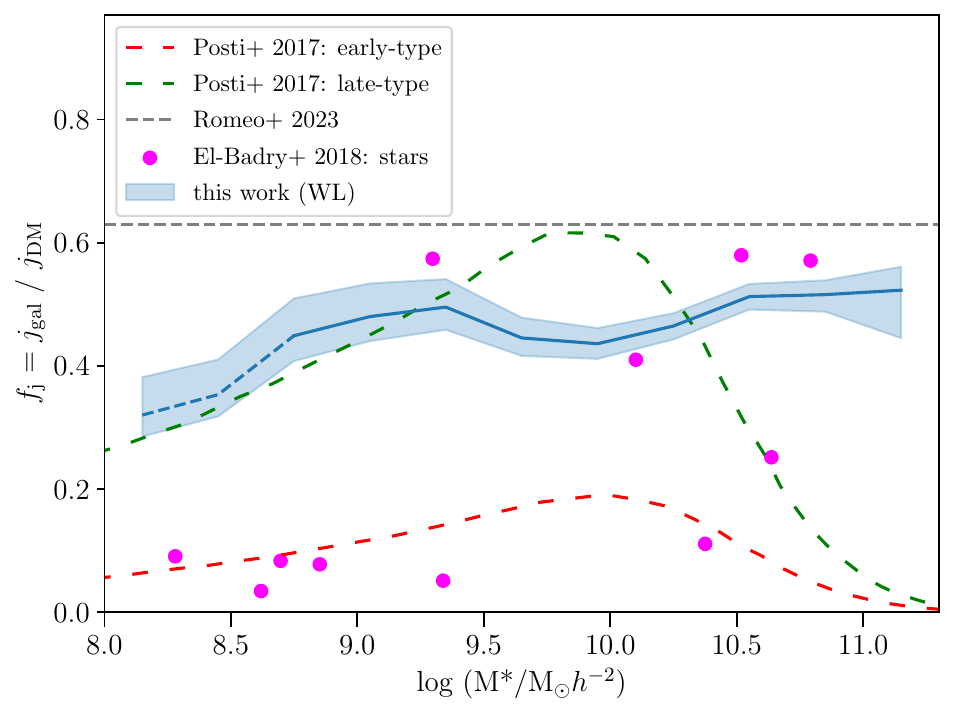}
    
    \caption{Fraction of halo angular momentum retained by galaxies ($f_{\rm j}$) as function of stellar mass estimated using combining our results on galaxy size-halo radius relation with simple disc galaxy formation model of \protect\cite{Mo1998}. Similar results from \protect\cite{Posti2018} on spirals (in green) and early-type (in red) galaxies are also shown in this figure. The pink dots display the fraction of halo angular momentum retained by stars in galaxies in FIRE simulation coming from work of \protect\cite{El-Badry2018}. The horizontal grey dashed line marks the median value of retained halo angular momentum in stars from the work of \protect\cite{Romeo2023} } 
    \label{fig8}
\end{figure}

One such possible modification in our canonical model of galaxy size could be that the angular momentum retention fraction is stellar mass dependent. Noting that $r_{1/2} \approx R_{\rm e} = 1.68 R_{\rm d}$, we can write $f_{\rm j} = (\sqrt{2}/1.68)$ . ($r_{1/2}$/$R_{\rm 200c}$) . $1/\langle \lambda \rangle$. Adopting a \preetish{value of $\langle \lambda \rangle$ = 0.035} following \cite{Maccio2007}, we can use our measured value of $r_{1/2}$/$R_{\rm 200c}$ to indirectly measure the fraction of halo angular momentum retained ($f_{j}$) by galaxies. In the left panel of Figure \ref{fig8}, we plot the angular momentum retention fraction $f_{j}$ as a function of galaxy stellar mass based on our results. The blue line denotes the median value of $f_{j}$ calculated using the procedure described above and the shaded region is the error on the median. If the canonical galaxy size prescription is valid, then this plot implies that retention of angular momentum in dwarfs is likely inefficient by as much as a factor of two as compared to high-mass galaxies. 

Using an analytical prescription to combine SMHMRs from literature and the Fall relation, \cite{Posti2018} have also reported that the retained fraction of angular momentum ($f_{\rm j}$) to be linked with masses of galaxies and their host dark matter mass. They find that the trend of $f_{\rm j}$ with either stellar mass or halo mass can be represented by a double power law with different normalizations for spiral and elliptical galaxies. We show their median results (derived using the \cite{Behroozi2013} SMHMR) for late and early-type galaxies as green and red dashed lines on Figure \ref{fig8}, respectively. We see that their predicted $f_{\rm j}$ shows an initial increase with stellar mass which is consistent with our results. We also observe rough quantitative agreement between our estimated $f_{\rm j}$ and the same predicted for late-type galaxies by \cite{Posti2018} at stellar masses below $\sim 10^{9.35} {\rm M}_{\odot}h^{-2}$. Even though our work has been done without regard to morphology, it is known that late-type galaxies dominate the lower stellar mass regime of the galaxy population in the low-redshift universe, making this a fair comparison. However, the angular momentum retention fraction shows a decreasing trend with increasing stellar mass for galaxies more massive than $\sim 10^{10.6} {\rm M}_{\odot}h^{-2}$ whereas we observe $f_{\rm j}$ to be roughly constant in this mass range.

Similar mass dependence of angular momentum retention fraction has been found by \cite{El-Badry2018} who studied the relation between the angular momentum of galaxies and their host dark matter haloes using suite of cosmological zoom-in simulations from the FIRE project. For comparison, we have displayed the fraction of halo angular momentum retained by stars in galaxies as found by \cite{El-Badry2018} as pink dots in the left panel of Figure \ref{fig8}. The massive galaxies from FIRE show values of $f_{j}$ which are consistent with our inferences from observations. Qualitatively, they also find a monotonic decrease in the angular momentum retention fractions with decreasing galaxy mass. According to \cite{El-Badry2018}, low-mass haloes accrete gas less efficiently compared to high-mass haloes, especially at a late cosmic time when the average specific angular momentum of accreted gas is highest. This leads to the formation of dwarfs galaxies with low values of halo angular momentum fraction.

\preetish{The stellar mass dependence of angular momentum retention fraction is, however, not supported by all works. In recent work, \cite{Romeo2023} have studied the retained fraction of stellar and gas angular momentum in galaxies using kinematic measurements. They find the median value of retained angular momentum in stars to be around 0.63 independent of the stellar mass. We have shown their result in Figure \ref{fig8} as a horizontal grey dashed line. Our results are in qualitative agreement with the results of \cite{Romeo2023} in intermediate to high mass ends but differ from those at the lowest mass end.}

\cite{Gomez2022} have studied the angular momentum of IllustrisTNG galaxies and its connection to kinematic and visual morphology. They find that early and late-type galaxies retain $\sim$10-20 and $\sim$50-60 percent of halo angular momentum, respectively. But within the stellar mass range $10^{9-11} {\rm M}_{\odot}$, the $f_{\rm j}$ for both early and late-type galaxies is largely independent of stellar mass. This is in contrast with the results of \citep{Posti2018} but however, qualitatively agrees with our results in the intermediate to the high stellar mass range. \cite{Gomez2022} have not studied galaxies less massive than $10^{9} {\rm M}_{\odot}$.

Considering the aforementioned results from the literature, along with our own paints a picture where there is a lack of consensus on the behavior of $f_{\rm j}$ as a function of stellar mass. At a lower stellar mass regime, 
our results indicate a mass-dependent halo angular momentum retention for dwarf galaxies. This is challenging for the models where galaxies get their sizes by acquiring a fixed fraction of halo spin. However, It must be noted that the simple models of \cite{Fall1980} and \cite{Mo1998} are strictly valid for pure disc galaxies living in isothermal haloes. Several refinements to this model have been made in the literature for predicting the size of disc galaxies in NFW haloes. This modification is usually in the form $R_{\rm d} = 1/\sqrt{2}. f_{\rm c}^{-1/2} . f_{\rm R}(\lambda, c, m_{\rm d}, j_{\rm d}) . f_{\rm j} . \lambda . R_{\rm 200c} $, where c is the halo concentration, $m_d$ is the ratio of the disc to halo mass and $j_{\rm d}$ is the ratio of the disc to halo total angular momentum. The additional factors $f_{\rm c}$ and $f_{\rm R}$ are complex functions (see their form in section 2.3 of \cite{Mo1998}) making it hard to directly infer the $f_{\rm j}$ as was done easily in a simplistic disc galaxy size model using our observations. 

On the other hand, our results are based on a morphological mixed galaxy sample. It is understandable that a one-to-one comparison of our results cannot be carried out with the prediction of such models. However, we see differences between observations and theoretical predictions for only galaxies having stellar masses less than $\sim 10^{9} {\rm M}_{\odot}h^{-2}$. Moreover, this is also the stellar mass range where the number density of disc-dominated galaxies in a low redshift universe is an order of magnitude higher than that of spheroidal galaxies \citep{Kelvin2014}. This implies that statistical results presented in our work for low-mass galaxies should be driven by disc-dominated galaxies, making our results in the dwarf sector important for putting constraints on galaxy formation models.

\section{Summary \& Conclusions} \label{sec6}

We have presented an observational inference of the relationship between the median galaxy size and dark matter halo radius. We create a sample of $\sim$ 38,000 low redshifts ($z<0.3$) galaxies common in GAMA and the Subaru HSC-SSP survey PDR2. Our sample of galaxies span a stellar mass range of 8$< \log (M_*/{\rm M}_{\odot}h^{-2})<11.3$. We measure the galaxy sizes using $i-$band images from HSC-SSP PDR2 and construct a median galaxy size-stellar mass relation. We divide our sample into a number of mass bins and obtained stacked weak lensing signals associated with each bin. This signal was then modeled in the framework of the conditional stellar mass function which allows us to infer the stellar mass - halo mass relation. We combine these median galaxy size-stellar mass and stellar mass-halo mass relation to link the median galaxy size to the dark matter halo radius. Our main findings are the following:

\begin{itemize}
\item \preetish{We infer a linear relation between the median galaxy size-halo radius relation over two orders of magnitude in stellar masses from $\sim 10^{9}  {\rm M}_{\odot}h^{-2}$. If we extrapolate the stellar mass-halo mass relation (SMHMR) to stellar masses lower than this, we see an indication of a departure from this linearity which manifests itself as decreasing galaxy size halo radius ratio with the decrease in stellar mass. The galaxy size halo radius ratio shows up to 30 percent smaller values compared to the linear relation for galaxies with stellar masses $\sim 10^{8}  {\rm M}_{\odot}h^{-2}$. The qualitative trend of stellar mass dependence of galaxy size-halo radius relation remains the same even if we use other existing SMHMRs from the literature.}

\item We compare our results on galaxy size-halo radius relation with similar results from past studies. Our results are in qualitative agreement with abundance matching results of \cite{Kravtsov2013, Huang2017} and \cite{Somerville2018} for galaxies more massive than \preetish{$\sim 10^{9} {\rm M}_{\odot}h^{-2}$.} We do not find consistent results with abundance matching work of \cite{Zhang2022} who predict a monotonic decreasing trend of size ratio with increasing halo masses. Our results on the relationship between galaxy size and halo radius as a function of stellar mass are in qualitative agreement with the halo concentration dependent galaxy size prescription by \cite{Jiang2019} at intermediate halo masses only. We observe a departure from this prediction at low and high halo mass ends in our analysis.

\item \preetish{The 30\% change in the ratio of galaxy-to-halo radius as a function of stellar mass in low-mass/dwarf galaxies have not been reported in past studies. Its existence (albeit based on the extrapolation of our halo mass estimates) either points to our lack of knowledge of dark matter content (which needs to be almost 0.5 dex higher for a universal galaxy-halo radius relation) of dwarf galaxies or else requires modification in simplistic models of galaxy formation predicting sizes of dwarf galaxies via assigning them a constant fraction of halo spin.}

\item In the case of the latter possibility, we show that the fraction of halo angular momentum retained by dwarf galaxies should also be stellar mass dependent. Similar results are also found in analytical work by \cite{Posti2018} and a study based on FIRE hydro-dynamical simulation by \cite{El-Badry2018}. Our predicted trend of angular momentum retention fraction with stellar mass qualitatively agrees with the results in these works in the low mass regime.
\end{itemize}

This work is one of the first studies to probe the relationship between the sizes of galaxies and dark matter haloes in a purely observational manner. In contrast to previous studies which rely on modeling assumptions to indirectly assign halo masses to galaxies, we are able to directly probe the dark matter content of our sample galaxies using weak gravitational lensing. This method has unveiled a possible stellar mass dependence of galaxy size-halo radius connection at the low mass end which has some interesting implications for the galaxy formation picture. Our observational results on galaxy size halo radius connection may prove to be useful for studies estimating sizes of galaxies based on just dark matter halo properties and, as well as for testing and constraining newer galaxy formation models. 

\section*{Acknowledgements}

\preetish{We thank the anonymous referee for insightful comments that have improved both the content and presentation of this paper.} We thank Lalitwadee Kawinwanichakij and John Silverman for providing us with data from \cite{Kawinwanichakij2021} as well as for useful discussion regarding this work. We also thank \preetish{Andrey Kravtsov}, Jingjing Shi, and Masahiro Takada for useful comments on this work. DR thank the University Grants Commission (UGC) of India, for providing financial support as a senior research fellow. We acknowledge the use of  the high performance computing facility - Pegasus at IUCAA.

The Hyper Suprime-Cam (HSC) collaboration includes the astronomical communities of Japan and Taiwan, and Princeton University. The HSC instrumentation and software were developed by the National Astronomical Observatory of Japan (NAOJ), the Kavli Institute for the Physics and Mathematics of the Universe (Kavli IPMU), the University of Tokyo, the High Energy Accelerator Research Organization (KEK), the Academia Sinica Institute for Astronomy and Astrophysics in Taiwan (ASIAA), and Princeton University. Funding was contributed by the FIRST program from the Japanese Cabinet Office, the Ministry of Education, Culture, Sports, Science and Technology (MEXT), the Japan Society for the Promotion of Science (JSPS), Japan Science and Technology Agency (JST), the Toray Science Foundation, NAOJ, Kavli IPMU, KEK, ASIAA, and Princeton University.

This paper is based in part on data collected at the Subaru Telescope and retrieved from the HSC data archive system, which is operated by the Subaru Telescope and Astronomy Data Center (ADC) at National Astronomical Observatory of Japan. Data analysis was in part carried out with the cooperation of Center for Computational Astrophysics (CfCA), National Astronomical Observatory of Japan. The Subaru Telescope is honored and grateful for the opportunity of observing the Universe from Maunakea, which has the cultural, historical and natural significance in Hawaii. 

GAMA is a joint European-Australasian project based around a spectroscopic campaign using the Anglo-Australian Telescope. The GAMA input catalogue is based on data taken from the Sloan Digital Sky Survey and the UKIRT Infrared Deep Sky Survey. Complementary imaging of the GAMA regions is being obtained by a number of independent survey programmes including GALEX MIS, VST KiDS, VISTA VIKING, WISE, Herschel-ATLAS, GMRT and ASKAP providing UV to radio coverage. GAMA is funded by the STFC (UK), the ARC (Australia), the AAO, and the participating institutions. The GAMA website is http://www.gama-survey.org/ .

\section*{Data Availability}
The measurements related to weak lensing analysis, stellar mass halo mass relation, galaxy size mass relation, and galaxy size halo radius relation are made available in tabular format at https://github.com/divyarana-cosmo/size\_mass\_gama\_hsc\_s16a


\bibliographystyle{mnras}
\bibliography{mnras_template} 

\begin{thebibliography}{}
\makeatletter
\relax
\def\mn@urlcharsother{\let\do\@makeother \do\$\do\&\do\#\do\^\do\_\do\%\do\~}
\def\mn@doi{\begingroup\mn@urlcharsother \@ifnextchar [ {\mn@doi@}
  {\mn@doi@[]}}
\def\mn@doi@[#1]#2{\def\@tempa{#1}\ifx\@tempa\@empty \href
  {http://dx.doi.org/#2} {doi:#2}\else \href {http://dx.doi.org/#2} {#1}\fi
  \endgroup}
\def\mn@eprint#1#2{\mn@eprint@#1:#2::\@nil}
\def\mn@eprint@arXiv#1{\href {http://arxiv.org/abs/#1} {{\tt arXiv:#1}}}
\def\mn@eprint@dblp#1{\href {http://dblp.uni-trier.de/rec/bibtex/#1.xml}
  {dblp:#1}}
\def\mn@eprint@#1:#2:#3:#4\@nil{\def\@tempa {#1}\def\@tempb {#2}\def\@tempc
  {#3}\ifx \@tempc \@empty \let \@tempc \@tempb \let \@tempb \@tempa \fi \ifx
  \@tempb \@empty \def\@tempb {arXiv}\fi \@ifundefined
  {mn@eprint@\@tempb}{\@tempb:\@tempc}{\expandafter \expandafter \csname
  mn@eprint@\@tempb\endcsname \expandafter{\@tempc}}}

\bibitem[\protect\citeauthoryear{{Aihara} et~al.,}{{Aihara}
  et~al.}{2018}]{Aihara2018}
{Aihara} H.,  et~al., 2018, \mn@doi [\pasj] {10.1093/pasj/psx066}, \href
  {https://ui.adsabs.harvard.edu/abs/2018PASJ...70S...4A} {70, S4}

\bibitem[\protect\citeauthoryear{{Aihara} et~al.,}{{Aihara}
  et~al.}{2019}]{Aihara2019}
{Aihara} H.,  et~al., 2019, \mn@doi [\pasj] {10.1093/pasj/psz103}, \href
  {https://ui.adsabs.harvard.edu/abs/2019PASJ...71..114A} {71, 114}

\bibitem[\protect\citeauthoryear{{Ardila} et~al.,}{{Ardila}
  et~al.}{2021}]{Ardila2021}
{Ardila} F.,  et~al., 2021, \mn@doi [\mnras] {10.1093/mnras/staa3215}, \href
  {https://ui.adsabs.harvard.edu/abs/2021MNRAS.500..432A} {500, 432}

\bibitem[\protect\citeauthoryear{{Astropy Collaboration} et~al.,}{{Astropy
  Collaboration} et~al.}{2022}]{astropy}
{Astropy Collaboration} et~al., 2022, \mn@doi [apj] {10.3847/1538-4357/ac7c74},
  \href {https://ui.adsabs.harvard.edu/abs/2022ApJ...935..167A} {935, 167}

\bibitem[\protect\citeauthoryear{{Baldry} et~al.,}{{Baldry}
  et~al.}{2012}]{Baldry2012}
{Baldry} I.~K.,  et~al., 2012, \mn@doi [\mnras]
  {10.1111/j.1365-2966.2012.20340.x}, \href
  {https://ui.adsabs.harvard.edu/abs/2012MNRAS.421..621B} {421, 621}

\bibitem[\protect\citeauthoryear{{Behroozi}, {Wechsler}  \&
  {Conroy}}{{Behroozi} et~al.}{2013}]{Behroozi2013}
{Behroozi} P.~S.,  {Wechsler} R.~H.,   {Conroy} C.,  2013, \mn@doi [\apj]
  {10.1088/0004-637X/770/1/57}, \href
  {https://ui.adsabs.harvard.edu/abs/2013ApJ...770...57B} {770, 57}

\bibitem[\protect\citeauthoryear{{Behroozi}, {Wechsler}, {Hearin}  \&
  {Conroy}}{{Behroozi} et~al.}{2019}]{Behroozi2019}
{Behroozi} P.,  {Wechsler} R.~H.,  {Hearin} A.~P.,   {Conroy} C.,  2019,
  \mn@doi [\mnras] {10.1093/mnras/stz1182}, \href
  {https://ui.adsabs.harvard.edu/abs/2019MNRAS.488.3143B} {488, 3143}

\bibitem[\protect\citeauthoryear{{Berlind} et~al.,}{{Berlind}
  et~al.}{2003}]{Berlind2003}
{Berlind} A.~A.,  et~al., 2003, \mn@doi [\apj] {10.1086/376517}, \href
  {https://ui.adsabs.harvard.edu/abs/2003ApJ...593....1B} {593, 1}

\bibitem[\protect\citeauthoryear{{Bernstein} \& {Jarvis}}{{Bernstein} \&
  {Jarvis}}{2002}]{Bernstein2002}
{Bernstein} G.~M.,  {Jarvis} M.,  2002, \mn@doi [\aj] {10.1086/338085}, \href
  {https://ui.adsabs.harvard.edu/abs/2002AJ....123..583B} {123, 583}

\bibitem[\protect\citeauthoryear{{Bradley} et~al.,}{{Bradley}
  et~al.}{2020}]{Bradley2020}
{Bradley} L.,  et~al., 2020, {astropy/photutils: 1.0.1}, Zenodo,
  \mn@doi{10.5281/zenodo.4049061}

\bibitem[\protect\citeauthoryear{{Bullock}, {Wechsler}  \&
  {Somerville}}{{Bullock} et~al.}{2002}]{Bullock2002}
{Bullock} J.~S.,  {Wechsler} R.~H.,   {Somerville} R.~S.,  2002, \mn@doi
  [\mnras] {10.1046/j.1365-8711.2002.04959.x}, \href
  {https://ui.adsabs.harvard.edu/abs/2002MNRAS.329..246B} {329, 246}

\bibitem[\protect\citeauthoryear{{Burkert} et~al.,}{{Burkert}
  et~al.}{2016}]{Burkert2016}
{Burkert} A.,  et~al., 2016, \mn@doi [\apj] {10.3847/0004-637X/826/2/214},
  \href {https://ui.adsabs.harvard.edu/abs/2016ApJ...826..214B} {826, 214}

\bibitem[\protect\citeauthoryear{{Cacciato}, {van den Bosch}, {More}, {Mo}  \&
  {Yang}}{{Cacciato} et~al.}{2013}]{2013_Cacciato}
{Cacciato} M.,  {van den Bosch} F.~C.,  {More} S.,  {Mo} H.,   {Yang} X.,
  2013, \mn@doi [\mnras] {10.1093/mnras/sts525}, \href
  {https://ui.adsabs.harvard.edu/abs/2013MNRAS.430..767C} {430, 767}

\bibitem[\protect\citeauthoryear{{Capaccioli}}{{Capaccioli}}{1989}]{Capaccioli1989}
{Capaccioli} M.,  1989, in {Corwin} Harold~G. J.,  {Bottinelli} L.,  eds, World
  of Galaxies (Le Monde des Galaxies). pp 208--227

\bibitem[\protect\citeauthoryear{{Cole}, {Aragon-Salamanca}, {Frenk}, {Navarro}
   \& {Zepf}}{{Cole} et~al.}{1994}]{Cole1994}
{Cole} S.,  {Aragon-Salamanca} A.,  {Frenk} C.~S.,  {Navarro} J.~F.,   {Zepf}
  S.~E.,  1994, \mn@doi [\mnras] {10.1093/mnras/271.4.781}, \href
  {https://ui.adsabs.harvard.edu/abs/1994MNRAS.271..781C} {271, 781}

\bibitem[\protect\citeauthoryear{{Conroy} \& {Wechsler}}{{Conroy} \&
  {Wechsler}}{2009}]{Conroy2009}
{Conroy} C.,  {Wechsler} R.~H.,  2009, \mn@doi [\apj]
  {10.1088/0004-637X/696/1/620}, \href
  {https://ui.adsabs.harvard.edu/abs/2009ApJ...696..620C} {696, 620}

\bibitem[\protect\citeauthoryear{{Conroy}, {Wechsler}  \& {Kravtsov}}{{Conroy}
  et~al.}{2006}]{Conroy2006}
{Conroy} C.,  {Wechsler} R.~H.,   {Kravtsov} A.~V.,  2006, \mn@doi [\apj]
  {10.1086/503602}, \href
  {https://ui.adsabs.harvard.edu/abs/2006ApJ...647..201C} {647, 201}

\bibitem[\protect\citeauthoryear{{Conselice}}{{Conselice}}{2014}]{Conselice2014}
{Conselice} C.~J.,  2014, \mn@doi [\araa]
  {10.1146/annurev-astro-081913-040037}, \href
  {https://ui.adsabs.harvard.edu/abs/2014ARA&A..52..291C} {52, 291}

\bibitem[\protect\citeauthoryear{{Cooray} \& {Sheth}}{{Cooray} \&
  {Sheth}}{2002}]{2002_Cooray}
{Cooray} A.,  {Sheth} R.,  2002, \mn@doi [\physrep]
  {10.1016/S0370-1573(02)00276-4}, \href
  {https://ui.adsabs.harvard.edu/abs/2002PhR...372....1C} {372, 1}

\bibitem[\protect\citeauthoryear{{Desmond}, {Mao}, {Wechsler}, {Crain}  \&
  {Schaye}}{{Desmond} et~al.}{2017}]{Desmond2017}
{Desmond} H.,  {Mao} Y.-Y.,  {Wechsler} R.~H.,  {Crain} R.~A.,   {Schaye} J.,
  2017, \mn@doi [\mnras] {10.1093/mnrasl/slx093}, \href
  {https://ui.adsabs.harvard.edu/abs/2017MNRAS.471L..11D} {471, L11}

\bibitem[\protect\citeauthoryear{{Diemer}}{{Diemer}}{2018}]{Diemer2018}
{Diemer} B.,  2018, \mn@doi [\apjs] {10.3847/1538-4365/aaee8c}, \href
  {https://ui.adsabs.harvard.edu/abs/2018ApJS..239...35D} {239, 35}

\bibitem[\protect\citeauthoryear{{Driver} et~al.,}{{Driver}
  et~al.}{2011}]{Driver2011}
{Driver} S.~P.,  et~al., 2011, \mn@doi [\mnras]
  {10.1111/j.1365-2966.2010.18188.x}, \href
  {https://ui.adsabs.harvard.edu/abs/2011MNRAS.413..971D} {413, 971}

\bibitem[\protect\citeauthoryear{{Driver} et~al.,}{{Driver}
  et~al.}{2022}]{Driver2022}
{Driver} S.~P.,  et~al., 2022, \mn@doi [\mnras] {10.1093/mnras/stac472}, \href
  {https://ui.adsabs.harvard.edu/abs/2022MNRAS.513..439D} {513, 439}

\bibitem[\protect\citeauthoryear{{Dvornik} et~al.,}{{Dvornik}
  et~al.}{2018}]{2018Dvornik}
{Dvornik} A.,  et~al., 2018, \mn@doi [\mnras] {10.1093/mnras/sty1502}, \href
  {https://ui.adsabs.harvard.edu/abs/2018MNRAS.479.1240D} {479, 1240}

\bibitem[\protect\citeauthoryear{{Dvornik} et~al.,}{{Dvornik}
  et~al.}{2020}]{Dvornik2020}
{Dvornik} A.,  et~al., 2020, \mn@doi [\aap] {10.1051/0004-6361/202038693},
  \href {https://ui.adsabs.harvard.edu/abs/2020A&A...642A..83D} {642, A83}

\bibitem[\protect\citeauthoryear{{E Greene} et~al.,}{{E Greene}
  et~al.}{2022}]{Greene2022}
{E Greene} J.,  et~al., 2022, \mn@doi [arXiv e-prints]
  {10.48550/arXiv.2204.11883}, \href
  {https://ui.adsabs.harvard.edu/abs/2022arXiv220411883E} {p. arXiv:2204.11883}

\bibitem[\protect\citeauthoryear{{El-Badry} et~al.,}{{El-Badry}
  et~al.}{2018}]{El-Badry2018}
{El-Badry} K.,  et~al., 2018, \mn@doi [\mnras] {10.1093/mnras/stx2482}, \href
  {https://ui.adsabs.harvard.edu/abs/2018MNRAS.473.1930E} {473, 1930}

\bibitem[\protect\citeauthoryear{{Fall} \& {Efstathiou}}{{Fall} \&
  {Efstathiou}}{1980}]{Fall1980}
{Fall} S.~M.,  {Efstathiou} G.,  1980, \mn@doi [\mnras]
  {10.1093/mnras/193.2.189}, \href
  {https://ui.adsabs.harvard.edu/abs/1980MNRAS.193..189F} {193, 189}

\bibitem[\protect\citeauthoryear{{Foreman-Mackey}, {Hogg}, {Lang}  \&
  {Goodman}}{{Foreman-Mackey} et~al.}{2013}]{2013_Mackey}
{Foreman-Mackey} D.,  {Hogg} D.~W.,  {Lang} D.,   {Goodman} J.,  2013, \mn@doi
  [\pasp] {10.1086/670067}, \href
  {https://ui.adsabs.harvard.edu/abs/2013PASP..125..306F} {125, 306}

\bibitem[\protect\citeauthoryear{{Gadotti}}{{Gadotti}}{2009}]{Gadotti2009}
{Gadotti} D.~A.,  2009, \mn@doi [\mnras] {10.1111/j.1365-2966.2008.14257.x},
  \href {https://ui.adsabs.harvard.edu/abs/2009MNRAS.393.1531G} {393, 1531}

\bibitem[\protect\citeauthoryear{{Genel} et~al.,}{{Genel}
  et~al.}{2014}]{Genel2014}
{Genel} S.,  et~al., 2014, \mn@doi [\mnras] {10.1093/mnras/stu1654}, \href
  {https://ui.adsabs.harvard.edu/abs/2014MNRAS.445..175G} {445, 175}

\bibitem[\protect\citeauthoryear{{Goodman} \& {Weare}}{{Goodman} \&
  {Weare}}{2010}]{2010_Goodman}
{Goodman} J.,  {Weare} J.,  2010, \mn@doi [Communications in Applied
  Mathematics and Computational Science] {10.2140/camcos.2010.5.65}, \href
  {https://ui.adsabs.harvard.edu/abs/2010CAMCS...5...65G} {5, 65}

\bibitem[\protect\citeauthoryear{{Grogin} et~al.,}{{Grogin}
  et~al.}{2011}]{Grogin2011}
{Grogin} N.~A.,  et~al., 2011, \mn@doi [\apjs] {10.1088/0067-0049/197/2/35},
  \href {https://ui.adsabs.harvard.edu/abs/2011ApJS..197...35G} {197, 35}

\bibitem[\protect\citeauthoryear{{Guo}, {White}, {Angulo}, {Henriques},
  {Lemson}, {Boylan-Kolchin}, {Thomas}  \& {Short}}{{Guo}
  et~al.}{2013}]{Guo2013}
{Guo} Q.,  {White} S.,  {Angulo} R.~E.,  {Henriques} B.,  {Lemson} G.,
  {Boylan-Kolchin} M.,  {Thomas} P.,   {Short} C.,  2013, \mn@doi [\mnras]
  {10.1093/mnras/sts115}, \href
  {https://ui.adsabs.harvard.edu/abs/2013MNRAS.428.1351G} {428, 1351}

\bibitem[\protect\citeauthoryear{{Hartlap}, {Simon}  \& {Schneider}}{{Hartlap}
  et~al.}{2007}]{2007_Hartlap}
{Hartlap} J.,  {Simon} P.,   {Schneider} P.,  2007, \mn@doi [\aap]
  {10.1051/0004-6361:20066170}, \href
  {https://ui.adsabs.harvard.edu/abs/2007A&A...464..399H} {464, 399}

\bibitem[\protect\citeauthoryear{{Henriques}, {White}, {Thomas}, {Angulo},
  {Guo}, {Lemson}, {Springel}  \& {Overzier}}{{Henriques}
  et~al.}{2015}]{Hernquues2015}
{Henriques} B. M.~B.,  {White} S. D.~M.,  {Thomas} P.~A.,  {Angulo} R.,  {Guo}
  Q.,  {Lemson} G.,  {Springel} V.,   {Overzier} R.,  2015, \mn@doi [\mnras]
  {10.1093/mnras/stv705}, \href
  {https://ui.adsabs.harvard.edu/abs/2015MNRAS.451.2663H} {451, 2663}

\bibitem[\protect\citeauthoryear{{Hirata} \& {Seljak}}{{Hirata} \&
  {Seljak}}{2003}]{Hirata2003}
{Hirata} C.,  {Seljak} U.,  2003, \mn@doi [\mnras]
  {10.1046/j.1365-8711.2003.06683.x}, \href
  {https://ui.adsabs.harvard.edu/abs/2003MNRAS.343..459H} {343, 459}

\bibitem[\protect\citeauthoryear{{Hirata} et~al.,}{{Hirata}
  et~al.}{2004}]{2004_Hirata}
{Hirata} C.~M.,  et~al., 2004, \mn@doi [\mnras]
  {10.1111/j.1365-2966.2004.08090.x}, \href
  {https://ui.adsabs.harvard.edu/abs/2004MNRAS.353..529H} {353, 529}

\bibitem[\protect\citeauthoryear{{Huang} et~al.,}{{Huang}
  et~al.}{2017}]{Huang2017}
{Huang} K.-H.,  et~al., 2017, \mn@doi [\apj] {10.3847/1538-4357/aa62a6}, \href
  {https://ui.adsabs.harvard.edu/abs/2017ApJ...838....6H} {838, 6}

\bibitem[\protect\citeauthoryear{{Hudson} et~al.,}{{Hudson}
  et~al.}{2015}]{Hudson2015}
{Hudson} M.~J.,  et~al., 2015, \mn@doi [\mnras] {10.1093/mnras/stu2367}, \href
  {https://ui.adsabs.harvard.edu/abs/2015MNRAS.447..298H} {447, 298}

\bibitem[\protect\citeauthoryear{{Huertas-Company}, {Shankar}, {Mei},
  {Bernardi}, {Aguerri}, {Meert}  \& {Vikram}}{{Huertas-Company}
  et~al.}{2013}]{Huertas-Company2013}
{Huertas-Company} M.,  {Shankar} F.,  {Mei} S.,  {Bernardi} M.,  {Aguerri}
  J.~A.~L.,  {Meert} A.,   {Vikram} V.,  2013, \mn@doi [\apj]
  {10.1088/0004-637X/779/1/29}, \href
  {https://ui.adsabs.harvard.edu/abs/2013ApJ...779...29H} {779, 29}

\bibitem[\protect\citeauthoryear{{Ilbert} et~al.,}{{Ilbert}
  et~al.}{2009}]{2009_Ilbert}
{Ilbert} O.,  et~al., 2009, \mn@doi [\apj] {10.1088/0004-637X/690/2/1236},
  \href {https://ui.adsabs.harvard.edu/abs/2009ApJ...690.1236I} {690, 1236}

\bibitem[\protect\citeauthoryear{{Jiang} et~al.,}{{Jiang}
  et~al.}{2019}]{Jiang2019}
{Jiang} F.,  et~al., 2019, \mn@doi [\mnras] {10.1093/mnras/stz1952}, \href
  {https://ui.adsabs.harvard.edu/abs/2019MNRAS.488.4801J} {488, 4801}

\bibitem[\protect\citeauthoryear{{Jones}, {Papastergis}, {Pandya}, {Leisman},
  {Romanowsky}, {Yung}, {Somerville}  \& {Adams}}{{Jones}
  et~al.}{2018}]{Jones2018}
{Jones} M.~G.,  {Papastergis} E.,  {Pandya} V.,  {Leisman} L.,  {Romanowsky}
  A.~J.,  {Yung} L.~Y.~A.,  {Somerville} R.~S.,   {Adams} E.~A.~K.,  2018,
  \mn@doi [\aap] {10.1051/0004-6361/201732409}, \href
  {https://ui.adsabs.harvard.edu/abs/2018A&A...614A..21J} {614, A21}

\bibitem[\protect\citeauthoryear{{Kauffmann}, {White}  \&
  {Guiderdoni}}{{Kauffmann} et~al.}{1993}]{Kauffmann1993}
{Kauffmann} G.,  {White} S.~D.~M.,   {Guiderdoni} B.,  1993, \mn@doi [\mnras]
  {10.1093/mnras/264.1.201}, \href
  {https://ui.adsabs.harvard.edu/abs/1993MNRAS.264..201K} {264, 201}

\bibitem[\protect\citeauthoryear{{Kawinwanichakij} et~al.,}{{Kawinwanichakij}
  et~al.}{2021}]{Kawinwanichakij2021}
{Kawinwanichakij} L.,  et~al., 2021, \mn@doi [\apj] {10.3847/1538-4357/ac1f21},
  \href {https://ui.adsabs.harvard.edu/abs/2021ApJ...921...38K} {921, 38}

\bibitem[\protect\citeauthoryear{{Kelvin} et~al.,}{{Kelvin}
  et~al.}{2014}]{Kelvin2014}
{Kelvin} L.~S.,  et~al., 2014, \mn@doi [\mnras] {10.1093/mnras/stu1507}, \href
  {https://ui.adsabs.harvard.edu/abs/2014MNRAS.444.1647K} {444, 1647}

\bibitem[\protect\citeauthoryear{{Kilbinger}}{{Kilbinger}}{2015}]{2015_Kilbinger}
{Kilbinger} M.,  2015, \mn@doi [Reports on Progress in Physics]
  {10.1088/0034-4885/78/8/086901}, \href
  {https://ui.adsabs.harvard.edu/abs/2015RPPh...78h6901K} {78, 086901}

\bibitem[\protect\citeauthoryear{{Komatsu} et~al.,}{{Komatsu}
  et~al.}{2011}]{Komatsu2011}
{Komatsu} E.,  et~al., 2011, \mn@doi [\apjs] {10.1088/0067-0049/192/2/18},
  \href {https://ui.adsabs.harvard.edu/abs/2011ApJS..192...18K} {192, 18}

\bibitem[\protect\citeauthoryear{{Kravtsov}}{{Kravtsov}}{2013}]{Kravtsov2013}
{Kravtsov} A.~V.,  2013, \mn@doi [\apjl] {10.1088/2041-8205/764/2/L31}, \href
  {https://ui.adsabs.harvard.edu/abs/2013ApJ...764L..31K} {764, L31}

\bibitem[\protect\citeauthoryear{{Kravtsov}, {Berlind}, {Wechsler}, {Klypin},
  {Gottl{\"o}ber}, {Allgood}  \& {Primack}}{{Kravtsov}
  et~al.}{2004}]{Kravtsov2004}
{Kravtsov} A.~V.,  {Berlind} A.~A.,  {Wechsler} R.~H.,  {Klypin} A.~A.,
  {Gottl{\"o}ber} S.,  {Allgood} B.,   {Primack} J.~R.,  2004, \mn@doi [\apj]
  {10.1086/420959}, \href
  {https://ui.adsabs.harvard.edu/abs/2004ApJ...609...35K} {609, 35}

\bibitem[\protect\citeauthoryear{{Lange} et~al.,}{{Lange}
  et~al.}{2015}]{Lange2015}
{Lange} R.,  et~al., 2015, \mn@doi [\mnras] {10.1093/mnras/stu2467}, \href
  {https://ui.adsabs.harvard.edu/abs/2015MNRAS.447.2603L} {447, 2603}

\bibitem[\protect\citeauthoryear{{Leauthaud} et~al.,}{{Leauthaud}
  et~al.}{2012}]{Leauthaud2012}
{Leauthaud} A.,  et~al., 2012, \mn@doi [\apj] {10.1088/0004-637X/744/2/159},
  \href {https://ui.adsabs.harvard.edu/abs/2012ApJ...744..159L} {744, 159}

\bibitem[\protect\citeauthoryear{{Leauthaud}, {Singh}, {Luo}, {Ardila},
  {Greco}, {Capak}, {Greene}  \& {Mayer}}{{Leauthaud}
  et~al.}{2020}]{Leauthaud2020}
{Leauthaud} A.,  {Singh} S.,  {Luo} Y.,  {Ardila} F.,  {Greco} J.~P.,  {Capak}
  P.,  {Greene} J.~E.,   {Mayer} L.,  2020, \mn@doi [Physics of the Dark
  Universe] {10.1016/j.dark.2020.100719}, \href
  {https://ui.adsabs.harvard.edu/abs/2020PDU....3000719L} {30, 100719}

\bibitem[\protect\citeauthoryear{{Lima Neto}, {Gerbal}  \& {M{\'a}rquez}}{{Lima
  Neto} et~al.}{1999}]{Neto1999}
{Lima Neto} G.~B.,  {Gerbal} D.,   {M{\'a}rquez} I.,  1999, \mn@doi [\mnras]
  {10.1046/j.1365-8711.1999.02849.x}, \href
  {https://ui.adsabs.harvard.edu/abs/1999MNRAS.309..481L} {309, 481}

\bibitem[\protect\citeauthoryear{{Macci{\`o}}, {Dutton}, {van den Bosch},
  {Moore}, {Potter}  \& {Stadel}}{{Macci{\`o}} et~al.}{2007}]{Maccio2007}
{Macci{\`o}} A.~V.,  {Dutton} A.~A.,  {van den Bosch} F.~C.,  {Moore} B.,
  {Potter} D.,   {Stadel} J.,  2007, \mn@doi [\mnras]
  {10.1111/j.1365-2966.2007.11720.x}, \href
  {https://ui.adsabs.harvard.edu/abs/2007MNRAS.378...55M} {378, 55}

\bibitem[\protect\citeauthoryear{{Mandelbaum}}{{Mandelbaum}}{2018}]{2018ARA&A_Mandelbaum}
{Mandelbaum} R.,  2018, \mn@doi [\araa] {10.1146/annurev-astro-081817-051928},
  \href {https://ui.adsabs.harvard.edu/abs/2018ARA&A..56..393M} {56, 393}

\bibitem[\protect\citeauthoryear{{Mandelbaum} et~al.,}{{Mandelbaum}
  et~al.}{2005}]{2005_Mandelbaum}
{Mandelbaum} R.,  et~al., 2005, \mn@doi [\mnras]
  {10.1111/j.1365-2966.2005.09282.x}, \href
  {https://ui.adsabs.harvard.edu/abs/2005MNRAS.361.1287M} {361, 1287}

\bibitem[\protect\citeauthoryear{{Mandelbaum}, {Seljak}, {Kauffmann}, {Hirata}
  \& {Brinkmann}}{{Mandelbaum} et~al.}{2006a}]{Mandelbaum2006}
{Mandelbaum} R.,  {Seljak} U.,  {Kauffmann} G.,  {Hirata} C.~M.,   {Brinkmann}
  J.,  2006a, \mn@doi [\mnras] {10.1111/j.1365-2966.2006.10156.x}, \href
  {https://ui.adsabs.harvard.edu/abs/2006MNRAS.368..715M} {368, 715}

\bibitem[\protect\citeauthoryear{{Mandelbaum}, {Seljak}, {Cool}, {Blanton},
  {Hirata}  \& {Brinkmann}}{{Mandelbaum} et~al.}{2006b}]{2006_Mandelbaum}
{Mandelbaum} R.,  {Seljak} U.,  {Cool} R.~J.,  {Blanton} M.,  {Hirata} C.~M.,
  {Brinkmann} J.,  2006b, \mn@doi [\mnras] {10.1111/j.1365-2966.2006.10906.x},
  \href {https://ui.adsabs.harvard.edu/abs/2006MNRAS.372..758M} {372, 758}

\bibitem[\protect\citeauthoryear{{Mandelbaum} et~al.,}{{Mandelbaum}
  et~al.}{2008}]{2008_Mandelbaum}
{Mandelbaum} R.,  et~al., 2008, \mn@doi [\mnras]
  {10.1111/j.1365-2966.2008.12947.x}, \href
  {https://ui.adsabs.harvard.edu/abs/2008MNRAS.386..781M} {386, 781}

\bibitem[\protect\citeauthoryear{{Mandelbaum}, {Slosar}, {Baldauf}, {Seljak},
  {Hirata}, {Nakajima}, {Reyes}  \& {Smith}}{{Mandelbaum}
  et~al.}{2013}]{2013_Mandelbaum}
{Mandelbaum} R.,  {Slosar} A.,  {Baldauf} T.,  {Seljak} U.,  {Hirata} C.~M.,
  {Nakajima} R.,  {Reyes} R.,   {Smith} R.~E.,  2013, \mn@doi [\mnras]
  {10.1093/mnras/stt572}, \href
  {https://ui.adsabs.harvard.edu/abs/2013MNRAS.432.1544M} {432, 1544}

\bibitem[\protect\citeauthoryear{{Mandelbaum} et~al.,}{{Mandelbaum}
  et~al.}{2018a}]{Mandelbaum2018}
{Mandelbaum} R.,  et~al., 2018a, \mn@doi [\pasj] {10.1093/pasj/psx130}, \href
  {https://ui.adsabs.harvard.edu/abs/2018PASJ...70S..25M} {70, S25}

\bibitem[\protect\citeauthoryear{{Mandelbaum} et~al.,}{{Mandelbaum}
  et~al.}{2018b}]{2018_Mandlebaum}
{Mandelbaum} R.,  et~al., 2018b, \mn@doi [\mnras] {10.1093/mnras/sty2420},
  \href {https://ui.adsabs.harvard.edu/abs/2018MNRAS.481.3170M} {481, 3170}

\bibitem[\protect\citeauthoryear{{Manwadkar} \& {Kravtsov}}{{Manwadkar} \&
  {Kravtsov}}{2022}]{Manwadkar2022}
{Manwadkar} V.,  {Kravtsov} A.~V.,  2022, \mn@doi [\mnras]
  {10.1093/mnras/stac2452}, \href
  {https://ui.adsabs.harvard.edu/abs/2022MNRAS.516.3944M} {516, 3944}

\bibitem[\protect\citeauthoryear{{Meert}, {Vikram}  \& {Bernardi}}{{Meert}
  et~al.}{2015}]{Meert2015}
{Meert} A.,  {Vikram} V.,   {Bernardi} M.,  2015, \mn@doi [\mnras]
  {10.1093/mnras/stu2333}, \href
  {https://ui.adsabs.harvard.edu/abs/2015MNRAS.446.3943M} {446, 3943}

\bibitem[\protect\citeauthoryear{{Miyatake} et~al.,}{{Miyatake}
  et~al.}{2015}]{2015_Miyatake}
{Miyatake} H.,  et~al., 2015, \mn@doi [\apj] {10.1088/0004-637X/806/1/1}, \href
  {https://ui.adsabs.harvard.edu/abs/2015ApJ...806....1M} {806, 1}

\bibitem[\protect\citeauthoryear{{Miyatake} et~al.,}{{Miyatake}
  et~al.}{2019}]{2019_Miyatake}
{Miyatake} H.,  et~al., 2019, \mn@doi [\apj] {10.3847/1538-4357/ab0af0}, \href
  {https://ui.adsabs.harvard.edu/abs/2019ApJ...875...63M} {875, 63}

\bibitem[\protect\citeauthoryear{{Mo}, {Mao}  \& {White}}{{Mo}
  et~al.}{1998}]{Mo1998}
{Mo} H.~J.,  {Mao} S.,   {White} S. D.~M.,  1998, \mn@doi [\mnras]
  {10.1046/j.1365-8711.1998.01227.x}, \href
  {https://ui.adsabs.harvard.edu/abs/1998MNRAS.295..319M} {295, 319}

\bibitem[\protect\citeauthoryear{{More}, {van den Bosch}, {Cacciato}, {Mo},
  {Yang}  \& {Li}}{{More} et~al.}{2009a}]{More2009}
{More} S.,  {van den Bosch} F.~C.,  {Cacciato} M.,  {Mo} H.~J.,  {Yang} X.,
  {Li} R.,  2009a, \mn@doi [\mnras] {10.1111/j.1365-2966.2008.14095.x}, \href
  {https://ui.adsabs.harvard.edu/abs/2009MNRAS.392..801M} {392, 801}

\bibitem[\protect\citeauthoryear{{More}, {van den Bosch}  \& {Cacciato}}{{More}
  et~al.}{2009b}]{More2009a}
{More} S.,  {van den Bosch} F.~C.,   {Cacciato} M.,  2009b, \mn@doi [\mnras]
  {10.1111/j.1365-2966.2008.14114.x}, \href
  {https://ui.adsabs.harvard.edu/abs/2009MNRAS.392..917M} {392, 917}

\bibitem[\protect\citeauthoryear{{More}, {van den Bosch}, {Cacciato}, {Skibba},
  {Mo}  \& {Yang}}{{More} et~al.}{2011}]{More2011}
{More} S.,  {van den Bosch} F.~C.,  {Cacciato} M.,  {Skibba} R.,  {Mo} H.~J.,
  {Yang} X.,  2011, \mn@doi [\mnras] {10.1111/j.1365-2966.2010.17436.x}, \href
  {https://ui.adsabs.harvard.edu/abs/2011MNRAS.410..210M} {410, 210}

\bibitem[\protect\citeauthoryear{{More}, {\VAN{vanDenBosch}{Van den}{van den}}
  Bosch, {Cacciato}, {More}, {Mo}  \& {Yang}}{{More} et~al.}{2013}]{2013_More}
{More} S.,  {\VAN{vanDenBosch}{Van den}{van den}} Bosch F.~C.,  {Cacciato} M.,
  {More} A.,  {Mo} H.,   {Yang} X.,  2013, \mn@doi [\mnras]
  {10.1093/mnras/sts697}, \href
  {https://ui.adsabs.harvard.edu/abs/2013MNRAS.430..747M} {430, 747}

\bibitem[\protect\citeauthoryear{{More}, {Miyatake}, {Mandelbaum}, {Takada},
  {Spergel}, {Brownstein}  \& {Schneider}}{{More} et~al.}{2015}]{2015_More}
{More} S.,  {Miyatake} H.,  {Mandelbaum} R.,  {Takada} M.,  {Spergel} D.~N.,
  {Brownstein} J.~R.,   {Schneider} D.~P.,  2015, \mn@doi [\apj]
  {10.1088/0004-637X/806/1/2}, \href
  {https://ui.adsabs.harvard.edu/abs/2015ApJ...806....2M} {806, 2}

\bibitem[\protect\citeauthoryear{{Moster}, {Naab}  \& {White}}{{Moster}
  et~al.}{2013}]{Moster2013}
{Moster} B.~P.,  {Naab} T.,   {White} S. D.~M.,  2013, \mn@doi [\mnras]
  {10.1093/mnras/sts261}, \href
  {https://ui.adsabs.harvard.edu/abs/2013MNRAS.428.3121M} {428, 3121}

\bibitem[\protect\citeauthoryear{{Murata} et~al.,}{{Murata}
  et~al.}{2019}]{2019_Murata}
{Murata} R.,  et~al., 2019, \mn@doi [\pasj] {10.1093/pasj/psz092}, \href
  {https://ui.adsabs.harvard.edu/abs/2019PASJ...71..107M} {71, 107}

\bibitem[\protect\citeauthoryear{{Nakajima}, {Mandelbaum}, {Seljak}, {Cohn},
  {Reyes}  \& {Cool}}{{Nakajima} et~al.}{2012}]{2012_Nakajima}
{Nakajima} R.,  {Mandelbaum} R.,  {Seljak} U.,  {Cohn} J.~D.,  {Reyes} R.,
  {Cool} R.,  2012, \mn@doi [\mnras] {10.1111/j.1365-2966.2011.20249.x}, \href
  {https://ui.adsabs.harvard.edu/abs/2012MNRAS.420.3240N} {420, 3240}

\bibitem[\protect\citeauthoryear{{Navarro}, {Frenk}  \& {White}}{{Navarro}
  et~al.}{1996}]{1996_Navarro}
{Navarro} J.~F.,  {Frenk} C.~S.,   {White} S. D.~M.,  1996, \mn@doi [\apj]
  {10.1086/177173}, \href
  {https://ui.adsabs.harvard.edu/abs/1996ApJ...462..563N} {462, 563}

\bibitem[\protect\citeauthoryear{{Nelson} et~al.,}{{Nelson}
  et~al.}{2019}]{Nelson2019}
{Nelson} D.,  et~al., 2019, \mn@doi [Computational Astrophysics and Cosmology]
  {10.1186/s40668-019-0028-x}, \href
  {https://ui.adsabs.harvard.edu/abs/2019ComAC...6....2N} {6, 2}

\bibitem[\protect\citeauthoryear{{Peacock} \& {Smith}}{{Peacock} \&
  {Smith}}{2000}]{Peacock2000}
{Peacock} J.~A.,  {Smith} R.~E.,  2000, \mn@doi [\mnras]
  {10.1046/j.1365-8711.2000.03779.x}, \href
  {https://ui.adsabs.harvard.edu/abs/2000MNRAS.318.1144P} {318, 1144}

\bibitem[\protect\citeauthoryear{{Peng}, {Ho}, {Impey}  \& {Rix}}{{Peng}
  et~al.}{2010}]{Peng2010}
{Peng} C.~Y.,  {Ho} L.~C.,  {Impey} C.~D.,   {Rix} H.-W.,  2010, \mn@doi [\aj]
  {10.1088/0004-6256/139/6/2097}, \href
  {https://ui.adsabs.harvard.edu/abs/2010AJ....139.2097P} {139, 2097}

\bibitem[\protect\citeauthoryear{{Posti}, {Pezzulli}, {Fraternali}  \& {Di
  Teodoro}}{{Posti} et~al.}{2018}]{Posti2018}
{Posti} L.,  {Pezzulli} G.,  {Fraternali} F.,   {Di Teodoro} E.~M.,  2018,
  \mn@doi [\mnras] {10.1093/mnras/stx3168}, \href
  {https://ui.adsabs.harvard.edu/abs/2018MNRAS.475..232P} {475, 232}

\bibitem[\protect\citeauthoryear{{Raveri} \& {Hu}}{{Raveri} \&
  {Hu}}{2019}]{2019_Raveri}
{Raveri} M.,  {Hu} W.,  2019, \mn@doi [\prd] {10.1103/PhysRevD.99.043506},
  \href {https://ui.adsabs.harvard.edu/abs/2019PhRvD..99d3506R} {99, 043506}

\bibitem[\protect\citeauthoryear{{Reyes}, {Mandelbaum}, {Gunn}, {Nakajima},
  {Seljak}  \& {Hirata}}{{Reyes} et~al.}{2012}]{2012_Reyes}
{Reyes} R.,  {Mandelbaum} R.,  {Gunn} J.~E.,  {Nakajima} R.,  {Seljak} U.,
  {Hirata} C.~M.,  2012, \mn@doi [\mnras] {10.1111/j.1365-2966.2012.21472.x},
  \href {https://ui.adsabs.harvard.edu/abs/2012MNRAS.425.2610R} {425, 2610}

\bibitem[\protect\citeauthoryear{{Robotham} et~al.,}{{Robotham}
  et~al.}{2011}]{Robotham2011}
{Robotham} A.~S.~G.,  et~al., 2011, \mn@doi [\mnras]
  {10.1111/j.1365-2966.2011.19217.x}, \href
  {https://ui.adsabs.harvard.edu/abs/2011MNRAS.416.2640R} {416, 2640}

\bibitem[\protect\citeauthoryear{{Rodriguez-Gomez} et~al.,}{{Rodriguez-Gomez}
  et~al.}{2019}]{Gomez2019}
{Rodriguez-Gomez} V.,  et~al., 2019, \mn@doi [\mnras] {10.1093/mnras/sty3345},
  \href {https://ui.adsabs.harvard.edu/abs/2019MNRAS.483.4140R} {483, 4140}

\bibitem[\protect\citeauthoryear{{Rodriguez-Gomez} et~al.,}{{Rodriguez-Gomez}
  et~al.}{2022}]{Gomez2022}
{Rodriguez-Gomez} V.,  et~al., 2022, \mn@doi [\mnras] {10.1093/mnras/stac806},
  \href {https://ui.adsabs.harvard.edu/abs/2022MNRAS.512.5978R} {512, 5978}

\bibitem[\protect\citeauthoryear{{Rodr{\'\i}guez-Puebla}, {Primack},
  {Avila-Reese}  \& {Faber}}{{Rodr{\'\i}guez-Puebla} et~al.}{2017}]{Puebla2017}
{Rodr{\'\i}guez-Puebla} A.,  {Primack} J.~R.,  {Avila-Reese} V.,   {Faber}
  S.~M.,  2017, \mn@doi [\mnras] {10.1093/mnras/stx1172}, \href
  {https://ui.adsabs.harvard.edu/abs/2017MNRAS.470..651R} {470, 651}

\bibitem[\protect\citeauthoryear{{Rodriguez}, {Montero-Dorta}, {Angulo},
  {Artale}  \& {Merch{\'a}n}}{{Rodriguez} et~al.}{2021}]{Rodriguez2021}
{Rodriguez} F.,  {Montero-Dorta} A.~D.,  {Angulo} R.~E.,  {Artale} M.~C.,
  {Merch{\'a}n} M.,  2021, \mn@doi [\mnras] {10.1093/mnras/stab1571}, \href
  {https://ui.adsabs.harvard.edu/abs/2021MNRAS.505.3192R} {505, 3192}

\bibitem[\protect\citeauthoryear{{Rohr} et~al.,}{{Rohr}
  et~al.}{2022}]{Rohr2022}
{Rohr} E.,  et~al., 2022, \mn@doi [\mnras] {10.1093/mnras/stab3625}, \href
  {https://ui.adsabs.harvard.edu/abs/2022MNRAS.510.3967R} {510, 3967}

\bibitem[\protect\citeauthoryear{{Romeo}, {Agertz}  \& {Renaud}}{{Romeo}
  et~al.}{2023}]{Romeo2023}
{Romeo} A.~B.,  {Agertz} O.,   {Renaud} F.,  2023, \mn@doi [\mnras]
  {10.1093/mnras/stac3074}, \href
  {https://ui.adsabs.harvard.edu/abs/2023MNRAS.518.1002R} {518, 1002}

\bibitem[\protect\citeauthoryear{{Rowe} et~al.,}{{Rowe}
  et~al.}{2015}]{2015_rowe}
{Rowe} B.~T.~P.,  et~al., 2015, \mn@doi [Astronomy and Computing]
  {10.1016/j.ascom.2015.02.002}, \href
  {https://ui.adsabs.harvard.edu/abs/2015A&C....10..121R} {10, 121}

\bibitem[\protect\citeauthoryear{{Roy} et~al.,}{{Roy} et~al.}{2018}]{Roy2018}
{Roy} N.,  et~al., 2018, \mn@doi [\mnras] {10.1093/mnras/sty1917}, \href
  {https://ui.adsabs.harvard.edu/abs/2018MNRAS.480.1057R} {480, 1057}

\bibitem[\protect\citeauthoryear{{Schaye} et~al.,}{{Schaye}
  et~al.}{2015}]{Schaye2015}
{Schaye} J.,  et~al., 2015, \mn@doi [\mnras] {10.1093/mnras/stu2058}, \href
  {https://ui.adsabs.harvard.edu/abs/2015MNRAS.446..521S} {446, 521}

\bibitem[\protect\citeauthoryear{{Seljak}}{{Seljak}}{2000}]{2000_Seljak}
{Seljak} U.,  2000, \mn@doi [\mnras] {10.1046/j.1365-8711.2000.03715.x}, \href
  {https://ui.adsabs.harvard.edu/abs/2000MNRAS.318..203S} {318, 203}

\bibitem[\protect\citeauthoryear{{Sheldon} et~al.,}{{Sheldon}
  et~al.}{2004}]{2004_Sheldon}
{Sheldon} E.~S.,  et~al., 2004, \mn@doi [\aj] {10.1086/383293}, \href
  {https://ui.adsabs.harvard.edu/abs/2004AJ....127.2544S} {127, 2544}

\bibitem[\protect\citeauthoryear{{Shen}, {Mo}, {White}, {Blanton}, {Kauffmann},
  {Voges}, {Brinkmann}  \& {Csabai}}{{Shen} et~al.}{2003}]{Shen2003}
{Shen} S.,  {Mo} H.~J.,  {White} S. D.~M.,  {Blanton} M.~R.,  {Kauffmann} G.,
  {Voges} W.,  {Brinkmann} J.,   {Csabai} I.,  2003, \mn@doi [\mnras]
  {10.1046/j.1365-8711.2003.06740.x}, \href
  {https://ui.adsabs.harvard.edu/abs/2003MNRAS.343..978S} {343, 978}

\bibitem[\protect\citeauthoryear{{Simard}, {Mendel}, {Patton}, {Ellison}  \&
  {McConnachie}}{{Simard} et~al.}{2011}]{Simard2011}
{Simard} L.,  {Mendel} J.~T.,  {Patton} D.~R.,  {Ellison} S.~L.,
  {McConnachie} A.~W.,  2011, \mn@doi [\apjs] {10.1088/0067-0049/196/1/11},
  \href {https://ui.adsabs.harvard.edu/abs/2011ApJS..196...11S} {196, 11}

\bibitem[\protect\citeauthoryear{{Singh}, {Mandelbaum}, {Seljak}, {Slosar}  \&
  {Vazquez Gonzalez}}{{Singh} et~al.}{2017}]{2017_Singh}
{Singh} S.,  {Mandelbaum} R.,  {Seljak} U.,  {Slosar} A.,   {Vazquez Gonzalez}
  J.,  2017, \mn@doi [\mnras] {10.1093/mnras/stx1828}, \href
  {https://ui.adsabs.harvard.edu/abs/2017MNRAS.471.3827S} {471, 3827}

\bibitem[\protect\citeauthoryear{{Somerville} \& {Primack}}{{Somerville} \&
  {Primack}}{1999}]{Somerville1999}
{Somerville} R.~S.,  {Primack} J.~R.,  1999, \mn@doi [\mnras]
  {10.1046/j.1365-8711.1999.03032.x}, \href
  {https://ui.adsabs.harvard.edu/abs/1999MNRAS.310.1087S} {310, 1087}

\bibitem[\protect\citeauthoryear{{Somerville} et~al.,}{{Somerville}
  et~al.}{2018}]{Somerville2018}
{Somerville} R.~S.,  et~al., 2018, \mn@doi [\mnras] {10.1093/mnras/stx2040},
  \href {https://ui.adsabs.harvard.edu/abs/2018MNRAS.473.2714S} {473, 2714}

\bibitem[\protect\citeauthoryear{{Sonnenfeld}, {Wang}  \&
  {Bahcall}}{{Sonnenfeld} et~al.}{2019}]{Sonnenfeld2019}
{Sonnenfeld} A.,  {Wang} W.,   {Bahcall} N.,  2019, \mn@doi [\aap]
  {10.1051/0004-6361/201834260}, \href
  {https://ui.adsabs.harvard.edu/abs/2019A&A...622A..30S} {622, A30}

\bibitem[\protect\citeauthoryear{{Tanaka}}{{Tanaka}}{2015}]{2015_Tanaka}
{Tanaka} M.,  2015, \mn@doi [\apj] {10.1088/0004-637X/801/1/20}, \href
  {https://ui.adsabs.harvard.edu/abs/2015ApJ...801...20T} {801, 20}

\bibitem[\protect\citeauthoryear{{Tanaka} et~al.,}{{Tanaka}
  et~al.}{2018}]{2018_Tanaka}
{Tanaka} M.,  et~al., 2018, \mn@doi [\pasj] {10.1093/pasj/psx077}, \href
  {https://ui.adsabs.harvard.edu/abs/2018PASJ...70S...9T} {70, S9}

\bibitem[\protect\citeauthoryear{{Taylor} et~al.,}{{Taylor}
  et~al.}{2011}]{Taylor2011}
{Taylor} E.~N.,  et~al., 2011, \mn@doi [\mnras]
  {10.1111/j.1365-2966.2011.19536.x}, \href
  {https://ui.adsabs.harvard.edu/abs/2011MNRAS.418.1587T} {418, 1587}

\bibitem[\protect\citeauthoryear{{Trujillo}, {Chamba}  \& {Knapen}}{{Trujillo}
  et~al.}{2020}]{Trujillio2020}
{Trujillo} I.,  {Chamba} N.,   {Knapen} J.~H.,  2020, \mn@doi [\mnras]
  {10.1093/mnras/staa236}, \href
  {https://ui.adsabs.harvard.edu/abs/2020MNRAS.493...87T} {493, 87}

\bibitem[\protect\citeauthoryear{{Vale} \& {Ostriker}}{{Vale} \&
  {Ostriker}}{2004}]{Vale2004}
{Vale} A.,  {Ostriker} J.~P.,  2004, \mn@doi [\mnras]
  {10.1111/j.1365-2966.2004.08059.x}, \href
  {https://ui.adsabs.harvard.edu/abs/2004MNRAS.353..189V} {353, 189}

\bibitem[\protect\citeauthoryear{{\VAN{VanDenBosch}{van den}{van den}}~Bosch,
  {More}, {Cacciato}, {Mo}  \& {Yang}}{{\VAN{VanDenBosch}{van den}{van
  den}}~Bosch et~al.}{2013}]{2013_bosch}
{\VAN{VanDenBosch}{van den}{van den}}~Bosch F.~C.,  {More} S.,  {Cacciato} M.,
  {Mo} H.,   {Yang} X.,  2013, \mn@doi [\mnras] {10.1093/mnras/sts006}, \href
  {https://ui.adsabs.harvard.edu/abs/2013MNRAS.430..725V} {430, 725}

\bibitem[\protect\citeauthoryear{{\VAN{VanDerWel}{van der}{van der}}~Wel
  et~al.,}{{\VAN{VanDerWel}{van der}{van der}}~Wel et~al.}{2012}]{Wel2012}
{\VAN{VanDerWel}{van der}{van der}}~Wel A.,  et~al., 2012, \mn@doi [\apjs]
  {10.1088/0067-0049/203/2/24}, \href
  {https://ui.adsabs.harvard.edu/abs/2012ApJS..203...24V} {203, 24}

\bibitem[\protect\citeauthoryear{{\VAN{VanUitert}{van}{van}}~Uitert
  et~al.,}{{\VAN{VanUitert}{van}{van}}~Uitert et~al.}{2016}]{Uitert2016}
{\VAN{VanUitert}{van}{van}}~Uitert E.,  et~al., 2016, \mn@doi [\mnras]
  {10.1093/mnras/stw747}, \href
  {https://ui.adsabs.harvard.edu/abs/2016MNRAS.459.3251V} {459, 3251}

\bibitem[\protect\citeauthoryear{{V{\'a}zquez-Mata} et~al.,}{{V{\'a}zquez-Mata}
  et~al.}{2020}]{Vazquez-Mata2020}
{V{\'a}zquez-Mata} J.~A.,  et~al., 2020, \mn@doi [\mnras]
  {10.1093/mnras/staa2889}, \href
  {https://ui.adsabs.harvard.edu/abs/2020MNRAS.499..631V} {499, 631}

\bibitem[\protect\citeauthoryear{{Velander} et~al.,}{{Velander}
  et~al.}{2014}]{Valender2014}
{Velander} M.,  et~al., 2014, \mn@doi [\mnras] {10.1093/mnras/stt2013}, \href
  {https://ui.adsabs.harvard.edu/abs/2014MNRAS.437.2111V} {437, 2111}

\bibitem[\protect\citeauthoryear{{Vogelsberger} et~al.,}{{Vogelsberger}
  et~al.}{2014}]{Vogelsberger2014}
{Vogelsberger} M.,  et~al., 2014, \mn@doi [\nat] {10.1038/nature13316}, \href
  {https://ui.adsabs.harvard.edu/abs/2014Natur.509..177V} {509, 177}

\bibitem[\protect\citeauthoryear{{Wang}, {Wang}, {Mo}, {van den Bosch}  \&
  {Yang}}{{Wang} et~al.}{2020}]{Wang2020}
{Wang} E.,  {Wang} H.,  {Mo} H.,  {van den Bosch} F.~C.,   {Yang} X.,  2020,
  \mn@doi [\apj] {10.3847/1538-4357/ab6217}, \href
  {https://ui.adsabs.harvard.edu/abs/2020ApJ...889...37W} {889, 37}

\bibitem[\protect\citeauthoryear{{Wang} et~al.,}{{Wang}
  et~al.}{2021}]{Wang2021}
{Wang} W.,  et~al., 2021, \mn@doi [\apj] {10.3847/1538-4357/ac0e38}, \href
  {https://ui.adsabs.harvard.edu/abs/2021ApJ...919...25W} {919, 25}

\bibitem[\protect\citeauthoryear{{Wechsler} \& {Tinker}}{{Wechsler} \&
  {Tinker}}{2018}]{Wechlser2018}
{Wechsler} R.~H.,  {Tinker} J.~L.,  2018, \mn@doi [\araa]
  {10.1146/annurev-astro-081817-051756}, \href
  {https://ui.adsabs.harvard.edu/abs/2018ARA&A..56..435W} {56, 435}

\bibitem[\protect\citeauthoryear{{White} \& {Frenk}}{{White} \&
  {Frenk}}{1991}]{White1991}
{White} S. D.~M.,  {Frenk} C.~S.,  1991, \mn@doi [\apj] {10.1086/170483}, \href
  {https://ui.adsabs.harvard.edu/abs/1991ApJ...379...52W} {379, 52}

\bibitem[\protect\citeauthoryear{{White} \& {Rees}}{{White} \&
  {Rees}}{1978}]{White1978}
{White} S.~D.~M.,  {Rees} M.~J.,  1978, \mn@doi [\mnras]
  {10.1093/mnras/183.3.341}, \href
  {https://ui.adsabs.harvard.edu/abs/1978MNRAS.183..341W} {183, 341}

\bibitem[\protect\citeauthoryear{Williams}{Williams}{2001}]{Williams2001}
Williams D.,  2001, Weighing the odds : a course in probability and statistics.
Cambridge University Press, Cambridge

\bibitem[\protect\citeauthoryear{{Williams} et~al.,}{{Williams}
  et~al.}{2016}]{Williams2016}
{Williams} R.~P.,  et~al., 2016, \mn@doi [\mnras] {10.1093/mnras/stw2185},
  \href {https://ui.adsabs.harvard.edu/abs/2016MNRAS.463.2746W} {463, 2746}

\bibitem[\protect\citeauthoryear{{Wright} et~al.,}{{Wright}
  et~al.}{2017}]{Wright2017}
{Wright} A.~H.,  et~al., 2017, \mn@doi [\mnras] {10.1093/mnras/stx1149}, \href
  {https://ui.adsabs.harvard.edu/abs/2017MNRAS.470..283W} {470, 283}

\bibitem[\protect\citeauthoryear{{Yang}, {Mo}  \& {van den Bosch}}{{Yang}
  et~al.}{2003}]{Yang2003}
{Yang} X.,  {Mo} H.~J.,   {van den Bosch} F.~C.,  2003, \mn@doi [\mnras]
  {10.1046/j.1365-8711.2003.06254.x}, \href
  {https://ui.adsabs.harvard.edu/abs/2003MNRAS.339.1057Y} {339, 1057}

\bibitem[\protect\citeauthoryear{{Yang}, {Mo}, {van den Bosch}, {Pasquali},
  {Li}  \& {Barden}}{{Yang} et~al.}{2007}]{Yang2007}
{Yang} X.,  {Mo} H.~J.,  {van den Bosch} F.~C.,  {Pasquali} A.,  {Li} C.,
  {Barden} M.,  2007, \mn@doi [\apj] {10.1086/522027}, \href
  {https://ui.adsabs.harvard.edu/abs/2007ApJ...671..153Y} {671, 153}

\bibitem[\protect\citeauthoryear{{Yang}, {Mo}  \& {van den Bosch}}{{Yang}
  et~al.}{2008}]{2008_Yang}
{Yang} X.,  {Mo} H.~J.,   {van den Bosch} F.~C.,  2008, \mn@doi [\apj]
  {10.1086/528954}, \href
  {https://ui.adsabs.harvard.edu/abs/2008ApJ...676..248Y} {676, 248}

\bibitem[\protect\citeauthoryear{{Yang}, {Mo}  \& {van den Bosch}}{{Yang}
  et~al.}{2009}]{2009_Yang}
{Yang} X.,  {Mo} H.~J.,   {van den Bosch} F.~C.,  2009, \mn@doi [\apj]
  {10.1088/0004-637X/695/2/900}, \href
  {https://ui.adsabs.harvard.edu/abs/2009ApJ...695..900Y} {695, 900}

\bibitem[\protect\citeauthoryear{{Zanisi} et~al.,}{{Zanisi}
  et~al.}{2020}]{Zanisi2020}
{Zanisi} L.,  et~al., 2020, \mn@doi [\mnras] {10.1093/mnras/stz3516}, \href
  {https://ui.adsabs.harvard.edu/abs/2020MNRAS.492.1671Z} {492, 1671}

\bibitem[\protect\citeauthoryear{{Zanisi} et~al.,}{{Zanisi}
  et~al.}{2021}]{Zanisi2021}
{Zanisi} L.,  et~al., 2021, \mn@doi [\mnras] {10.1093/mnras/staa3864}, \href
  {https://ui.adsabs.harvard.edu/abs/2021MNRAS.501.4359Z} {501, 4359}

\bibitem[\protect\citeauthoryear{{Zaritsky} \& {Behroozi}}{{Zaritsky} \&
  {Behroozi}}{2023}]{Zaritsky2023}
{Zaritsky} D.,  {Behroozi} P.,  2023, \mn@doi [\mnras]
  {10.1093/mnras/stac3610}, \href
  {https://ui.adsabs.harvard.edu/abs/2023MNRAS.519..871Z} {519, 871}

\bibitem[\protect\citeauthoryear{{Zhang}, {Yang}  \& {Guo}}{{Zhang}
  et~al.}{2022}]{Zhang2022}
{Zhang} Y.,  {Yang} X.,   {Guo} H.,  2022, \mn@doi [\mnras]
  {10.1093/mnras/stac2934}, \href
  {https://ui.adsabs.harvard.edu/abs/2022MNRAS.517.3579Z} {517, 3579}

\bibitem[\protect\citeauthoryear{Zonca, Singer, Lenz, Reinecke, Rosset, Hivon
  \& Gorski}{Zonca et~al.}{2019}]{Healpy}
Zonca A.,  Singer L.,  Lenz D.,  Reinecke M.,  Rosset C.,  Hivon E.,   Gorski
  K.,  2019, \mn@doi [Journal of Open Source Software] {10.21105/joss.01298},
  4, 1298

\bibitem[\protect\citeauthoryear{{Zu} \& {Mandelbaum}}{{Zu} \&
  {Mandelbaum}}{2015}]{Zu2015}
{Zu} Y.,  {Mandelbaum} R.,  2015, \mn@doi [\mnras] {10.1093/mnras/stv2062},
  \href {https://ui.adsabs.harvard.edu/abs/2015MNRAS.454.1161Z} {454, 1161}

\bibitem[\protect\citeauthoryear{{de Graaff}, {Trayford}, {Franx}, {Schaller},
  {Schaye}  \& {van der Wel}}{{de Graaff} et~al.}{2022}]{Graaff2022}
{de Graaff} A.,  {Trayford} J.,  {Franx} M.,  {Schaller} M.,  {Schaye} J.,
  {van der Wel} A.,  2022, \mn@doi [\mnras] {10.1093/mnras/stab3510}, \href
  {https://ui.adsabs.harvard.edu/abs/2022MNRAS.511.2544D} {511, 2544}

\bibitem[\protect\citeauthoryear{{van der Wel} et~al.,}{{van der Wel}
  et~al.}{2023}]{Wel2023}
{van der Wel} A.,  et~al., 2023, \mn@doi [arXiv e-prints]
  {10.48550/arXiv.2307.03264}, \href
  {https://ui.adsabs.harvard.edu/abs/2023arXiv230703264V} {p. arXiv:2307.03264}

\makeatother
\end{thebibliography}




\appendix
\section{Effect of size conversion on galaxy size-halo radius relation}
\begin{figure*}
    \centering
    \includegraphics[width = 0.48\textwidth]{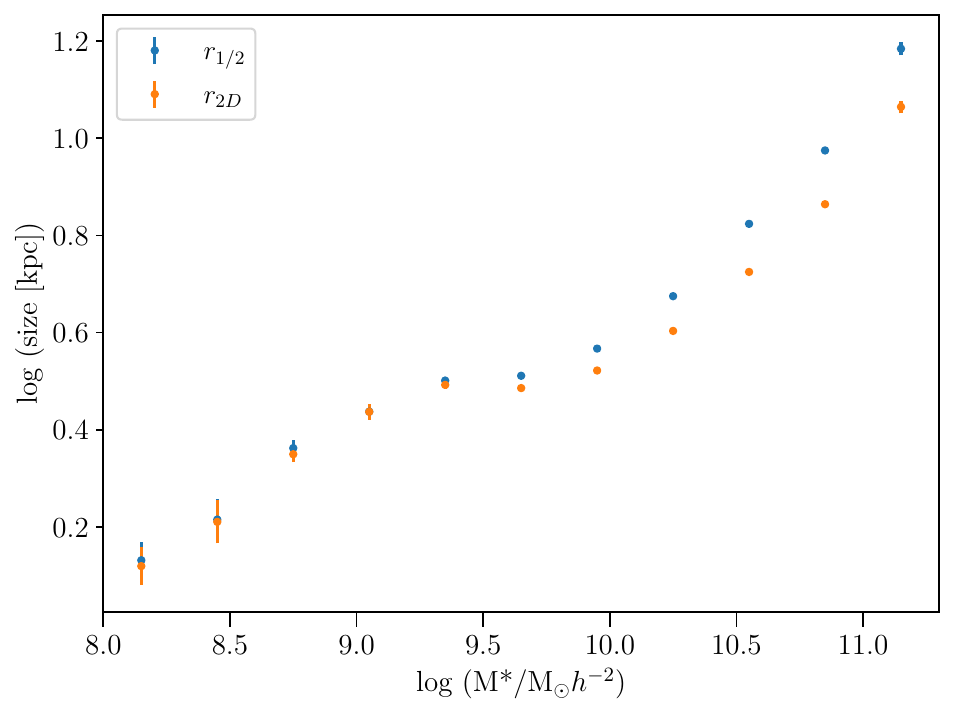}
    \includegraphics[width = 0.48\textwidth]{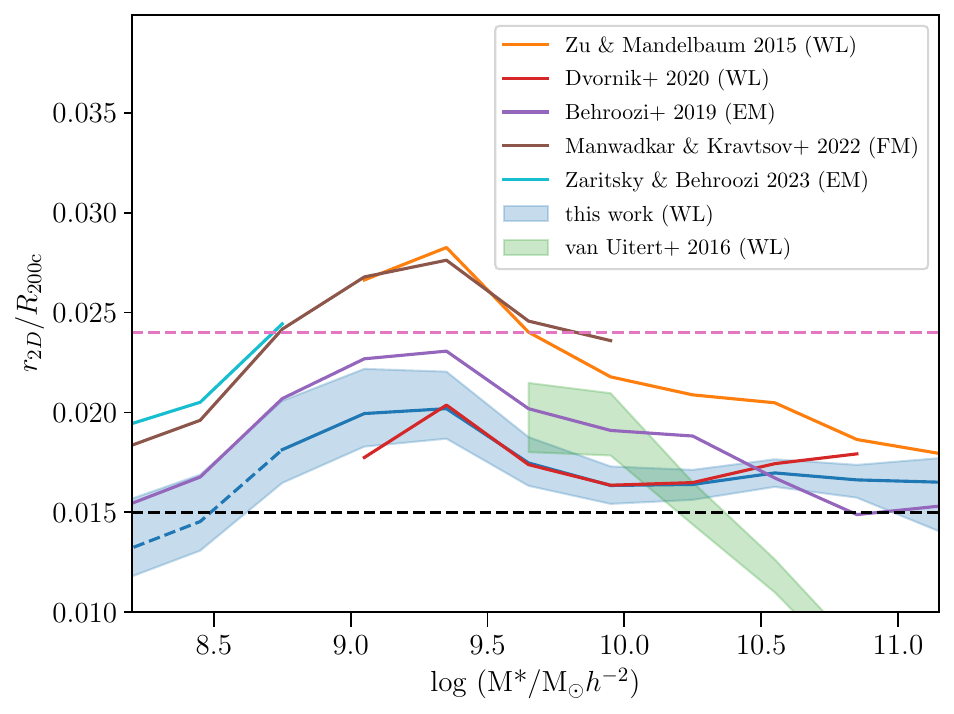}
    
    \caption{\preetish{Left:} Comparison between median 3D vs median 2D galaxy size-mass relation. \preetish{Right:} The stellar mass dependence of $r_{\rm 2D}/R_{\rm 200c}$ computer using our derived and other stellar mass-halo mass relations from the literature. The convention of colours and markers remains the same as in previous figures } 
    \label{2D}
\end{figure*}

\preetish{Here we discuss how the definition of galaxy size may impact the galaxy size-halo radius connection. In this work, we converted the observed sky projected two-dimensional (2D) half-light radius to the three-dimensional (3d) half-light radius ($r_{1/2}$) following the prescription given at the beginning of section \ref{sec4}. The 3D galaxy size-mass relation was combined with our SMHMR to obtain the galaxy size-halo radius relation. In Figure, we show how this relation change when we use 2D half-light radius ($r_{2D}$) as galaxy size. The left panel of Figure \ref{2D} shows a comparison of 3D and 2D size-mass relation. Notice that the inflection point of the size-mass relation at the stellar mass of $\sim 10^{9.35} {\rm M}_{\odot}h^{-2}$ becomes more prominently visible in the 2D case. The right panel of Figure \ref{2D} shows the ratio of galaxy 2D size to halo radius as a function of stellar mass when we use our and other different stellar mass-halo mass relations (SMHMRs) from the literature. With our SMHMR, we find that moving from lowest to highest stellar mass, the ratio $r_{\rm 2D}/R_{\rm 200c}$ first increases then decreases a bit and stays constant thereafter. For a few other SMHMRs, this ratio shows a monotonous decrease with an increase in stellar mass after $\sim 10^{9.35} {\rm M}_{\odot}h^{-2}$.} 

\section{Weak lensing covariance}
\label{apx_wl_cov}
\begin{figure}
    \centering
    \includegraphics[width =\columnwidth]{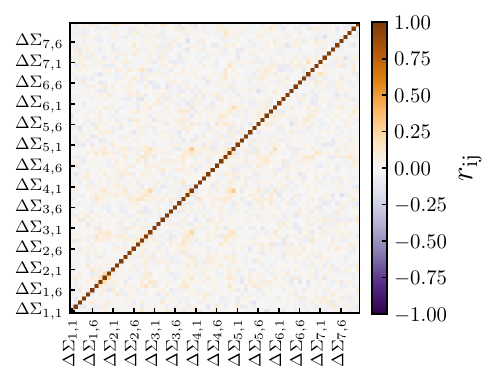}
    \caption{{\it Weak lensing profile covariance:} The above plot shows the correlation coefficient of the shapenoise covariance between the radial bins for seven different stellar mass bins of our lensing sample. The labels $\Delta \Sigma_{m,n}$ indicated the $n^{\rm th}$ radial bin of the $m^{\rm th}$ stellar mass bin. We use 500 times randomly rotations on the source galaxies to compute the shapenoise covariance. } 
    \label{fig:wl_cov}
\end{figure}

In Figure \ref{fig:wl_cov} we shows the correlation coefficient $r_{\rm ij}$ for the shapenoise covariance used in our weak lensing analysis. We calculate $r_{\rm ij}$ using the covariance $C_{\rm ij}$ as

\begin{equation}
    r_{\rm ij} = \frac{C_{\rm ij}}{\sqrt{C_{\rm ii}C_{\rm ij}}}\,,
\end{equation}
where $C_{\rm ij}$ represents the covariance between $i^{\rm th}$ and $j^{\rm th}$ radial bins. The $\Delta\Sigma_{m,n}$ on both x and y axis denotes the $n^{\rm th}$ radial bin of $m^{\rm th}$ stellar mass bin. We describe the details on the estimation of the covariance in Section \ref{sec_esd}.  

We have also estimated the large-scale structure (LSS) covariance using the jackknife technique. For this purpose, we divided the overlapping area between GAMA and HSC S16A into 71 jackknife regions having roughly $ 1.4 \,{\rm deg}^2$ area each. We tested the impact of the large-scale structure (LSS) covariance by rerunning our analysis using only the radial bins where the shape noise dominates the covariance. Within each bin we find these radial ranges by comparing the shape noise covariance with the jackknife estimate. We found that the constraints we obtain agree with those from our fiducial analysis using only the shape noise covariance for all the radial bins. This shows that the LSS contribution to the covariance has no significant impact on our inference. 
 

\section{Systematics for weak gravitational lensing}
\label{apx_null_wl}
\begin{figure}
    \centering
    \includegraphics[width =\columnwidth]{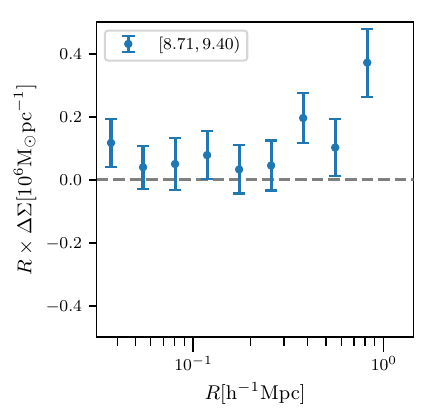}
    \caption{{\it Weak lensing signal around random points:} The blue data points with errors represents the signal around the random points for the stellar mass bin $\log(M^*/{\rm M}_{\odot}h^{-2}) \in (8.71, 9.40)$. The errorbars denotes the error over mean computed using 100 different random realizations.} 
    \label{fig:rand_bin_1}
\end{figure}

We used the cross component of ESD measurement for individual stellar mass bins and agreed well with the expected null signal. We found the signal measurements around the random points also to be consistent with zero in the scales we consider in this work, except for the first stellar mass bin. As shown with blue points in Figure \ref{fig:rand_bin_1}, we observe a systematic signal around the random points for the stellar mass bin $\log(M^*/{\rm M}_{\odot}h^{-2}) \in (8.71, 9.40)$  which increases as we go to larger radial bins. The errors in this case are computed using 100 different random realizations. This systematic also manifests itself in the $\Delta \Sigma$ measurements in this stellar mass bin. In our analysis, we correct for this systematic by subtracting the signal around randoms from the ESD measurements for our lensing sample.

\section{Halo model Parameter degeneracies}
\begin{figure*}
    \centering
    \includegraphics[width =\textwidth]{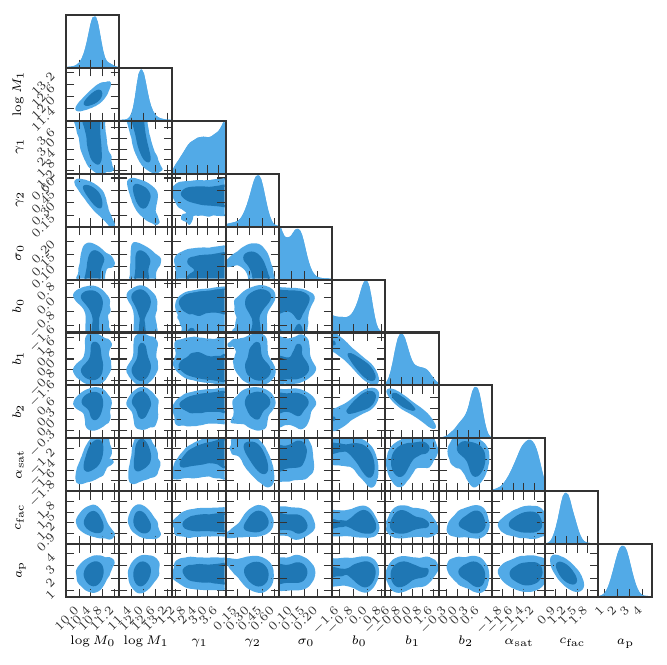}
    \caption{{\it Weak lensing parameter posterior:} The above plot shows the degeneracies in the parameters from the joint modelling of the weak lensing signals around the galaxies split in seven different stellar mass bins.} 
    \label{fig:wl_corner}
\end{figure*}

The degeneracies between the conditional stellar mass function modelling parameters for the weak lensing analysis are shown in Figure \ref{fig:wl_corner}. We also present the correlation between factors scaling the concentration-mass relation $c_{\rm fac}$ and the $a_{\rm p}$ for the baryonic contribution. The individual parameters in our halo model are described in detail in Section \ref{sec_esd}, and constraints along with the assumed priors are provided in Table \ref{tab:modparam}. 
\label{apx_wl_corner}

\bsp	
\label{lastpage}
\end{document}